%% file: main.tex
\documentclass{article}

\usepackage[utf8]{inputenc}
\usepackage[T1]{fontenc}
\usepackage{microtype}
\usepackage{graphicx}
\usepackage{booktabs}
\usepackage{multirow}
\usepackage{array}
\usepackage{amsmath,amssymb,amsthm}
\usepackage{xcolor}
\usepackage{xspace}
\usepackage{etoolbox}
\usepackage{enumitem}
\usepackage{tikz}
\usetikzlibrary{arrows.meta,positioning,fit,backgrounds,calc,shapes.geometric}
\usepackage{xurl}
\usepackage[hidelinks]{hyperref}
\hypersetup{hypertexnames=false}
\usepackage[capitalise,nameinlink]{cleveref}

\usepackage[accepted]{mlsys2025}

\input{macros}
\input{generated/facts}
\factdef{F-T476}{0\%}{testbed}{}
\factdef{F-T487}{16\%}{testbed}{}
\factdef{F-T488}{100\%}{testbed}{}
\factdef{F-T283}{+46~pp}{testbed}{}
\factdef{F-T485}{+68~pp}{testbed}{}
\factdef{F-T477}{+96~pp}{testbed}{}
\factdef{F-T493}{+82~pp}{testbed}{}

\mlsystitlerunning{Airlock: Bounding the Multi-Agent Blast Radius via Selective Online Containment}

\begin{document}
\hypersetup{pdfsubject={}}

\twocolumn[
\mlsystitle{Airlock: Bounding the Multi-Agent Blast Radius\\ via Selective Online Containment}

\mlsyssetsymbol{equal}{*}
\begin{mlsysauthorlist}
\mlsysauthor{Zhonghao Zhan}{equal,icl,oxf}
\mlsysauthor{Javad Forough}{equal,icl}
\mlsysauthor{Yefan Zhang}{equal,ind}
\mlsysauthor{Krinos Li}{icl}
\mlsysauthor{Xiao Ma}{icl}
\mlsysauthor{Hamed Haddadi}{oxf}
\end{mlsysauthorlist}
\mlsysaffiliation{icl}{Imperial College London}
\mlsysaffiliation{oxf}{University of Oxford}
\mlsysaffiliation{ind}{Independent Researcher}
\mlsyskeywords{Machine Learning Systems, Multi-Agent Systems, Containment, Provenance}

\vskip 0.3in
\begin{abstract}
\input{sections/00-abstract}
\end{abstract}
\vskip 0.2in
]

\printAffiliationsAndNotice{\mlsysEqualContribution}

\input{sections/s1-intro}
\input{sections/s2-application}
\input{sections/s3-design}
\input{sections/s4-impl}
\input{sections/s5-eval}
\input{sections/s6-related}

\bibliography{refs}
\IfFileExists{mlsys2025.bst}{\bibliographystyle{mlsys2025}}{\bibliographystyle{plainnat}}

\clearpage
\appendix
\crefalias{section}{appendix}
\renewcommand{\dbltopfraction}{0.95}\setcounter{dbltopnumber}{3}\renewcommand{\dblfloatpagefraction}{0.6}\renewcommand{\topfraction}{0.9}
\counterwithin{figure}{section}\counterwithin{table}{section}

\input{sections/app-a-theory}

\section{Runtime Artifact, Baselines, and Per-Cell Results}
\label{app:baselines}
\input{sections/app-b-baselines}

\section{Pre-registration, Scoring, and Amendments}
\label{app:prereg}
\input{sections/app-c-summary}We recorded each experimental hypothesis and its decision thresholds in the project's internal decision log on the date shown in \cref{tab:hyp}, prior to running the corresponding evaluation grid (no external registry was used).
In the registration records, the system's internal name (``cut mode'' in H15) denotes \OC, ``Trace \& quarantine'' (``quarantine mode'') denotes \FQ, and ``RT-full'' / ``RT-bare'' denote \RT with and without the provenance and containment layer.
\input{sections/app-c-prereg}

\section{Workloads and Detection Delays}
\label{app:workloads}
\input{sections/app-d-workloads}

\section{Simulation}
\label{app:sim}
\input{sections/app-e-simulation}

\section{Cost Model, Scaling, Case Study, and Scope}
\label{app:limitations}
\input{sections/app-f-limitations}

\section{Native Controls of Agent Frameworks}
\label{app:frameworks}
\input{sections/app-e-frameworks}

\end{document}

%% file: macros.tex
\newif\ifdraftnotes
\draftnotesfalse

\definecolor{factgray}{gray}{0.45}
\definecolor{factsim}{RGB}{200,110,0}
\definecolor{factlocal}{RGB}{120,60,160}
\definecolor{facttheory}{RGB}{0,120,60}
\definecolor{tbdred}{RGB}{200,0,0}
\definecolor{noteblue}{RGB}{0,70,170}
\definecolor{mistblue}{RGB}{96,126,149}
\definecolor{mistlight}{RGB}{168,195,214}
\definecolor{sandbrown}{RGB}{176,150,120}
\definecolor{sandtaupe}{RGB}{184,174,166}
\definecolor{sandbeige}{RGB}{226,208,188}
\definecolor{silklite}{RGB}{243,238,232}
\definecolor{inkdark}{RGB}{52,56,60}
\definecolor{inkgray}{RGB}{110,114,118}

\definecolor{mutedgreen}{RGB}{91,127,94}
\definecolor{mutedred}{RGB}{158,82,78}

\colorlet{songred}{mutedred}
\colorlet{songteal}{mistblue}
\colorlet{songgreen}{mutedgreen}
\colorlet{songochre}{sandbrown}
\colorlet{songgold}{sandbrown}
\colorlet{songsand}{sandbeige}
\colorlet{songink}{inkdark}
\colorlet{songtaupe}{sandtaupe}
\colorlet{songsilk}{silklite}
\colorlet{taint}{mutedred}
\colorlet{clean}{inkdark}
\colorlet{ctrl}{mistblue}
\colorlet{quar}{sandbrown}
\colorlet{release}{mutedgreen}

\newcommand{\factdef}[4]{%
  \expandafter\def\csname fact@val@#1\endcsname{#2}%
  \expandafter\def\csname fact@tier@#1\endcsname{#3}%
  \expandafter\def\csname fact@label@#1\endcsname{#4}}
\newcommand{\factv}[1]{%
  \ifcsname fact@val@#1\endcsname\csname fact@val@#1\endcsname
  \else\textcolor{tbdred}{\textbf{??#1}}\PackageWarning{facts}{Undefined fact #1}\fi}
\newcommand{\facttag}[1]{%
  \ifcsname fact@tier@#1\endcsname
    \ifcsstring{fact@label@#1}{pre-calibration}%
      {\textsuperscript{\tiny\textcolor{tbdred}{#1 PRE-CAL}}}%
      {\ifcsstring{fact@label@#1}{pilot}{\textsuperscript{\tiny\textcolor{tbdred}{#1 PILOT}}}%
      {\ifcsstring{fact@tier@#1}{simulation}{\textsuperscript{\tiny\textcolor{factsim}{#1 sim}}}%
        {\ifcsstring{fact@tier@#1}{local}{\textsuperscript{\tiny\textcolor{factlocal}{#1 local}}}%
          {\ifcsstring{fact@tier@#1}{theory-check}{\textsuperscript{\tiny\textcolor{facttheory}{#1}}}%
            {\textsuperscript{\tiny\textcolor{factgray}{#1}}}}}}}%
  \fi}
\newcommand{\fact}[1]{\factv{#1}\ifdraftnotes\facttag{#1}\fi}

\newcommand{\conf}[2]{#1\ifdraftnotes\textsuperscript{\tiny\textcolor{factgray}{#2}}\fi}

\newcommand{\FB}{\textsc{Airlock}\xspace}
\newcommand{\NoC}{No containment\xspace}
\newcommand{\GSafe}{G-Safeguard\xspace}
\newcommand{\SinkG}{Sink guard\xspace}

\definecolor{fbstep}{HTML}{595959}
\DeclareRobustCommand{\circled}[1]{%
  \tikz[baseline=(fbm.base)]\node(fbm)[fill=fbstep,text=white,
    font=\scriptsize\bfseries,inner sep=0pt,minimum size=0.8em]{#1};}

\newcommand{\FQ}{full quarantine\xspace}
\newcommand{\FQc}{Full quarantine\xspace}
\newcommand{\OC}{optimized cut\xspace}
\newcommand{\OCc}{Optimized cut\xspace}

\newcommand{\EGc}{Exit guard\xspace}
\newcommand{\TS}{Trace \& stop\xspace}
\newcommand{\CutND}{Cut without deadlines\xspace}
\newcommand{\Greedy}{Greedy cut\xspace}
\newcommand{\StopTeam}{Stop team\xspace}
\newcommand{\RejSrc}{Reject source\xspace}
\newcommand{\StopDown}{Stop downstream\xspace}

\newcommand{\RT}{\textsc{EdgeCitadel}\xspace}
\newcommand{\HomeSvc}{HomeSvc\xspace}
\newcommand{\sited}{site daemon\xspace}

\newcommand{\GP}{G_P}
\newcommand{\GC}{G_C}
\newcommand{\Act}{\mathcal{A}}
\newcommand{\Rem}{\mathrm{Rem}}

\newcommand{\tone}{\tau_1}
\newcommand{\Ba}{B_a}
\newcommand{\Bone}{B_1}
\newcommand{\Btwo}{B_2}
\newcommand{\Scope}{\mathrm{Scope}}
\newcommand{\Grant}{\mathrm{Grant}}
\newcommand{\taint}{\mathrm{taint}}
\newcommand{\Reach}{\mathrm{Reach}}
\newcommand{\cond}[1]{\textup{(#1)}}

\theoremstyle{plain}
\newtheorem{theorem}{Theorem}
\newtheorem{proposition}[theorem]{Proposition}

\theoremstyle{definition}

\theoremstyle{remark}

\setlist{nosep,leftmargin=*}
\RequirePackage{adjustbox}
\newcommand{\fitcol}[1]{\adjustbox{max width=\columnwidth}{#1}}

%% file: generated/facts.tex
\factdef{F-E001}{6}{scaled-testbed}{}
\factdef{F-E002}{2}{scaled-testbed}{}
\factdef{F-E003}{4}{scaled-testbed}{}
\factdef{F-E004}{10}{scaled-testbed}{}
\factdef{F-E005}{3}{scaled-testbed}{}
\factdef{F-E006}{7}{scaled-testbed}{}
\factdef{F-E007}{2}{scaled-testbed}{}
\factdef{F-E008}{14}{scaled-testbed}{}
\factdef{F-E009}{16}{scaled-testbed}{}
\factdef{F-E010}{3}{scaled-testbed}{}
\factdef{F-E011}{5}{scaled-testbed}{}
\factdef{F-E012}{0}{scaled-testbed}{}
\factdef{F-E013}{5}{scaled-testbed}{}
\factdef{F-E014}{6}{scaled-testbed}{}
\factdef{F-E015}{43}{scaled-testbed}{}
\factdef{F-E016}{0}{scaled-testbed}{}
\factdef{F-E017}{40.8 s}{scaled-testbed}{}
\factdef{F-E018}{74.9 s}{scaled-testbed}{}
\factdef{F-E019}{109.4 s}{scaled-testbed}{}
\factdef{F-E020}{42.1 s}{scaled-testbed}{}
\factdef{F-E021}{71.6 s}{scaled-testbed}{}
\factdef{F-E022}{90.8 s}{scaled-testbed}{}
\factdef{F-E023}{28.6 s}{scaled-testbed}{}
\factdef{F-E024}{83.0 s}{scaled-testbed}{}
\factdef{F-E025}{127.1 s}{scaled-testbed}{}
\factdef{F-E026}{33.2 s}{scaled-testbed}{}
\factdef{F-E027}{67.7 s}{scaled-testbed}{}
\factdef{F-E028}{85.2 s}{scaled-testbed}{}
\factdef{F-E029}{6}{scaled-testbed}{}
\factdef{F-E030}{5}{scaled-testbed}{}
\factdef{F-E031}{7}{scaled-testbed}{}
\factdef{F-E032}{0}{scaled-testbed}{}
\factdef{F-E033}{4}{scaled-testbed}{}
\factdef{F-E034}{2}{scaled-testbed}{}
\factdef{F-E035}{4}{scaled-testbed}{}
\factdef{F-E036}{2}{scaled-testbed}{}
\factdef{F-E037}{2}{scaled-testbed}{}
\factdef{F-E038}{3}{scaled-testbed}{}
\factdef{F-E039}{0}{scaled-testbed}{}
\factdef{F-E040}{0.35 s}{scaled-testbed}{}
\factdef{F-E041}{0.49 s}{scaled-testbed}{}
\factdef{F-E042}{2}{scaled-testbed}{}
\factdef{F-E043}{750 ms}{scaled-testbed}{}
\factdef{F-E044}{90}{scaled-testbed}{}
\factdef{F-E045}{182}{scaled-testbed}{}
\factdef{F-E046}{50.9 s}{scaled-testbed}{}
\factdef{F-E047}{92.8 s}{scaled-testbed}{}
\factdef{F-E048}{119.4 s}{scaled-testbed}{}
\factdef{F-E049}{50.9 s}{scaled-testbed}{}
\factdef{F-E050}{62.6 s}{scaled-testbed}{}
\factdef{F-E051}{0.96}{scaled-testbed}{}
\factdef{F-E052}{21.7 s}{scaled-testbed}{}
\factdef{F-E053}{22.5 s}{scaled-testbed}{}
\factdef{F-E054}{0.38}{scaled-testbed}{}
\factdef{F-E055}{38}{scaled-testbed}{}
\factdef{F-E056}{126}{scaled-testbed}{}
\factdef{F-E057}{96}{scaled-testbed}{}
\factdef{F-E058}{0.48}{scaled-testbed}{}
\factdef{F-E059}{0.56}{scaled-testbed}{}
\factdef{F-E060}{0.43}{scaled-testbed}{}
\factdef{F-E061}{0.68}{scaled-testbed}{}
\factdef{F-E062}{41}{scaled-testbed}{}
\factdef{F-E063}{278}{scaled-testbed}{}
\factdef{F-E064}{37}{scaled-testbed}{}
\factdef{F-E065}{194}{scaled-testbed}{}
\factdef{F-E066}{203}{scaled-testbed}{}
\factdef{F-E067}{245}{scaled-testbed}{}
\factdef{F-E068}{0}{scaled-testbed}{post-hoc}
\factdef{F-E069}{0}{scaled-testbed}{post-hoc}
\factdef{F-E070}{2}{scaled-testbed}{post-hoc}
\factdef{F-E071}{4}{scaled-testbed}{post-hoc}
\factdef{F-E072}{0}{scaled-testbed}{post-hoc}
\factdef{F-E073}{0}{scaled-testbed}{post-hoc}
\factdef{F-E074}{3}{scaled-testbed}{post-hoc}
\factdef{F-E075}{7}{scaled-testbed}{post-hoc}
\factdef{F-E076}{0}{scaled-testbed}{post-hoc}
\factdef{F-E077}{2}{scaled-testbed}{post-hoc}
\factdef{F-E078}{4}{scaled-testbed}{post-hoc}
\factdef{F-E079}{0}{scaled-testbed}{post-hoc}
\factdef{F-E080}{0.49}{scaled-testbed}{}
\factdef{F-E081}{6}{scaled-testbed}{post-hoc}
\factdef{F-E082}{6}{scaled-testbed}{post-hoc}
\factdef{F-E083}{6}{scaled-testbed}{post-hoc}
\factdef{F-E084}{0}{scaled-testbed}{post-hoc}
\factdef{F-E085}{0}{scaled-testbed}{post-hoc}
\factdef{F-E086}{10}{scaled-testbed}{post-hoc}
\factdef{F-E087}{7}{scaled-testbed}{post-hoc}
\factdef{F-E088}{7}{scaled-testbed}{post-hoc}
\factdef{F-E089}{0}{scaled-testbed}{post-hoc}
\factdef{F-E090}{0}{scaled-testbed}{post-hoc}
\factdef{F-E091}{6}{scaled-testbed}{post-hoc}
\factdef{F-E092}{4}{scaled-testbed}{post-hoc}
\factdef{F-E093}{4}{scaled-testbed}{post-hoc}
\factdef{F-E094}{0}{scaled-testbed}{post-hoc}
\factdef{F-E095}{2}{scaled-testbed}{post-hoc}
\factdef{F-E096}{6}{scaled-testbed}{}
\factdef{F-E097}{0}{scaled-testbed}{}
\factdef{F-E098}{6}{scaled-testbed}{}
\factdef{F-E099}{0}{scaled-testbed}{}
\factdef{F-E100}{6}{scaled-testbed}{}
\factdef{F-E101}{1}{scaled-testbed}{}
\factdef{F-E102}{9}{scaled-testbed}{}
\factdef{F-E103}{0}{scaled-testbed}{}
\factdef{F-E104}{0}{scaled-testbed}{}
\factdef{F-E105}{3}{scaled-testbed}{}
\factdef{F-E106}{3}{scaled-testbed}{}
\factdef{F-E107}{0}{scaled-testbed}{}
\factdef{F-E108}{0}{scaled-testbed}{}
\factdef{F-E109}{0.89}{scaled-testbed}{}
\factdef{F-E110}{0.87}{scaled-testbed}{}
\factdef{F-E111}{0.92}{scaled-testbed}{}
\factdef{F-E112}{0.45}{scaled-testbed}{}
\factdef{F-E113}{0.61}{scaled-testbed}{}
\factdef{F-E114}{39}{scaled-testbed}{}
\factdef{F-E115}{308}{scaled-testbed}{}
\factdef{F-E116}{287}{scaled-testbed}{}
\factdef{F-E117}{26.7}{scaled-testbed}{}
\factdef{F-E118}{35.3}{scaled-testbed}{}
\factdef{F-E119}{80.6}{scaled-testbed}{}
\factdef{F-E120}{10}{scaled-testbed}{}
\factdef{F-E121}{0}{scaled-testbed}{}
\factdef{F-E122}{10}{scaled-testbed}{}
\factdef{F-E123}{0}{scaled-testbed}{}
\factdef{F-E124}{10}{scaled-testbed}{}
\factdef{F-E125}{0}{scaled-testbed}{}
\factdef{F-E126}{59}{scaled-testbed}{}
\factdef{F-E127}{0}{scaled-testbed}{}
\factdef{F-E128}{0}{scaled-testbed}{}
\factdef{F-E129}{5}{scaled-testbed}{}
\factdef{F-E130}{5}{scaled-testbed}{}
\factdef{F-E131}{0}{scaled-testbed}{}
\factdef{F-E132}{0}{scaled-testbed}{}
\factdef{F-E133}{0.92}{scaled-testbed}{}
\factdef{F-E134}{0.86}{scaled-testbed}{}
\factdef{F-E135}{1.00}{scaled-testbed}{}
\factdef{F-E136}{0.40}{scaled-testbed}{}
\factdef{F-E137}{0.58}{scaled-testbed}{}
\factdef{F-E138}{37}{scaled-testbed}{}
\factdef{F-E139}{141}{scaled-testbed}{}
\factdef{F-E140}{236}{scaled-testbed}{}
\factdef{F-E141}{32.5}{scaled-testbed}{}
\factdef{F-E142}{70.0}{scaled-testbed}{}
\factdef{F-E143}{90.9}{scaled-testbed}{}
\factdef{F-E144}{6}{scaled-testbed}{}
\factdef{F-E145}{0}{scaled-testbed}{}
\factdef{F-E146}{6}{scaled-testbed}{}
\factdef{F-E147}{0}{scaled-testbed}{}
\factdef{F-E148}{6}{scaled-testbed}{}
\factdef{F-E149}{0}{scaled-testbed}{}
\factdef{F-E150}{11}{scaled-testbed}{}
\factdef{F-E151}{0}{scaled-testbed}{}
\factdef{F-E152}{1}{scaled-testbed}{}
\factdef{F-E153}{4}{scaled-testbed}{}
\factdef{F-E154}{4}{scaled-testbed}{}
\factdef{F-E155}{0}{scaled-testbed}{}
\factdef{F-E156}{0}{scaled-testbed}{}
\factdef{F-E157}{0.98}{scaled-testbed}{}
\factdef{F-E158}{0.93}{scaled-testbed}{}
\factdef{F-E159}{1.00}{scaled-testbed}{}
\factdef{F-E160}{0.43}{scaled-testbed}{}
\factdef{F-E161}{14.59}{scaled-testbed}{}
\factdef{F-E162}{38}{scaled-testbed}{}
\factdef{F-E163}{111}{scaled-testbed}{}
\factdef{F-E164}{284}{scaled-testbed}{}
\factdef{F-E165}{52.4}{scaled-testbed}{}
\factdef{F-E166}{90.8}{scaled-testbed}{}
\factdef{F-E167}{116.0}{scaled-testbed}{}
\factdef{F-E168}{22}{scaled-testbed}{}
\factdef{F-E169}{0}{scaled-testbed}{}
\factdef{F-E170}{0}{scaled-testbed}{}
\factdef{F-E171}{1}{scaled-testbed}{}
\factdef{F-E172}{79}{scaled-testbed}{}
\factdef{F-E173}{0}{scaled-testbed}{}
\factdef{F-E174}{1}{scaled-testbed}{}
\factdef{F-E175}{12}{scaled-testbed}{}
\factdef{F-E176}{0}{scaled-testbed}{}
\factdef{F-E177}{0}{scaled-testbed}{}
\factdef{F-E178}{0.48}{scaled-testbed}{}
\factdef{F-E179}{1}{scaled-testbed}{}
\factdef{F-E180}{20.5 s}{scaled-testbed}{}
\factdef{F-E181}{56.5 s}{scaled-testbed}{}
\factdef{F-E182}{12}{scaled-testbed}{post-hoc}
\factdef{F-E183}{12}{scaled-testbed}{post-hoc}
\factdef{F-E184}{1,690}{scaled-testbed}{post-hoc}
\factdef{F-E185}{1,690}{scaled-testbed}{post-hoc}
\factdef{F-E186}{0}{scaled-testbed}{post-hoc}
\factdef{F-L001}{44 of 45}{local}{}
\factdef{F-L002}{3.3 s}{local}{}
\factdef{F-L003}{102}{local}{}
\factdef{F-L004}{2.4 s}{local}{}
\factdef{F-L005}{0}{local}{}
\factdef{F-L006}{33 of 33}{local}{}
\factdef{F-L007}{12}{local}{dry-run}
\factdef{F-L008}{12}{local}{dry-run}
\factdef{F-L009}{54 ms}{local}{dry-run}
\factdef{F-L010}{905 ms}{local}{dry-run}
\factdef{F-L011}{328 ms}{local}{dry-run}
\factdef{F-L012}{0}{local}{dry-run}
\factdef{F-L013}{8}{local}{dry-run}
\factdef{F-L101}{13 of 13}{local}{}
\factdef{F-L102}{13 of 13}{local}{}
\factdef{F-L103}{0}{local}{}
\factdef{F-L104}{23}{local}{}
\factdef{F-L105}{22}{local}{}
\factdef{F-L106}{8}{local}{}
\factdef{F-L107}{10}{local}{}
\factdef{F-L108}{4}{local}{}
\factdef{F-L109}{3}{local}{}
\factdef{F-L110}{3}{local}{}
\factdef{F-L111}{5–8 ms}{local}{}
\factdef{F-L112}{0.022–0.031 ms}{local}{}
\factdef{F-L113}{59863 ms}{local}{}
\factdef{F-L114}{21}{local}{}
\factdef{F-L115}{22}{local}{}
\factdef{F-L116}{11}{local}{}
\factdef{F-L117}{3}{local}{}
\factdef{F-L118}{3.2\%–9.1\%}{local}{}
\factdef{F-L119}{0.5\%–0.8\%}{local}{}
\factdef{F-L120}{516 ms}{local}{}
\factdef{F-L121}{16646–42433 ms}{local}{}
\factdef{F-M001}{200}{theory-check}{}
\factdef{F-M002}{200}{theory-check}{}
\factdef{F-M003}{300}{theory-check}{}
\factdef{F-M004}{200}{theory-check}{}
\factdef{F-M005}{300}{theory-check}{}
\factdef{F-M006}{200}{theory-check}{}
\factdef{F-M007}{200}{theory-check}{}
\factdef{F-M008}{150}{theory-check}{}
\factdef{F-M009}{150}{theory-check}{}
\factdef{F-M010}{148}{theory-check}{}
\factdef{F-M011}{150}{theory-check}{}
\factdef{F-M012}{2}{theory-check}{}
\factdef{F-M013}{300}{theory-check}{}
\factdef{F-M014}{50}{theory-check}{}
\factdef{F-M015}{400}{theory-check}{}
\factdef{F-M016}{31}{theory-check}{}
\factdef{F-M017}{0}{theory-check}{}
\factdef{F-M018}{26}{theory-check}{}
\factdef{F-M019}{4}{theory-check}{}
\factdef{F-M020}{21}{theory-check}{}
\factdef{F-M021}{0}{theory-check}{}
\factdef{F-M022}{11}{theory-check}{}
\factdef{F-M023}{101}{theory-check}{}
\factdef{F-M024}{1}{theory-check}{}
\factdef{F-M025}{0}{theory-check}{}
\factdef{F-M026}{2}{theory-check}{}
\factdef{F-S001}{0\%}{simulation}{}
\factdef{F-S002}{100\%}{simulation}{}
\factdef{F-S003}{4\%}{simulation}{}
\factdef{F-S004}{62\%}{simulation}{}
\factdef{F-S005}{98–99\%}{simulation}{}
\factdef{F-S006}{80\%}{simulation}{}
\factdef{F-S007}{0}{simulation}{}
\factdef{F-S008}{83.9}{simulation}{}
\factdef{F-S009}{13}{simulation}{}
\factdef{F-S010}{4}{simulation}{}
\factdef{F-S011}{0.5–2.2 pp}{simulation}{}
\factdef{F-S012}{36}{simulation}{}
\factdef{F-S013}{50}{simulation}{}
\factdef{F-S014}{5}{simulation}{}
\factdef{F-S015}{9.7}{simulation}{}
\factdef{F-S016}{12.6}{simulation}{}
\factdef{F-S017}{3.9}{simulation}{}
\factdef{F-S018}{31}{simulation}{}
\factdef{F-S019}{5}{simulation}{}
\factdef{F-S020}{0}{simulation}{}
\factdef{F-S021}{11}{simulation}{}
\factdef{F-S022}{1}{simulation}{}
\factdef{F-S023}{12}{simulation}{}
\factdef{F-S024}{0}{simulation}{}
\factdef{F-S025}{2,206}{simulation}{}
\factdef{F-S026}{8}{simulation}{}
\factdef{F-S027}{2,198}{simulation}{}
\factdef{F-S028}{0}{simulation}{}
\factdef{F-S029}{16.0x}{simulation}{}
\factdef{F-S030}{0}{simulation}{}
\factdef{F-S031}{0}{simulation}{}
\factdef{F-S032}{18–31 pp}{simulation}{}
\factdef{F-S033}{+0.4 to +10.7 pp}{simulation}{}
\factdef{F-S034}{+23.5 pp}{simulation}{}
\factdef{F-S035}{50.7}{simulation}{}
\factdef{F-S036}{6.6}{simulation}{}
\factdef{F-S037}{93.0}{simulation}{}
\factdef{F-S038}{81.2}{simulation}{}
\factdef{F-S039}{32.2}{simulation}{}
\factdef{F-S040}{198.4}{simulation}{}
\factdef{F-S041}{2.5x}{simulation}{}
\factdef{F-S042}{7.6x}{simulation}{}
\factdef{F-S043}{6.2x}{simulation}{}
\factdef{F-S044}{14.0x}{simulation}{}
\factdef{F-S045}{413}{simulation}{}
\factdef{F-S046}{396}{simulation}{}
\factdef{F-S047}{17}{simulation}{}
\factdef{F-S048}{0}{simulation}{}
\factdef{F-S049}{93.3\%}{simulation}{}
\factdef{F-S050}{96.5\%}{simulation}{}
\factdef{F-S051}{72.5\%}{simulation}{}
\factdef{F-S052}{88.8\%}{simulation}{}
\factdef{F-S053}{37\%}{simulation}{}
\factdef{F-S054}{97\%}{simulation}{}
\factdef{F-S055}{87\%}{simulation}{}
\factdef{F-S056}{51\%}{simulation}{}
\factdef{F-S057}{8.5 s}{simulation}{}
\factdef{F-S058}{10.4}{simulation}{}
\factdef{F-S059}{17.5}{simulation}{}
\factdef{F-S060}{9.6x}{simulation}{}
\factdef{F-S061}{105.1}{simulation}{}
\factdef{F-S062}{273.5}{simulation}{}
\factdef{F-S063}{4.3}{simulation}{}
\factdef{F-S064}{6.3}{simulation}{}
\factdef{F-S065}{84.5}{simulation}{}
\factdef{F-S066}{19.7x}{simulation}{}
\factdef{F-S067}{24.4x}{simulation}{}
\factdef{F-S068}{43.4x}{simulation}{}
\factdef{F-S069}{99.2\%}{simulation}{}
\factdef{F-S070}{92.9\%}{simulation}{}
\factdef{F-S071}{75\%}{simulation}{}
\factdef{F-S072}{97.5\%}{simulation}{}
\factdef{F-S073}{88.6\%}{simulation}{}
\factdef{F-S074}{79.4\%}{simulation}{}
\factdef{F-S075}{66.3\%}{simulation}{}
\factdef{F-S076}{90–98\%}{simulation}{}
\factdef{F-S077}{100\%}{simulation}{}
\factdef{F-S078}{92.4\%}{simulation}{}
\factdef{F-S079}{240}{simulation}{}
\factdef{F-S080}{19}{simulation}{}
\factdef{F-S081}{71.9\%}{simulation}{}
\factdef{F-S082}{310}{simulation}{}
\factdef{F-S083}{8,164}{simulation}{}
\factdef{F-S084}{311}{simulation}{}
\factdef{F-S085}{371}{simulation}{}
\factdef{F-S086}{7.1 ms}{simulation}{}
\factdef{F-S087}{15.5 ms}{simulation}{}
\factdef{F-S088}{23 ms}{simulation}{}
\factdef{F-S089}{1,639}{simulation}{}
\factdef{F-S090}{97.8\%}{simulation}{}
\factdef{F-S091}{80.3\%}{simulation}{}
\factdef{F-S092}{0}{simulation}{}
\factdef{F-S093}{100\%}{simulation}{}
\factdef{F-S094}{0\%}{simulation}{}
\factdef{F-S095}{0.083\%}{simulation}{}
\factdef{F-S096}{100\%}{simulation}{}
\factdef{F-S097}{28,964}{simulation}{}
\factdef{F-S098}{24}{simulation}{}
\factdef{F-S099}{0.087 s}{simulation}{}
\factdef{F-S100}{165}{simulation}{}
\factdef{F-S101}{145}{simulation}{}
\factdef{F-S102}{5.8 ms}{simulation}{}
\factdef{F-S103}{8.1 ms}{simulation}{}
\factdef{F-S104}{14 ms}{simulation}{}
\factdef{F-S105}{2,000}{simulation}{}
\factdef{F-S106}{2,000}{simulation}{}
\factdef{F-S107}{0}{simulation}{post-hoc}
\factdef{F-S108}{0\%}{simulation}{post-hoc}
\factdef{F-S109}{100\%}{simulation}{post-hoc}
\factdef{F-S110}{13,556}{simulation}{post-hoc}
\factdef{F-S111}{120}{simulation}{post-hoc}
\factdef{F-S112}{+0.91}{simulation}{post-hoc}
\factdef{F-S113}{+0.48}{simulation}{post-hoc}
\factdef{F-S114}{+1.34}{simulation}{post-hoc}
\factdef{F-S115}{17}{simulation}{post-hoc}
\factdef{F-S116}{0}{simulation}{post-hoc}
\factdef{F-S117}{$-$0.76}{simulation}{post-hoc}
\factdef{F-S118}{$-$1.00}{simulation}{post-hoc}
\factdef{F-S119}{$-$0.52}{simulation}{post-hoc}
\factdef{F-S120}{6}{simulation}{post-hoc}
\factdef{F-S121}{82}{simulation}{post-hoc}
\factdef{F-S122}{1,860}{simulation}{post-hoc}
\factdef{F-S123}{48}{simulation}{post-hoc}
\factdef{F-S124}{1}{simulation}{post-hoc}
\factdef{F-S125}{$-$0.52}{simulation}{post-hoc}
\factdef{F-S126}{+0.63}{simulation}{post-hoc}
\factdef{F-S127}{2,700}{simulation}{post-hoc}
\factdef{F-S128}{840}{simulation}{post-hoc}
\factdef{F-S129}{1}{simulation}{post-hoc}
\factdef{F-S130}{+3.25}{simulation}{post-hoc}
\factdef{F-S131}{13}{simulation}{post-hoc}
\factdef{F-S132}{$-$1.06}{simulation}{post-hoc}
\factdef{F-S133}{23}{simulation}{post-hoc}
\factdef{F-S134}{$-$6.2}{simulation}{post-hoc}
\factdef{F-S135}{$-$5.3}{simulation}{post-hoc}
\factdef{F-S136}{839}{simulation}{post-hoc}
\factdef{F-S137}{$-$0.12}{simulation}{post-hoc}
\factdef{F-S138}{+0.26}{simulation}{post-hoc}
\factdef{F-S139}{120}{simulation}{post-hoc}
\factdef{F-S140}{120}{simulation}{post-hoc}
\factdef{F-S141}{2,723}{simulation}{post-hoc}
\factdef{F-S142}{2,726}{simulation}{post-hoc}
\factdef{F-S143}{18,920}{simulation}{}
\factdef{F-S144}{5,346}{simulation}{}
\factdef{F-S145}{24,266}{simulation}{}
\factdef{F-T001}{18 ms}{testbed}{}
\factdef{F-T002}{82 ms}{testbed}{}
\factdef{F-T003}{44 ms}{testbed}{}
\factdef{F-T004}{117 ms}{testbed}{}
\factdef{F-T005}{33–75 ms}{testbed}{pre-calibration}
\factdef{F-T006}{47–161 ms}{testbed}{pre-calibration}
\factdef{F-T007}{53–173 ms}{testbed}{pre-calibration}
\factdef{F-T008}{4 of 4}{testbed}{pre-calibration}
\factdef{F-T009}{146–155 ms}{testbed}{pre-calibration}
\factdef{F-T010}{1 of 1}{testbed}{pre-calibration}
\factdef{F-T011}{0.43 ms}{testbed}{}
\factdef{F-T012}{0.90 ms}{testbed}{}
\factdef{F-T013}{1.6 ms}{testbed}{}
\factdef{F-T014}{1.4 ms}{testbed}{}
\factdef{F-T015}{15–105 ms}{testbed}{pre-calibration}
\factdef{F-T016}{35–170 ms}{testbed}{pre-calibration}
\factdef{F-T017}{124 ms}{testbed}{}
\factdef{F-T020}{169}{testbed}{pre-calibration}
\factdef{F-T021}{14}{testbed}{pre-calibration}
\factdef{F-T022}{14}{testbed}{pre-calibration}
\factdef{F-T023}{0.31 s}{testbed}{pre-calibration}
\factdef{F-T024}{5}{testbed}{pre-calibration}
\factdef{F-T025}{12}{testbed}{pre-calibration}
\factdef{F-T026}{2}{testbed}{pre-calibration}
\factdef{F-T027}{22}{testbed}{pre-calibration}
\factdef{F-T028}{29}{testbed}{pre-calibration}
\factdef{F-T029}{24}{testbed}{pre-calibration}
\factdef{F-T031}{133 ms}{testbed}{}
\factdef{F-T032}{398 ms}{testbed}{}
\factdef{F-T033}{49 ms}{testbed}{}
\factdef{F-T034}{884 ms}{testbed}{}
\factdef{F-T035}{169 of 169}{testbed}{pre-calibration}
\factdef{F-T036}{0}{testbed}{pre-calibration}
\factdef{F-T037}{84}{testbed}{pre-calibration}
\factdef{F-T038}{84}{testbed}{pre-calibration}
\factdef{F-T040}{63}{testbed}{pre-calibration}
\factdef{F-T041}{0}{testbed}{pre-calibration}
\factdef{F-T042}{6}{testbed}{pre-calibration}
\factdef{F-T043}{6}{testbed}{pre-calibration}
\factdef{F-T044}{100\%}{testbed}{pre-calibration}
\factdef{F-T045}{0}{testbed}{pre-calibration}
\factdef{F-T046}{0}{testbed}{pre-calibration}
\factdef{F-T047}{0}{testbed}{pre-calibration}
\factdef{F-T048}{0}{testbed}{pre-calibration}
\factdef{F-T050}{21}{testbed}{pre-calibration}
\factdef{F-T051}{0}{testbed}{pre-calibration}
\factdef{F-T052}{33 of 33}{testbed}{pre-calibration}
\factdef{F-T053}{0}{testbed}{pre-calibration}
\factdef{F-T054}{0}{testbed}{pre-calibration}
\factdef{F-T055}{43}{testbed}{pre-calibration}
\factdef{F-T056}{0}{testbed}{pre-calibration}
\factdef{F-T057}{45 of 45}{testbed}{pre-calibration}
\factdef{F-T058}{1.78 s}{testbed}{pre-calibration}
\factdef{F-T059}{0.29 s}{testbed}{pre-calibration}
\factdef{F-T060}{9}{testbed}{pre-calibration}
\factdef{F-T061}{20}{testbed}{pre-calibration}
\factdef{F-T062}{13}{testbed}{pre-calibration}
\factdef{F-T063}{0}{testbed}{pre-calibration}
\factdef{F-T064}{45 of 45}{testbed}{pre-calibration}
\factdef{F-T065}{33 of 33}{testbed}{pre-calibration}
\factdef{F-T066}{23}{testbed}{pre-calibration}
\factdef{F-T067}{23}{testbed}{pre-calibration}
\factdef{F-T068}{1.42 s}{testbed}{pre-calibration}
\factdef{F-T069}{0.25 s}{testbed}{pre-calibration}
\factdef{F-T070}{0}{testbed}{pre-calibration}
\factdef{F-T071}{0}{testbed}{pre-calibration}
\factdef{F-T072}{0}{testbed}{pre-calibration}
\factdef{F-T073}{45 ms}{testbed}{pilot}
\factdef{F-T074}{187 ms}{testbed}{pilot}
\factdef{F-T075}{187 ms}{testbed}{pilot}
\factdef{F-T076}{55 ms}{testbed}{pilot}
\factdef{F-T077}{329 ms}{testbed}{pilot}
\factdef{F-T078}{347 ms}{testbed}{pilot}
\factdef{F-T079}{58 ms}{testbed}{pilot}
\factdef{F-T080}{307 ms}{testbed}{pilot}
\factdef{F-T081}{321 ms}{testbed}{pilot}
\factdef{F-T082}{338 ms}{testbed}{pilot}
\factdef{F-T083}{11 ms}{testbed}{pilot}
\factdef{F-T084}{30.7$\times$}{testbed}{pilot}
\factdef{F-T085}{339 ms}{testbed}{pilot}
\factdef{F-T086}{3,212}{testbed}{pilot}
\factdef{F-T087}{9.4 ms}{testbed}{pilot}
\factdef{F-T088}{14.9 ms}{testbed}{pilot}
\factdef{F-T089}{132}{testbed}{pilot}
\factdef{F-T090}{0}{testbed}{pilot}
\factdef{F-T091}{18}{testbed}{pilot}
\factdef{F-T092}{0}{testbed}{pilot}
\factdef{F-T093}{40}{testbed}{pilot}
\factdef{F-T094}{40}{testbed}{pilot}
\factdef{F-T095}{40}{testbed}{pilot}
\factdef{F-T096}{0.20 s}{testbed}{pilot}
\factdef{F-T097}{0.95 s}{testbed}{pilot}
\factdef{F-T098}{33}{testbed}{pilot}
\factdef{F-T099}{33}{testbed}{pilot}
\factdef{F-T100}{134}{testbed}{pilot}
\factdef{F-T101}{135}{testbed}{pilot}
\factdef{F-T102}{99}{testbed}{pilot}
\factdef{F-T103}{99}{testbed}{pilot}
\factdef{F-T104}{0}{testbed}{pilot}
\factdef{F-T105}{88}{testbed}{pilot}
\factdef{F-T106}{99}{testbed}{pilot}
\factdef{F-T107}{0}{testbed}{pilot}
\factdef{F-T108}{84}{testbed}{pilot}
\factdef{F-T109}{39}{testbed}{pilot}
\factdef{F-T110}{27}{testbed}{pilot}
\factdef{F-T111}{154}{testbed}{pilot}
\factdef{F-T112}{65}{testbed}{pilot}
\factdef{F-T113}{370}{testbed}{pilot}
\factdef{F-T114}{277}{testbed}{pilot}
\factdef{F-T115}{462}{testbed}{pilot}
\factdef{F-T116}{1,486}{testbed}{pilot}
\factdef{F-T117}{584 ms}{testbed}{pilot}
\factdef{F-T118}{9,013 ms}{testbed}{pilot}
\factdef{F-T119}{0}{testbed}{pilot}
\factdef{F-T120}{0}{testbed}{pilot}
\factdef{F-T121}{621}{testbed}{}
\factdef{F-T122}{54}{testbed}{}
\factdef{F-T123}{54}{testbed}{}
\factdef{F-T124}{54}{testbed}{}
\factdef{F-T125}{54}{testbed}{}
\factdef{F-T126}{54}{testbed}{}
\factdef{F-T127}{54}{testbed}{}
\factdef{F-T128}{54}{testbed}{}
\factdef{F-T129}{54}{testbed}{}
\factdef{F-T130}{15}{testbed}{}
\factdef{F-T131}{12}{testbed}{}
\factdef{F-T132}{3}{testbed}{}
\factdef{F-T133}{0}{testbed}{}
\factdef{F-T134}{162}{testbed}{}
\factdef{F-T135}{0}{testbed}{}
\factdef{F-T136}{93}{testbed}{}
\factdef{F-T137}{129}{testbed}{}
\factdef{F-T138}{0.18 s}{testbed}{}
\factdef{F-T139}{0.55 s}{testbed}{}
\factdef{F-T140}{3.50 s}{testbed}{}
\factdef{F-T141}{0.098 s}{testbed}{}
\factdef{F-T142}{12}{testbed}{}
\factdef{F-T143}{12}{testbed}{}
\factdef{F-T144}{12}{testbed}{}
\factdef{F-T145}{16}{testbed}{}
\factdef{F-T146}{5}{testbed}{}
\factdef{F-T147}{1}{testbed}{}
\factdef{F-T148}{9}{testbed}{}
\factdef{F-T149}{2}{testbed}{}
\factdef{F-T150}{0}{testbed}{}
\factdef{F-T151}{48}{testbed}{}
\factdef{F-T152}{+4.3 pp}{testbed}{}
\factdef{F-T153}{$-$1.8}{testbed}{}
\factdef{F-T154}{+10.6}{testbed}{}
\factdef{F-T155}{0.185}{testbed}{}
\factdef{F-T156}{3}{testbed}{}
\factdef{F-T157}{2}{testbed}{}
\factdef{F-T158}{11}{testbed}{}
\factdef{F-T159}{2}{testbed}{}
\factdef{F-T160}{2}{testbed}{}
\factdef{F-T161}{11}{testbed}{}
\factdef{F-T162}{3}{testbed}{}
\factdef{F-T163}{2299 ms}{testbed}{}
\factdef{F-T164}{676 ms}{testbed}{}
\factdef{F-T165}{356 ms}{testbed}{}
\factdef{F-T166}{204}{testbed}{}
\factdef{F-T167}{6,801}{testbed}{}
\factdef{F-T168}{233}{testbed}{}
\factdef{F-T169}{2}{testbed}{}
\factdef{F-T170}{1}{testbed}{}
\factdef{F-T171}{0}{testbed}{}
\factdef{F-T172}{10}{testbed}{}
\factdef{F-T173}{10}{testbed}{}
\factdef{F-T174}{0}{testbed}{}
\factdef{F-T175}{0}{testbed}{}
\factdef{F-T176}{0}{testbed}{}
\factdef{F-T177}{0}{testbed}{}
\factdef{F-T178}{25}{testbed}{}
\factdef{F-T179}{50}{testbed}{}
\factdef{F-T180}{50}{testbed}{}
\factdef{F-T181}{136 ms}{testbed}{}
\factdef{F-T182}{1,149 ms}{testbed}{}
\factdef{F-T183}{10}{testbed}{}
\factdef{F-T184}{10}{testbed}{}
\factdef{F-T185}{10}{testbed}{}
\factdef{F-T186}{0}{testbed}{}
\factdef{F-T187}{0}{testbed}{}
\factdef{F-T188}{0}{testbed}{}
\factdef{F-T189}{100.0\%}{testbed}{}
\factdef{F-T190}{54.0\%}{testbed}{}
\factdef{F-T191}{25.4\%}{testbed}{}
\factdef{F-T192}{3}{testbed}{}
\factdef{F-T193}{54}{testbed}{}
\factdef{F-T194}{54}{testbed}{}
\factdef{F-T195}{54}{testbed}{}
\factdef{F-T196}{3.0\%}{testbed}{}
\factdef{F-T197}{+51.5}{testbed}{}
\factdef{F-T198}{$-$14.8}{testbed}{}
\factdef{F-T199}{$-$30.9}{testbed}{}
\factdef{F-T200}{3}{testbed}{}
\factdef{F-T201}{299 ms}{testbed}{}
\factdef{F-T202}{11.7 s}{testbed}{}
\factdef{F-T203}{12.6 s}{testbed}{}
\factdef{F-T204}{13.7 s}{testbed}{}
\factdef{F-T205}{19.5 s}{testbed}{}
\factdef{F-T206}{0.13}{testbed}{}
\factdef{F-T207}{$-$0.03}{testbed}{}
\factdef{F-T208}{0.28}{testbed}{}
\factdef{F-T209}{172}{testbed}{}
\factdef{F-T210}{$-$0.05}{testbed}{}
\factdef{F-T211}{64}{testbed}{}
\factdef{F-T212}{0.59}{testbed}{}
\factdef{F-T213}{0.33}{testbed}{}
\factdef{F-T214}{0.28}{testbed}{}
\factdef{F-T215}{57}{testbed}{}
\factdef{F-T216}{51}{testbed}{}
\factdef{F-T217}{4}{testbed}{}
\factdef{F-T218}{6.3}{testbed}{}
\factdef{F-T219}{19}{testbed}{}
\factdef{F-T220}{0.42}{testbed}{}
\factdef{F-T221}{1.3}{testbed}{}
\factdef{F-T222}{4.4}{testbed}{}
\factdef{F-T223}{18}{testbed}{}
\factdef{F-T224}{811}{testbed}{}
\factdef{F-T225}{270 s}{testbed}{}
\factdef{F-T226}{846 s}{testbed}{}
\factdef{F-T227}{32\%}{testbed}{}
\factdef{F-T228}{51}{testbed}{}
\factdef{F-T229}{0.7}{testbed}{model-calculation}
\factdef{F-T230}{0.72}{testbed}{model-calculation}
\factdef{F-T231}{527}{testbed}{model-calculation}
\factdef{F-T232}{0}{testbed}{model-calculation}
\factdef{F-T233}{72}{testbed}{model-calculation}
\factdef{F-T234}{110}{testbed}{model-calculation}
\factdef{F-T235}{115}{testbed}{}
\factdef{F-T236}{110}{testbed}{}
\factdef{F-T237}{0}{testbed}{}
\factdef{F-T238}{0}{testbed}{}
\factdef{F-T239}{45}{testbed}{}
\factdef{F-T240}{45}{testbed}{}
\factdef{F-T241}{9.0}{testbed}{}
\factdef{F-T242}{9.0}{testbed}{}
\factdef{F-T243}{1}{testbed}{}
\factdef{F-T244}{1}{testbed}{}
\factdef{F-T245}{5}{testbed}{}
\factdef{F-T246}{35.6}{testbed}{}
\factdef{F-T247}{32.6}{testbed}{}
\factdef{F-T248}{0}{testbed}{}
\factdef{F-T249}{0}{testbed}{}
\factdef{F-T250}{55}{testbed}{}
\factdef{F-T251}{45}{testbed}{}
\factdef{F-T252}{13}{testbed}{}
\factdef{F-T253}{13}{testbed}{}
\factdef{F-T254}{10}{testbed}{}
\factdef{F-T255}{10}{testbed}{}
\factdef{F-T256}{0}{testbed}{}
\factdef{F-T257}{20}{testbed}{}
\factdef{F-T258}{10}{testbed}{}
\factdef{F-T259}{10}{testbed}{}
\factdef{F-T260}{10}{testbed}{}
\factdef{F-T261}{0}{testbed}{}
\factdef{F-T262}{0}{testbed}{}
\factdef{F-T263}{54.0\%}{testbed}{}
\factdef{F-T264}{54.0\%}{testbed}{}
\factdef{F-T265}{54.0\%}{testbed}{}
\factdef{F-T266}{100.0\%}{testbed}{}
\factdef{F-T267}{92.4\%}{testbed}{}
\factdef{F-T268}{100.0\%}{testbed}{}
\factdef{F-T269}{94.3\%}{testbed}{}
\factdef{F-T270}{100.0\%}{testbed}{}
\factdef{F-T271}{20}{testbed}{}
\factdef{F-T272}{+0.235}{testbed}{}
\factdef{F-T273}{+0.133}{testbed}{}
\factdef{F-T274}{+0.335}{testbed}{}
\factdef{F-T275}{12}{testbed}{}
\factdef{F-T276}{7}{testbed}{}
\factdef{F-T277}{1}{testbed}{}
\factdef{F-T278}{0.0010}{testbed}{}
\factdef{F-T279}{+0.010}{testbed}{}
\factdef{F-T280}{$-$0.035}{testbed}{}
\factdef{F-T281}{+0.051}{testbed}{}
\factdef{F-T282}{1.00}{testbed}{}
\factdef{F-T283}{+0.460}{testbed}{}
\factdef{F-T284}{0.0019}{testbed}{}
\factdef{F-T285}{10}{testbed}{}
\factdef{F-T286}{16}{testbed}{}
\factdef{F-T287}{16}{testbed}{}
\factdef{F-T288}{21.5}{testbed}{}
\factdef{F-T289}{30}{testbed}{}
\factdef{F-T290}{0.10 s}{testbed}{}
\factdef{F-T291}{0.08 s}{testbed}{}
\factdef{F-T292}{0.33 s}{testbed}{}
\factdef{F-T293}{0.21 s}{testbed}{}
\factdef{F-T294}{36}{testbed}{}
\factdef{F-T295}{35}{testbed}{}
\factdef{F-T296}{199.5 ms}{testbed}{}
\factdef{F-T297}{179 ms}{testbed}{}
\factdef{F-T298}{649 ms}{testbed}{}
\factdef{F-T299}{0.046 s}{testbed}{}
\factdef{F-T300}{12}{testbed}{}
\factdef{F-T301}{12}{testbed}{}
\factdef{F-T302}{0}{testbed}{}
\factdef{F-T303}{0}{testbed}{}
\factdef{F-T304}{1.0 s}{testbed}{}
\factdef{F-T305}{24}{testbed}{}
\factdef{F-T306}{0}{testbed}{}
\factdef{F-T307}{0}{testbed}{}
\factdef{F-T308}{0.500}{testbed}{}
\factdef{F-T309}{0.499}{testbed}{}
\factdef{F-T310}{42}{testbed}{}
\factdef{F-T311}{0}{testbed}{}
\factdef{F-T312}{12}{testbed}{}
\factdef{F-T313}{42}{testbed}{}
\factdef{F-T314}{0\%}{testbed}{}
\factdef{F-T315}{42}{testbed}{}
\factdef{F-T316}{100\%}{testbed}{}
\factdef{F-T317}{72.6\%}{testbed}{}
\factdef{F-T318}{36}{testbed}{}
\factdef{F-T319}{2.3e-13}{testbed}{}
\factdef{F-T320}{9.3e-10}{testbed}{}
\factdef{F-T321}{+0.812}{testbed}{}
\factdef{F-T322}{0.000}{testbed}{}
\factdef{F-T323}{+0.0291}{testbed}{}
\factdef{F-T324}{+2.9 pp}{testbed}{}
\factdef{F-T325}{0.46}{testbed}{}
\factdef{F-T326}{3}{testbed}{}
\factdef{F-T327}{15}{testbed}{}
\factdef{F-T328}{9}{testbed}{}
\factdef{F-T329}{12}{testbed}{}
\factdef{F-T330}{42}{testbed}{}
\factdef{F-T331}{0}{testbed}{}
\factdef{F-T332}{0}{testbed}{}
\factdef{F-T333}{10}{testbed}{}
\factdef{F-T334}{0\%}{testbed}{}
\factdef{F-T335}{100\%}{testbed}{}
\factdef{F-T336}{5.4e-6}{testbed}{}
\factdef{F-T337}{1,653}{testbed}{}
\factdef{F-T338}{1,535}{testbed}{}
\factdef{F-T339}{10}{testbed}{}
\factdef{F-T340}{9}{testbed}{}
\factdef{F-T341}{44.6}{testbed}{}
\factdef{F-T342}{45}{testbed}{}
\factdef{F-T343}{32.6}{testbed}{}
\factdef{F-T344}{33}{testbed}{}
\factdef{F-T345}{43}{testbed}{}
\factdef{F-T346}{32}{testbed}{}
\factdef{F-T347}{0}{testbed}{}
\factdef{F-T348}{0}{testbed}{}
\factdef{F-T349}{0.19 s}{testbed}{}
\factdef{F-T350}{0.88 s}{testbed}{}
\factdef{F-T351}{10}{testbed}{}
\factdef{F-T352}{5}{testbed}{}
\factdef{F-T353}{48\%}{testbed}{}
\factdef{F-T354}{2}{testbed}{}
\factdef{F-T355}{2}{testbed}{}
\factdef{F-T356}{0.20}{testbed}{}
\factdef{F-T357}{2}{testbed}{}
\factdef{F-T358}{1}{testbed}{}
\factdef{F-T359}{76}{testbed}{}
\factdef{F-T360}{81}{testbed}{}
\factdef{F-T361}{15}{testbed}{}
\factdef{F-T362}{7}{testbed}{}
\factdef{F-T363}{13}{testbed}{}
\factdef{F-T364}{7.6 s}{testbed}{}
\factdef{F-T365}{1}{testbed}{}
\factdef{F-T366}{20}{testbed}{}
\factdef{F-T367}{2.4 s}{testbed}{}
\factdef{F-T368}{704}{testbed}{}
\factdef{F-T369}{0.088 s}{testbed}{}
\factdef{F-T370}{1}{testbed}{}
\factdef{F-T371}{0.087 s}{testbed}{}
\factdef{F-T372}{37}{testbed}{}
\factdef{F-T373}{1}{testbed}{}
\factdef{F-T374}{13}{testbed}{}
\factdef{F-T375}{0.689 s}{testbed}{}
\factdef{F-T376}{4}{testbed}{}
\factdef{F-T377}{0}{testbed}{}
\factdef{F-T378}{4.5 s}{testbed}{}
\factdef{F-T379}{4.56 s}{testbed}{}
\factdef{F-T380}{0.21 s}{testbed}{}
\factdef{F-T381}{16}{testbed}{}
\factdef{F-T382}{26.5}{testbed}{}
\factdef{F-T383}{12}{testbed}{}
\factdef{F-T384}{12}{testbed}{}
\factdef{F-T385}{17}{testbed}{}
\factdef{F-T386}{25.5}{testbed}{}
\factdef{F-T387}{21}{testbed}{}
\factdef{F-T388}{34}{testbed}{}
\factdef{F-T389}{0.098 s}{testbed}{}
\factdef{F-T390}{0.087 s}{testbed}{}
\factdef{F-T391}{0.18 s}{testbed}{}
\factdef{F-T392}{54}{testbed}{}
\factdef{F-T393}{0}{testbed}{}
\factdef{F-T394}{84}{testbed}{}
\factdef{F-T395}{87}{testbed}{}
\factdef{F-T396}{33}{testbed}{}
\factdef{F-T397}{36}{testbed}{}
\factdef{F-T398}{0.93 s}{testbed}{}
\factdef{F-T399}{0.13 s}{testbed}{}
\factdef{F-T400}{10.91 s}{testbed}{}
\factdef{F-T401}{15.13 s}{testbed}{}
\factdef{F-T402}{0.26 s}{testbed}{}
\factdef{F-T403}{132}{testbed}{}
\factdef{F-T404}{72}{testbed}{}
\factdef{F-T405}{72}{testbed}{}
\factdef{F-T406}{0}{testbed}{}
\factdef{F-T407}{0}{testbed}{}
\factdef{F-T408}{72}{testbed}{}
\factdef{F-T409}{0}{testbed}{}
\factdef{F-T410}{0}{testbed}{}
\factdef{F-T411}{12}{testbed}{}
\factdef{F-T412}{11}{testbed}{}
\factdef{F-T413}{12.4 s}{testbed}{}
\factdef{F-T414}{1}{testbed}{}
\factdef{F-T415}{0.12 s}{testbed}{}
\factdef{F-T416}{1.15 s}{testbed}{}
\factdef{F-T417}{6.17 s}{testbed}{}
\factdef{F-T418}{1.000}{testbed}{}
\factdef{F-T419}{0.983}{testbed}{}
\factdef{F-T420}{1.000}{testbed}{}
\factdef{F-T421}{0}{testbed}{}
\factdef{F-T422}{p = 1.0}{testbed}{}
\factdef{F-T423}{1}{testbed}{}
\factdef{F-T424}{1}{testbed}{}
\factdef{F-T425}{p = 0.5}{testbed}{}
\factdef{F-T426}{+1.1 pp}{testbed}{}
\factdef{F-T427}{$-$0.7 pp}{testbed}{}
\factdef{F-T428}{+1.7 pp}{testbed}{}
\factdef{F-T429}{12}{testbed}{}
\factdef{F-T430}{101}{testbed}{}
\factdef{F-T431}{101}{testbed}{}
\factdef{F-T432}{101}{testbed}{}
\factdef{F-T433}{101}{testbed}{}
\factdef{F-T434}{342 ms}{testbed}{}
\factdef{F-T435}{183 ms}{testbed}{}
\factdef{F-T436}{4 ms}{testbed}{}
\factdef{F-T437}{15 ms}{testbed}{}
\factdef{F-T438}{165 ms}{testbed}{}
\factdef{F-T439}{146 ms}{testbed}{}
\factdef{F-T440}{1 ms}{testbed}{}
\factdef{F-T441}{7 ms}{testbed}{}
\factdef{F-T442}{327 ms}{testbed}{}
\factdef{F-T443}{186 ms}{testbed}{}
\factdef{F-T444}{4 ms}{testbed}{}
\factdef{F-T445}{25 ms}{testbed}{}
\factdef{F-T446}{836 ms}{testbed}{}
\factdef{F-T447}{582 ms}{testbed}{}
\factdef{F-T448}{84 ms}{testbed}{}
\factdef{F-T449}{50 ms}{testbed}{}
\factdef{F-T450}{12.2$\times$}{testbed}{}
\factdef{F-T451}{20.8$\times$}{testbed}{}
\factdef{F-T452}{7.6$\times$}{testbed}{}
\factdef{F-T453}{46$\times$}{testbed}{}
\factdef{F-T454}{146$\times$}{testbed}{}
\factdef{F-T455}{46$\times$}{testbed}{}
\factdef{F-T456}{+87\%}{testbed}{}
\factdef{F-T457}{+13\%}{testbed}{}
\factdef{F-T458}{+76\%}{testbed}{}
\factdef{F-T459}{+44\%}{testbed}{}
\factdef{F-T460}{+19 ms}{testbed}{}
\factdef{F-T461}{0.38}{testbed}{}
\factdef{F-T462}{$-$9.2\%}{testbed}{}
\factdef{F-T463}{+3.8\%}{testbed}{}
\factdef{F-T464}{$-$9.1\%}{testbed}{}
\factdef{F-T465}{+3.4\%}{testbed}{}
\factdef{F-T466}{+0.7\%}{testbed}{}
\factdef{F-T467}{+4.5\%}{testbed}{}
\factdef{F-T468}{$-$8.5\%}{testbed}{}
\factdef{F-T469}{+8.5\%}{testbed}{}
\factdef{F-T470}{+23\%}{testbed}{}
\factdef{F-T471}{+71\%}{testbed}{}
\factdef{F-T472}{4$\times$}{testbed}{}
\factdef{F-T473}{17$\times$}{testbed}{}
\factdef{F-T474}{18}{testbed}{}
\factdef{F-T475}{18}{testbed}{}
\factdef{F-T476}{0.00}{testbed}{}
\factdef{F-T477}{+0.96}{testbed}{}
\factdef{F-T478}{6.1e-5}{testbed}{}
\factdef{F-T479}{6}{testbed}{}
\factdef{F-T480}{0.00}{testbed}{}
\factdef{F-T481}{0.67}{testbed}{}
\factdef{F-T482}{12}{testbed}{}
\factdef{F-T483}{0}{testbed}{}
\factdef{F-T484}{2.4e-4}{testbed}{}
\factdef{F-T485}{+0.68}{testbed}{}
\factdef{F-T486}{18}{testbed}{}
\factdef{F-T487}{0.16}{testbed}{}
\factdef{F-T488}{1.00}{testbed}{}
\factdef{F-T489}{1.00}{testbed}{}
\factdef{F-T490}{1.07$\times$}{testbed}{}
\factdef{F-T491}{111}{testbed}{}
\factdef{F-T492}{25}{testbed}{}
\factdef{F-T493}{+0.82}{testbed}{}
\factdef{F-T494}{0}{testbed}{}
\factdef{F-T495}{42}{testbed}{}
\factdef{F-T496}{3}{testbed}{}
\factdef{F-T497}{10}{testbed}{}
\factdef{F-T498}{0.036}{testbed}{}
\factdef{F-T499}{6}{testbed}{}
\factdef{F-T500}{6}{testbed}{}
\factdef{F-T501}{6}{testbed}{}
\factdef{F-T502}{0.00}{testbed}{}
\factdef{F-T503}{6}{testbed}{}
\factdef{F-T504}{0.08}{testbed}{}
\factdef{F-T505}{6}{testbed}{}
\factdef{F-T506}{6}{testbed}{}
\factdef{F-T507}{5}{testbed}{}
\factdef{F-T508}{33}{testbed}{}
\factdef{F-T509}{279}{testbed}{}
\factdef{F-T510}{0}{testbed}{}
\factdef{F-T511}{162}{testbed}{}
\factdef{F-T512}{30}{testbed}{}
\factdef{F-T513}{15}{testbed}{}
\factdef{F-T514}{72}{testbed}{}
\factdef{F-T515}{0}{testbed}{}
\factdef{F-T516}{36}{testbed}{}
\factdef{F-T517}{1}{testbed}{}
\factdef{F-T518}{0}{testbed}{}
\factdef{F-T519}{0}{testbed}{}
\factdef{F-T520}{+24.0\% to +31.3\%}{testbed}{}
\factdef{F-T521}{2}{testbed}{}
\factdef{F-T522}{+40\% to +85\%}{testbed}{}
\factdef{F-T523}{+96\% to +118\%}{testbed}{}
\factdef{F-T524}{86.4 MB}{testbed}{}
\factdef{F-T525}{4.09 tasks/s}{testbed}{}
\factdef{F-T526}{10.72 tasks/s}{testbed}{}
\factdef{F-T527}{0.38}{testbed}{}
\factdef{F-T528}{4.89–24.29 s}{testbed}{}
\factdef{F-T529}{635–673 MB}{testbed}{}
\factdef{F-T530}{2.24–11.71 s}{testbed}{}
\factdef{F-T531}{541–553 MB}{testbed}{}
\factdef{F-T532}{0.49–4.60 s}{testbed}{}
\factdef{F-T533}{692–699 MB}{testbed}{}
\factdef{F-T534}{0.60–5.28 s}{testbed}{}
\factdef{F-T535}{701–707 MB}{testbed}{}
\factdef{F-T536}{$-$8.5\% to +8.5\%}{testbed}{}
\factdef{F-T537}{115}{testbed}{}
\factdef{F-T538}{100}{testbed}{}
\factdef{F-T539}{269}{testbed}{}
\factdef{F-T540}{42}{testbed}{}
\factdef{F-T541}{129}{testbed}{}
\factdef{F-T542}{1,428}{testbed}{}
\factdef{F-T543}{1,832}{testbed}{}
\factdef{F-T544}{100.0\%}{testbed}{}
\factdef{F-T545}{100.0\%}{testbed}{}
\factdef{F-T546}{100.0\%}{testbed}{}
\factdef{F-T547}{100.0\%}{testbed}{}
\factdef{F-T548}{100.0\%}{testbed}{}
\factdef{F-T549}{100.0\%}{testbed}{}
\factdef{F-T550}{100.0\%}{testbed}{}
\factdef{F-T551}{27.8\%}{testbed}{}
\factdef{F-T552}{22.2\%}{testbed}{}
\factdef{F-T553}{5.6\%}{testbed}{}
\factdef{F-T554}{0.0\%}{testbed}{}
\factdef{F-T555}{100.0\%}{testbed}{}
\factdef{F-T556}{0.0\%}{testbed}{}
\factdef{F-T557}{28.6\%}{testbed}{}
\factdef{F-T558}{100.0\%}{testbed}{}
\factdef{F-T559}{100.0\%}{testbed}{}
\factdef{F-T560}{33.3\%}{testbed}{}
\factdef{F-T561}{100.0\%}{testbed}{}
\factdef{F-T562}{0.0\%}{testbed}{}
\factdef{F-T563}{86}{testbed}{}
\factdef{F-T564}{80}{testbed}{}
\factdef{F-T565}{30}{testbed}{}
\factdef{F-T566}{30}{testbed}{}
\factdef{F-T567}{0}{testbed}{}
\factdef{F-T568}{0.164 s}{testbed}{}
\factdef{F-T569}{0.402 s}{testbed}{}
\factdef{F-T570}{30}{testbed}{}
\factdef{F-T571}{30}{testbed}{}
\factdef{F-T572}{0}{testbed}{}
\factdef{F-T573}{0.276 s}{testbed}{}
\factdef{F-T574}{0.270 s}{testbed}{}
\factdef{F-T575}{24}{testbed}{}
\factdef{F-T576}{0}{testbed}{}
\factdef{F-T577}{4.50}{testbed}{}
\factdef{F-T578}{4.63}{testbed}{}
\factdef{F-T579}{0}{testbed}{}
\factdef{F-T580}{6}{testbed}{}
\factdef{F-T581}{1}{testbed}{}
\factdef{F-T582}{14}{testbed}{}
\factdef{F-T583}{0}{testbed}{}
\factdef{F-T584}{2,592}{testbed}{}
\factdef{F-T585}{15}{testbed}{}
\factdef{F-T586}{15}{testbed}{}
\factdef{F-T587}{21}{testbed}{}
\factdef{F-T588}{0}{testbed}{}
\factdef{F-T589}{0.168 s}{testbed}{}
\factdef{F-T590}{0}{testbed}{}
\factdef{F-T591}{0}{testbed}{}
\factdef{F-T592}{15}{testbed}{}
\factdef{F-T593}{15}{testbed}{}
\factdef{F-T594}{19}{testbed}{}
\factdef{F-T595}{3}{testbed}{}
\factdef{F-T596}{10}{testbed}{}
\factdef{F-T597}{0}{testbed}{}
\factdef{F-T598}{0}{testbed}{}
\factdef{F-T599}{0.550 s}{testbed}{}
\factdef{F-T600}{0.565 s}{testbed}{}
\factdef{F-T601}{0.550 s}{testbed}{}
\factdef{F-T602}{0}{testbed}{}
\factdef{F-T603}{5}{testbed}{}
\factdef{F-T604}{1}{testbed}{}
\factdef{F-T605}{6}{testbed}{}
\factdef{F-T606}{24.0}{testbed}{}
\factdef{F-T607}{0}{testbed}{}
\factdef{F-T608}{14}{testbed}{}
\factdef{F-T609}{\$0.62}{testbed}{}
\factdef{F-T610}{15}{testbed}{}
\factdef{F-T611}{15}{testbed}{}
\factdef{F-T612}{20}{testbed}{}
\factdef{F-T613}{3}{testbed}{}
\factdef{F-T614}{10}{testbed}{}
\factdef{F-T615}{0}{testbed}{}
\factdef{F-T616}{0}{testbed}{}
\factdef{F-T617}{0.218 s}{testbed}{}
\factdef{F-T618}{0.352 s}{testbed}{}
\factdef{F-T619}{0.249 s}{testbed}{}
\factdef{F-T620}{0}{testbed}{}
\factdef{F-T621}{5}{testbed}{}
\factdef{F-T622}{1}{testbed}{}
\factdef{F-T623}{6}{testbed}{}
\factdef{F-T624}{8.2}{testbed}{}
\factdef{F-T625}{0}{testbed}{}
\factdef{F-T626}{14}{testbed}{}
\factdef{F-T627}{\$0.56}{testbed}{}
\factdef{F-T628}{15}{testbed}{}
\factdef{F-T629}{15}{testbed}{}
\factdef{F-T630}{15}{testbed}{}
\factdef{F-T631}{3}{testbed}{}
\factdef{F-T632}{10}{testbed}{}
\factdef{F-T633}{0}{testbed}{}
\factdef{F-T634}{0}{testbed}{}
\factdef{F-T635}{0.481 s}{testbed}{}
\factdef{F-T636}{0.516 s}{testbed}{}
\factdef{F-T637}{0.503 s}{testbed}{}
\factdef{F-T638}{0}{testbed}{}
\factdef{F-T639}{5}{testbed}{}
\factdef{F-T640}{1}{testbed}{}
\factdef{F-T641}{6}{testbed}{}
\factdef{F-T642}{19.0}{testbed}{}
\factdef{F-T643}{0}{testbed}{}
\factdef{F-T644}{14}{testbed}{}
\factdef{F-T645}{\$2.59}{testbed}{}
\factdef{F-T646}{15}{testbed}{}
\factdef{F-T647}{15}{testbed}{}
\factdef{F-T648}{17}{testbed}{}
\factdef{F-T649}{3}{testbed}{}
\factdef{F-T650}{10}{testbed}{}
\factdef{F-T651}{0}{testbed}{}
\factdef{F-T652}{0}{testbed}{}
\factdef{F-T653}{0.136 s}{testbed}{}
\factdef{F-T654}{1.057 s}{testbed}{}
\factdef{F-T655}{0.369 s}{testbed}{}
\factdef{F-T656}{0}{testbed}{}
\factdef{F-T657}{5}{testbed}{}
\factdef{F-T658}{1}{testbed}{}
\factdef{F-T659}{6}{testbed}{}
\factdef{F-T660}{6.0}{testbed}{}
\factdef{F-T661}{0}{testbed}{}
\factdef{F-T662}{14}{testbed}{}
\factdef{F-T663}{\$0.59}{testbed}{}
\factdef{F-T664}{3}{testbed}{}
\factdef{F-T665}{0}{testbed}{}
\factdef{F-T666}{0.201 s}{testbed}{}
\factdef{F-T667}{0.210 s}{testbed}{}
\factdef{F-T668}{5}{testbed}{}
\factdef{F-T669}{6}{testbed}{}
\factdef{F-T670}{11.0}{testbed}{}
\factdef{F-T671}{0}{testbed}{}
\factdef{F-T672}{10}{testbed}{}
\factdef{F-T673}{3}{testbed}{}
\factdef{F-T674}{3}{testbed}{}
\factdef{F-T675}{48}{testbed}{}
\factdef{F-T676}{0}{testbed}{}
\factdef{F-T677}{0.307 s}{testbed}{}
\factdef{F-T678}{12.0}{testbed}{}
\factdef{F-T679}{6.0}{testbed}{}
\factdef{F-T680}{6.4}{testbed}{}
\factdef{F-T681}{6.5}{testbed}{}
\factdef{F-T682}{7.4}{testbed}{}
\factdef{F-T683}{5}{testbed}{}
\factdef{F-T684}{5}{testbed}{}
\factdef{F-T685}{0}{testbed}{}
\factdef{F-T686}{6}{testbed}{}
\factdef{F-T687}{0}{testbed}{}
\factdef{F-T688}{161}{testbed}{}
\factdef{F-T689}{13}{testbed}{}
\factdef{F-T690}{20}{testbed}{}
\factdef{F-T691}{20}{testbed}{}
\factdef{F-T692}{20}{testbed}{}
\factdef{F-T693}{1.8}{testbed}{}
\factdef{F-T694}{10.0}{testbed}{}
\factdef{F-T695}{4,252}{testbed}{}
\factdef{F-T696}{1,660}{testbed}{}
\factdef{F-T697}{0}{testbed}{}
\factdef{F-T698}{41}{testbed}{}
\factdef{F-T699}{41}{testbed}{}
\factdef{F-T700}{0}{testbed}{}
\factdef{F-T701}{0}{testbed}{post-hoc}
\factdef{F-T702}{3}{testbed}{post-hoc}
\factdef{F-T703}{$-$3 ms}{testbed}{post-hoc}
\factdef{F-T704}{161}{testbed}{}
\factdef{F-T705}{98}{testbed}{}
\factdef{F-T706}{0}{testbed}{}
\factdef{F-T707}{75}{testbed}{}
\factdef{F-T708}{\$4.36}{testbed}{}
\factdef{F-T709}{36}{testbed}{}
\factdef{F-T710}{2}{testbed}{}
\factdef{F-T711}{2}{testbed}{}
\factdef{F-T712}{0}{testbed}{}
\factdef{F-T713}{36}{testbed}{}
\factdef{F-T714}{0}{testbed}{}
\factdef{F-T715}{9}{testbed}{}
\factdef{F-T716}{9}{testbed}{}
\factdef{F-T717}{9}{testbed}{}
\factdef{F-T718}{0}{testbed}{}
\factdef{F-T719}{6}{testbed}{}
\factdef{F-T720}{18}{testbed}{}
\factdef{F-T721}{24}{testbed}{}
\factdef{F-T722}{12}{testbed}{}
\factdef{F-T723}{0}{testbed}{}
\factdef{F-T724}{20}{testbed}{}
\factdef{F-T725}{13}{testbed}{}
\factdef{F-T726}{13}{testbed}{}
\factdef{F-T727}{17}{testbed}{}
\factdef{F-T728}{11}{testbed}{}
\factdef{F-T729}{45}{testbed}{}
\factdef{F-T730}{45}{testbed}{}
\factdef{F-T731}{45}{testbed}{}
\factdef{F-T732}{33}{testbed}{}
\factdef{F-T733}{33}{testbed}{}
\factdef{F-T734}{33}{testbed}{}
\factdef{F-T735}{33}{testbed}{}
\factdef{F-T736}{13}{testbed}{}
\factdef{F-T737}{13}{testbed}{}
\factdef{F-T738}{13}{testbed}{}
\factdef{F-T739}{13}{testbed}{}
\factdef{F-T740}{10}{testbed}{}
\factdef{F-T741}{10}{testbed}{}
\factdef{F-T742}{10}{testbed}{}
\factdef{F-T743}{10}{testbed}{}
\factdef{F-T744}{$-$0.01 s}{testbed}{}
\factdef{F-T745}{0.06 s}{testbed}{}
\factdef{F-T746}{$-$0.02 s}{testbed}{}
\factdef{F-T747}{0.10 s}{testbed}{}
\factdef{F-T748}{0.00 s}{testbed}{}
\factdef{F-T749}{0.01 s}{testbed}{}
\factdef{F-T750}{0.24 s}{testbed}{}
\factdef{F-T751}{0}{testbed}{}
\factdef{F-T752}{0}{testbed}{}
\factdef{F-T753}{100\%}{testbed}{}
\factdef{F-T754}{94\%}{testbed}{}
\factdef{F-T755}{100\%}{testbed}{}
\factdef{F-T756}{42}{testbed}{}
\factdef{F-T757}{83\%}{testbed}{}
\factdef{F-T758}{0.99}{testbed}{}
\factdef{F-T759}{0.38}{testbed}{}
\factdef{F-T760}{0.65}{testbed}{}
\factdef{F-T761}{0.94}{testbed}{}
\factdef{F-T762}{0.83}{testbed}{}
\factdef{F-T763}{38\%}{testbed}{}
\factdef{F-T764}{2}{testbed}{}
\factdef{F-T765}{0}{testbed}{}
\factdef{F-T766}{1}{testbed}{}
\factdef{F-T767}{7}{testbed}{}
\factdef{F-T768}{7}{testbed}{}
\factdef{F-T769}{17}{testbed}{}
\factdef{F-T770}{0.062 s}{testbed}{}
\factdef{F-T771}{97\%}{testbed}{}
\factdef{F-T772}{0}{testbed}{}
\factdef{F-T773}{0}{testbed}{}
\factdef{F-T774}{1.0}{testbed}{}
\factdef{F-T775}{1.0}{testbed}{}
\factdef{F-T776}{1.0}{testbed}{}
\factdef{F-T777}{1.0}{testbed}{}
\factdef{F-T778}{1.0}{testbed}{}
\factdef{F-T779}{1.00}{testbed}{}
\factdef{F-T780}{0.63}{testbed}{}
\factdef{F-T781}{1.00}{testbed}{}
\factdef{F-T782}{0}{testbed}{}
\factdef{F-T783}{5}{testbed}{}
\factdef{F-T784}{5}{testbed}{}
\factdef{F-T785}{153}{testbed}{post-hoc}
\factdef{F-T786}{9}{testbed}{post-hoc}
\factdef{F-T787}{0}{testbed}{post-hoc}
\factdef{F-T788}{30}{testbed}{post-hoc}
\factdef{F-T789}{0}{testbed}{post-hoc}
\factdef{F-T790}{0}{testbed}{post-hoc}
\factdef{F-T791}{10}{testbed}{post-hoc}
\factdef{F-T792}{5}{testbed}{post-hoc}
\factdef{F-T793}{0}{testbed}{post-hoc}
\factdef{F-T794}{60}{testbed}{post-hoc}
\factdef{F-T795}{12}{testbed}{post-hoc}
\factdef{F-T796}{0}{testbed}{post-hoc}
\factdef{F-T797}{253}{testbed}{post-hoc}
\factdef{F-T798}{26}{testbed}{post-hoc}
\factdef{F-T799}{0}{testbed}{post-hoc}
\factdef{F-T800}{279}{testbed}{post-hoc}
\factdef{F-T801}{279}{testbed}{post-hoc}
\factdef{F-T802}{21}{testbed}{post-hoc}
\factdef{F-T803}{14 ms}{testbed}{post-hoc}
\factdef{F-T804}{4,091 ms}{testbed}{post-hoc}
\factdef{F-T805}{5}{testbed}{post-hoc}
\factdef{F-T806}{56}{testbed}{post-hoc}
\factdef{F-T807}{56}{testbed}{post-hoc}
\factdef{F-T808}{$-$6,234 ms}{testbed}{post-hoc}
\factdef{F-T809}{24}{testbed}{post-hoc}
\factdef{F-T810}{0}{testbed}{post-hoc}
\factdef{F-T811}{0}{testbed}{post-hoc}
\factdef{F-T812}{22}{testbed}{}
\factdef{F-T813}{8}{testbed}{}
\factdef{F-T814}{2}{testbed}{}
\factdef{F-T815}{22}{testbed}{}
\factdef{F-T816}{8}{testbed}{}
\factdef{F-T817}{22}{testbed}{}
\factdef{F-T818}{8}{testbed}{}
\factdef{F-T819}{2}{testbed}{}
\factdef{F-T820}{22}{testbed}{}
\factdef{F-T821}{8}{testbed}{}
\factdef{F-T822}{16}{testbed}{}
\factdef{F-T823}{8}{testbed}{}
\factdef{F-T824}{14}{testbed}{}
\factdef{F-T825}{7}{testbed}{}
\factdef{F-T826}{2}{testbed}{}
\factdef{F-T827}{14}{testbed}{}
\factdef{F-T828}{5}{testbed}{}
\factdef{F-T829}{2}{testbed}{}
\factdef{F-T830}{14}{testbed}{}
\factdef{F-T831}{6}{testbed}{}
\factdef{F-T832}{2}{testbed}{}
\factdef{F-T833}{10}{testbed}{}
\factdef{F-T834}{5}{testbed}{}
\factdef{F-T835}{2}{testbed}{}
\factdef{F-T836}{10}{testbed}{}
\factdef{F-T837}{7}{testbed}{}
\factdef{F-T838}{2}{testbed}{}
\factdef{F-T839}{52}{testbed}{}
\factdef{F-T840}{23}{testbed}{}
\factdef{F-T841}{9}{testbed}{}
\factdef{F-T842}{1}{testbed}{}
\factdef{F-T843}{10}{testbed}{}
\factdef{F-T844}{101}{testbed}{}
\factdef{F-T845}{621}{testbed}{post-hoc}
\factdef{F-T846}{182}{testbed}{post-hoc}
\factdef{F-T847}{0}{testbed}{post-hoc}
\factdef{F-T848}{0}{testbed}{post-hoc}
\factdef{F-T849}{0}{testbed}{post-hoc}
\factdef{F-T850}{115}{testbed}{post-hoc}
\factdef{F-T851}{70}{testbed}{post-hoc}
\factdef{F-T852}{0}{testbed}{post-hoc}
\factdef{F-T853}{0}{testbed}{post-hoc}
\factdef{F-T854}{0}{testbed}{post-hoc}
\factdef{F-T855}{100}{testbed}{post-hoc}
\factdef{F-T856}{26}{testbed}{post-hoc}
\factdef{F-T857}{0}{testbed}{post-hoc}
\factdef{F-T858}{0}{testbed}{post-hoc}
\factdef{F-T859}{0}{testbed}{post-hoc}
\factdef{F-T860}{132}{testbed}{post-hoc}
\factdef{F-T861}{37}{testbed}{post-hoc}
\factdef{F-T862}{0}{testbed}{post-hoc}
\factdef{F-T863}{0}{testbed}{post-hoc}
\factdef{F-T864}{0}{testbed}{post-hoc}
\factdef{F-T865}{36}{testbed}{post-hoc}
\factdef{F-T866}{11}{testbed}{post-hoc}
\factdef{F-T867}{0}{testbed}{post-hoc}
\factdef{F-T868}{0}{testbed}{post-hoc}
\factdef{F-T869}{0}{testbed}{post-hoc}
\factdef{F-T870}{1,004}{testbed}{post-hoc}
\factdef{F-T871}{326}{testbed}{post-hoc}
\factdef{F-T872}{0}{testbed}{post-hoc}
\factdef{F-T873}{0}{testbed}{post-hoc}
\factdef{F-T874}{0}{testbed}{post-hoc}
\factdef{F-T875}{161}{testbed}{post-hoc}
\factdef{F-T876}{1,870}{testbed}{post-hoc}
\factdef{F-T877}{0}{testbed}{post-hoc}
\factdef{F-T878}{0}{testbed}{post-hoc}
\factdef{F-T879}{12}{testbed}{post-hoc}
\factdef{F-T880}{122}{testbed}{post-hoc}
\factdef{F-T881}{0}{testbed}{post-hoc}
\factdef{F-T882}{0}{testbed}{post-hoc}
\factdef{F-T883}{1,992}{testbed}{post-hoc}
\factdef{F-T884}{4,252}{testbed}{post-hoc}
\factdef{F-T885}{2,382}{testbed}{post-hoc}
\factdef{F-T886}{1,362}{testbed}{post-hoc}
\factdef{F-T887}{704}{testbed}{post-hoc}
\factdef{F-T888}{0}{testbed}{post-hoc}
\factdef{F-T889}{1,310}{testbed}{post-hoc}
\factdef{F-T890}{0}{testbed}{post-hoc}
\factdef{F-T891}{362}{testbed}{post-hoc}
\factdef{F-T892}{560}{testbed}{post-hoc}
\factdef{F-T893}{0}{testbed}{post-hoc}
\factdef{F-T894}{342}{testbed}{post-hoc}
\factdef{F-T895}{0.77 s}{testbed}{post-hoc}
\factdef{F-T896}{1.07 s}{testbed}{post-hoc}
\factdef{F-T897}{1.32 s}{testbed}{post-hoc}

%% file: sections/00-abstract.tex
When multi-agent systems act on the external world, a single agent's faulty output can spread across tasks before detection, creating a system-wide \emph{blast radius}.
Single-agent guardrails are confined to individual harnesses, while existing multi-agent controls either cancel only the faulty source and miss downstream consumers or halt and roll back the entire workflow at the cost of healthy work.
We present \FB, a multi-agent runtime and guardrail for selective post-alert fault containment.
Its tree-structured runtime, \RT, narrows permissions vertically from parent to child agents to establish node boundaries and control switches.
Meanwhile, an append-only provenance DAG records cross-agent data consumption horizontally.
Upon an alert, \FB traces forward on this DAG to isolate tainted executions and pending actions, and backward to narrow root causes.
At the site switches, a default full quarantine tier freezes affected branches without a solver, while an assured optimized cut tier uses fence-then-prune to enforce a deadline-aware minimum cut minimizing collateral disruption.
Across \fact{F-T543} testbed runs, \fact{F-T704} physical smart-home LLM runs, and \fact{F-S145} simulation records, \FB blocks all post-alert faulty actuations on our main testbed and smart-home grids with zero post-convergence leaks.
Whereas native controls in A2A and LangGraph either contain at most \fact{F-T560} of faults or retain at most \fact{F-T487} of work without redo, \FB retains \fact{F-T485} to \fact{F-T477} more healthy work, and its optimized cut retains \fact{F-T283} more than full quarantine under dense spread.
These results show that co-designing runtime authority boundaries with online provenance enables surgical fault containment without sacrificing healthy work (\href{https://github.com/EdgeCitadelTeam/EdgeCitadel}{code}: \cref{app:baselines}).

%% file: sections/s1-intro.tex
\section{Introduction}
\label{sec:intro}

\input{figures/fig1-overview}

Large language models and tool-using agents are increasingly composed into multi-agent systems that act on the physical and digital world: orchestrators delegate subtasks across hosts holding distinct credentials, coding assistants collaborate with automated runners, and service agents control smart-home and laboratory devices~\citep{cemri2025mast,huang2025resilience,zhan2026hearthnet,forough2026agents}.
In such deployments, a single agent's faulty output rarely stays local: it crosses task and host boundaries without any downstream agent exceeding its own permissions, polluting shared state and subtasks before an anomaly is detected and creating a system-wide \emph{blast radius} when downstream consumers issue external actions.
The July 2026 OpenAI--Hugging Face incident shows such a blast radius at scale: during internal cybersecurity evaluations, agents meant to be isolated from one another turned shared package-repository credentials they legitimately held into a message board.
Monitoring flagged the activity, but its significance was not apparent and the run continued; halting the experiments and wiping the repository did not end it, as within two days newly started agents had built a new board on which roughly 1,200 agents exchanged over 70,000 messages and files, and about 700 then joined an attack on Hugging Face~\citep{openai2026hfreport,metr2026hfinvestigation}.
More broadly, this containment gap arises routinely from stale state, malformed tool outputs, or prompt injections.

Existing defenses and agent frameworks cannot contain this cross-agent blast radius without sacrificing healthy concurrent work.
Single-agent guardrails and planner-level information-flow controls inspect one model's prompts, context, or tool calls inside a single harness~\citep{costa2025fides,debenedetti2025camel}, with no visibility once a bad output has been consumed by peers.
Meanwhile, runtime controls in multi-agent frameworks act either on one agent or on the entire workflow: A2A cancels individual tasks~\citep{a2a2025}, LangGraph interrupts the graph or rolls back to a checkpoint and re-executes~\citep{langgraph2026}, and topological safeguards sever a flagged agent's future outgoing edges~\citep{wang2025gsafeguard,zhou2025guardian}.
Once a faulty output has been read by peers, cancelling or isolating the producer alone misses contaminated consumers heading toward external sinks, whereas halting the workflow or rolling back state destroys healthy work.

To bridge this gap, we ask: \emph{once a fault has spread across agents, can a runtime intercept all reachable unsafe external actions while allowing unaffected concurrent work to continue?}
We answer in the affirmative by co-designing a heterogeneous multi-agent runtime (\RT~\citep{zhan2026edgecitadel}) in the lower layer with the \FB guardrail controller in the upper layer (\cref{fig:overview}).
In the lower layer, \RT organizes execution as a tree hierarchy: permissions are delegated vertically from parent to child agents, monotonically narrowing each subtask's scope and enforcing hard authority boundaries and soft resource limits (leases, deadlines, and quotas) at every node (\cref{sec:arch:runtime}).
Each host's local \sited denies out-of-scope dispatches, terminates tasks exceeding their limits, mediates dispatches, messages, results, and lock-fenced gateway actions in a per-site ordered journal, and exposes uniform per-object switches (hold and revoke) across LLM harnesses, coding assistants, and CLIs.

In the upper layer, the \FB controller consumes the per-site journals inside a central coordinator to maintain an append-only provenance DAG and derive boundary alerts (\cref{sec:design:trace}).
When an alert fires, \FB uses the provenance DAG to \emph{look} and the live site switches to \emph{act}: it traces \emph{forward} to identify tainted executions and pending actions (or \emph{backward} to narrow candidate root causes) and projects live executions and switches into a control graph.
On this graph, \FB provides two containment tiers: a default zero-solver full quarantine that pauses tainted executions and revokes their egress (\cref{sec:arch:fq}), and an assured optimized cut that casts boundary selection as a deadline-aware minimum cut crediting only actions that land before each effect's deadline (\cref{sec:design:choose}), solvable in a single max-flow when active switches are vertex-local and nested (\cref{sec:design:fastpath}).
To enforce the cut while agents continue executing without a global log order, fence-then-prune (\cref{sec:design:hold}) fences the frontier and source-side egress until all actions reach journal-fenced convergence ($\tone$), then prunes temporary holds under a separation guard to guarantee zero post-$\tone$ leaks.

\noindent Our main contributions are as follows:
\begin{itemize}[leftmargin=*,nosep]
\item \textbf{A two-layer runtime and provenance architecture} (\cref{sec:arch}): We co-design \RT, which enforces vertical permission narrowing, per-site ordered journaling, lock-fenced gateways, and per-object switches, with \FB, which maintains an append-only provenance DAG for forward/backward tracing and full quarantine.
\item \textbf{Deadline-aware minimum cut and fence-then-prune} (\cref{sec:design}): We cast selective containment on a live control graph as a deadline-aware minimum cut with a laminar max-flow fast path (\cref{thm:fastpath}), enforced by fence-then-prune for zero post-$\tone$ leaks (\cref{thm:safety}).
\item \textbf{Physical and distributed evaluation} (\cref{sec:eval}): Across \fact{F-T543} testbed runs, \fact{F-T704} smart-home runs on six LLM deployments, and \fact{F-S145} simulation records, \FB blocks all post-alert faulty actuations on our main grids with zero post-$\tone$ leaks, retains \fact{F-T485} to \fact{F-T477} more healthy work without redo than leak-free baselines, and stays within \fact{F-T468} to \fact{F-T469} of A2A's makespan.
\end{itemize}

%% file: figures/fig1-overview.tex
\begin{figure*}[t]
\centering
\includegraphics[width=\textwidth]{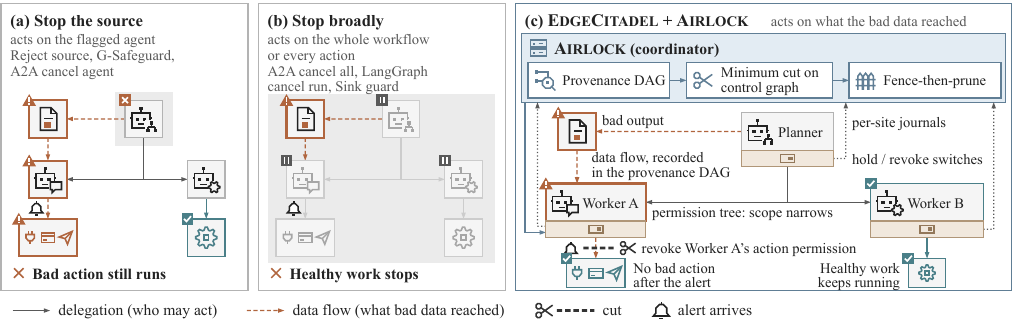}\par
\caption{Containment after bad data has been read. (a)~Stopping the flagged agent misses agents that already read its output. (b)~Stopping the whole workflow, or every action, also stops healthy work. (c)~\RT records what each agent read; \FB cuts only the actions the bad data reached (\cref{sec:arch,sec:design}).}
\label{fig:overview}
\end{figure*}

%% file: sections/s2-application.tex
\section{Background and Motivation}
\label{sec:app}

\input{figures/fig1-svc}

\subsection{Multi-Agent Topologies and Blast Radius}
\label{sec:app:topology}

Even when each agent invocation receives its permissions from a single parent, data consumption cuts across branches: an artifact written to a shared workspace or conversational memory by one agent is read by peers in other branches to drive downstream tool and device calls.
Flat substrates such as shared message boards or shared credential caches provide neither per-task authorization boundaries nor cross-agent provenance, leaving operators to choose between cancelling a single agent and shutting down the whole deployment.
To quantify the systems gap between stopping a faulty source and containing what its output has already reached, we examine a running smart-home service (\HomeSvc, \cref{fig:svc}) alongside our workflow benchmarks (\cref{sec:eval}).

\subsection{Three Empirical Observations in a Running Service}
\label{sec:app:obs}

\paragraph{The running service (\HomeSvc).}
In \HomeSvc, users query a long-lived assistant at \conf{0.5}{svc.json} requests/s across three distributed sites; \conf{30\%}{svc.json} of requests actuate a desk lamp via Home Assistant~\citep{homeassistant2026}, and shared conversational memory links consecutive turns.
A root planner supplies a device schedule once at startup, while an independent automation agent actuates two other devices every \conf{5~s}{svc.json} without reading the schedule.
User replies are configured as unprotected outputs ($c=0$), whereas device commands are \emph{protected effects}: tainted computation must never actuate them.
When the planner's schedule carries a corrupted entry, reading it taints every subsequent assistant turn through shared memory (\cref{fig:svc}).
We evaluate alerts that flag the faulty schedule either \emph{early} (before the assistant reads it) or \emph{late} (\conf{15~s}{svc.json} after the assistant begins serving).

\paragraph{O1: By the time an alert fires, taint has already spread to unissued future effects.}
In uncontained \HomeSvc runs at the late alert, \fact{F-T361} of \fact{F-T360} tasks are already tainted and \fact{F-T362} tainted lamp commands have committed, while \fact{F-T363} further tainted lamp commands remain preventable, the first arriving \fact{F-T364} later (medians).
\begin{table}[!t]
\centering\footnotesize\setlength{\tabcolsep}{3.2pt}
\caption{\HomeSvc on the testbed (early / late alert, means over 5 runs; best excl.\ \NoC \textbf{bolded}, worst \underline{underlined}).}
\label{tab:svc}
\vspace{2pt}
\fitcol{%
\begin{tabular}{@{}ll|cc|cc@{}}
\toprule
\multicolumn{2}{@{}l}{Control} & \multicolumn{2}{c}{Harm $\downarrow$} & \multicolumn{2}{c@{}}{Service $\uparrow$} \\
\cmidrule(lr){3-4}\cmidrule(l){5-6}
Family & \multicolumn{1}{l}{Policy} & Lamp cmds & \multicolumn{1}{c}{Tainted ans.} & Correct & \multicolumn{1}{c@{}}{Autom.} \\
\midrule
\multicolumn{2}{@{}l|}{\NoC} & \fact{F-T724}\,/\,\fact{F-T725} & \fact{F-T729}\,/\,\fact{F-T732} & 0\,/\,0 & \fact{F-T736}\,/\,\fact{F-T740} \\
\midrule\midrule
\multirow{2}{*}{\textbf{Source}}
 & - \RejSrc & \textbf{0}\,/\,\underline{\fact{F-T726}} & \textbf{0}\,/\,\underline{\fact{F-T733}} & \textbf{\fact{F-T240}}\,/\,\underline{0} & \textbf{\fact{F-T738}}\,/\,\textbf{\fact{F-T742}} \\
 & - \GSafe & \underline{\fact{F-T727}}\,/\,\fact{F-T728} & \underline{\fact{F-T731}}\,/\,\underline{\fact{F-T735}} & \underline{0}\,/\,\underline{0} & \textbf{\fact{F-T739}}\,/\,\textbf{\fact{F-T743}} \\
\cmidrule(r){1-2}
\multirow{2}{*}{\textbf{Coarse}}
 & - \StopTeam & \textbf{0}\,/\,\textbf{0} & \textbf{0}\,/\,\textbf{0} & \underline{0}\,/\,\underline{0} & \textbf{\fact{F-T737}}\,/\,\textbf{\fact{F-T741}} \\
 & - \SinkG & \textbf{0}\,/\,\textbf{0} & \underline{\fact{F-T730}}\,/\,\underline{\fact{F-T734}} & \underline{0}\,/\,\underline{0} & \underline{\fact{F-T256}}\,/\,\underline{0} \\
\cmidrule(r){1-2}
\multirow{2}{*}{\textbf{\FB}}
 & - \FQ & \textbf{\fact{F-T238}}\,/\,\textbf{\fact{F-T238}} & \textbf{\fact{F-T248}}\,/\,\textbf{\fact{F-T248}} & \fact{F-T242}\,/\,\underline{0} & \textbf{\fact{F-T252}}\,/\,\textbf{\fact{F-T254}} \\
 & - \OC & \textbf{\fact{F-T237}}\,/\,\textbf{\fact{F-T237}} & \fact{F-T246}\,/\,\fact{F-T247} & \fact{F-T241}\,/\,\underline{0} & \textbf{\fact{F-T252}}\,/\,\textbf{\fact{F-T254}} \\
\bottomrule
\end{tabular}}
\end{table}%
Across the seven workflow benchmarks in \cref{sec:eval}, the median late-alert cell has \fact{F-T352} tainted tasks (\fact{F-T353} of the fault's eventual reach) and \fact{F-T354} preventable protected effects.
Crucially, almost none of those preventable effects have been issued when the alert fires (\fact{F-T358} of \fact{F-T359}): they will be emitted by downstream work that has not yet executed, so cancelling queued effect calls alone cannot stop them, requiring online provenance to trace forward from a flagged artifact to reachable future executions (\cref{sec:design:trace}).

\paragraph{O2: Source-only controls leak downstream effects, while global controls destroy healthy work.}
On the testbed (\cref{tab:svc}), taking no action allows \fact{F-T725} tainted lamp commands per run to commit after the late alert.
Blocking the faulty output at its source (\RejSrc) keeps all \fact{F-T240} of \fact{F-T239} answers correct under an early alert, but leaks \fact{F-T726} commands per run once the schedule is read (\fact{F-T728} when severing the source's future edges via \GSafe~\citep{wang2025gsafeguard}).
Conversely, stopping the faulty agent's team (\StopTeam) protects the lamps only by shutting down the assistant, and blocking all device sinks (\SinkG) leaves the independent automation agent with \fact{F-T256} commands.
Provenance-guided containment blocks all tainted lamp commands (\fact{F-T237} leaked) while preserving all \fact{F-T254} of \fact{F-T255} late-alert automation commands: full quarantine halts tainted assistant turns, whereas the optimized cut revokes only the lamp grant and keeps unprotected replies open (\fact{F-T246}\,/\,\fact{F-T247} tainted answers; \cref{sec:eval:app}), motivating per-object switches that decouple execution holds from selective egress revocations (\cref{sec:arch:runtime,sec:design:choose}).

\paragraph{O3: Containment actions race against physical and external effect deadlines.}
A containment action prevents harm only if it lands at the target site before the protected actuation commits.
The preventable lamp commands above begin seconds after the alert, whereas a signed control command to the busiest site took up to \fact{F-T163} at the 99th percentile in our main runs; ignoring control latency can pick actions that arrive too late, requiring deadline-aware cut selection (\cref{sec:design:choose}) and fenced convergence before pruning holds (\cref{sec:design:hold}).

\subsection{Threat Model}
\label{sec:app:scope}

We target faults and misbehaving agents that do not adaptively evade \FB's runtime mediation: corrupted or stale outputs, prompt-injected tool reads, out-of-scope dispatches, runaway resource usage, or undeclared actions within an agent's credentials.
\FB takes the alert's initial origin set $X_0$ as given (\cref{thm:safety}; candidate root-cause narrowing for symptom-only alerts is evaluated in \cref{app:baselines}).
The operating system, per-site daemon, message broker, coordinator, and effect gateway form the trusted computing base; they may experience crash or network delays with bounded clock skew, and enforce \emph{complete mediation}~\citep{saltzer1975protection} over registered dispatches, messages, results, and device effects.

%% file: figures/fig1-svc.tex
\begin{figure}[t]
\centering
\includegraphics[width=\columnwidth]{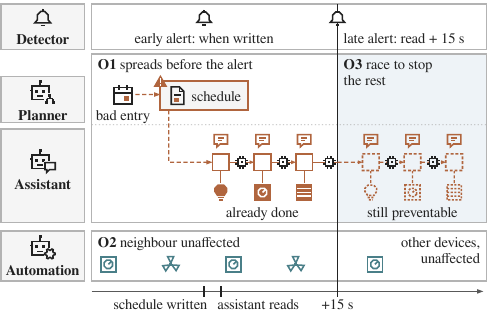}\par
\caption{Fault spread in \HomeSvc: a bad schedule entry taints the assistant's memory (O1), spares automation (O2), and races remaining lamp commands after the alert (O3; \cref{sec:app:obs}).}
\label{fig:svc}
\end{figure}

%% file: sections/s3-design.tex
\section{Architecture: Tree Hierarchy and Provenance DAG}
\label{sec:arch}
\label{sec:impl}

\input{figures/fig2-arch}

\RT~\citep{zhan2026edgecitadel} and \FB realize a two-dimensional structural principle (\cref{fig:arch}): \emph{authority flows vertically down a delegation tree, while data influence flows horizontally across an append-only provenance DAG}.
The delegation tree gives every execution unit an explicit authority and resource boundary with per-object site switches; the provenance DAG tracks how outputs propagate across tasks.

\subsection{Tree-Structured Heterogeneous Runtime (\RT)}
\label{sec:arch:runtime}

\paragraph{Parent-to-child delegation tree and boundary enforcement.}
\RT organizes multi-agent execution into a cross-site delegation tree.
Each host site runs a privileged \sited that manages agent identities, durable task state, a local provenance journal, and the NATS JetStream~\citep{nats2026} broker connection.
An invocation (one execution attempt of a task on an agent) delegates work by asking its local \sited to dispatch a child task; the \sited checks permissions, logs the provenance event, and enqueues the dispatch in a single atomic SQLite transaction.
Authority narrows monotonically down the tree: a child $c$ dispatched by parent invocation $p$ receives scope $\Scope(c) \subseteq \Grant(\mathrm{agent}(c)) \cap \Scope(p)$, where $\Grant$ is administrator policy and each permission rule pairs an effect class (e.g., file write, HTTP request, or device command) with an explicit target, bounded by subtree leases, deadlines, and quotas.
At each node boundary, the \sited directly enforces both hard authority scopes and soft resource limits: it denies out-of-scope dispatches, terminates tasks that exceed their lease, deadline, or quota, and writes these events into its per-site journal.

\paragraph{Heterogeneous attachment, lock-fenced gateway, and per-object switches.}
Agents attach to their local \sited through three adapter modes: managed workers (LLM harnesses and deterministic workflow runners), interactive connector sessions for third-party coding assistants, and wrapped command-line tools.
Because agents never hold broker or device credentials directly~\citep{forough2026agents} and pass all inter-agent data and external actions through the \sited's local API, every adapter traverses the same permission checks and per-site journal without relying on agent self-reporting.
For registered physical and external actuations, the \sited runs a built-in lock-fenced effect gateway: registered effects are re-checked at commit time under the site's control lock (the same lock acquired by incoming containment commands) and committed to the local journal before invoking the downstream device driver (e.g., Home Assistant).
Rather than coarse whole-subtree primitives, each \sited exposes six per-object control switches: reversible \emph{holds} on a task, its egress (label, under which the controller decides whether each emitted event is released), or an event (queued message or scheduled effect), and permanent \emph{revocations} of a grant (invocation egress), admission (ingress filter on a credential--target pair or agent), or credential.

\subsection{The Append-Only Provenance DAG ($\GP$)}
\label{sec:design:trace}

\paragraph{Per-site ordered journaling and DAG construction.}
Each \sited records task dispatches, mediated data reads and writes, and effect commits in a per-site ordered journal (ordered within each site, not globally across sites) and streams them over JetStream to the central coordinator.
Inside the coordinator, the \FB controller consumes the journal stream to maintain the provenance graph $\GP$ as an append-only directed acyclic graph, while deriving authority-overreach and resource-limit alerts from logged boundary violations alongside external content or edge anomaly alerts~\citep{forough2023anomaly,wang2025gsafeguard}.
Edges in $\GP$ point in the direction of influence across three channels: \emph{data} edges from mediated artifact and message flows (joined across sites by content-hashed version IDs; unless an invocation declares which mediated inputs it used, the daemon conservatively attributes all delivered inputs), \emph{session} edges linking consecutive invocations in a conversation or shared-memory session, and \emph{authority} edges from a parent invocation to its child task.

\paragraph{Forward isolation and backward root-cause narrowing.}
While the historical structure of $\GP$ is append-only, taint state evolves as journals arrive.
Taint propagation on $\GP$ is time-respecting: given initial contamination origins $X_0$, a vertex $v$ is tainted iff $v \in X_0$ or it consumed a tainted predecessor $u$ after $u$ became tainted, $\taint(v) \iff v \in X_0 \lor \exists (u \to v): \taint(u) \land t_{\taint}(u) \le t_{\mathrm{use}}(u \to v)$, evaluated over data and session edges using per-site timestamps.
When an alert flags a faulty invocation or artifact, the controller sets $X_0$ to that output version (or the violating invocation for a boundary alert) and traces \emph{forward} along $\GP$ to compute the live tainted reach (Steps~\circled{1}--\circled{3} in \cref{fig:walk}); when an alert reports only a downstream symptom, it first traverses $\GP$ \emph{backward} to narrow candidate ancestor root causes to a short list before tracing forward (\cref{app:baselines}).

\subsection{Default Containment Tier: Full Quarantine}
\label{sec:arch:fq}

Built directly on $\GP$ and the per-site switches, \FB's default tier, full quarantine, requires no optimization solver: as soon as forward tracing on $\GP$ marks an active unit tainted, it pauses the unit, revokes its egress, and permanently holds every artifact version and pending effect in the tainted reach (requiring no $\tone$ convergence barrier).
In \HomeSvc, full quarantine halts each tainted assistant invocation as it starts while leaving automation running (\cref{fig:walk}).

%% file: figures/fig2-arch.tex
\begin{figure}[t]
\centering
\includegraphics[width=\columnwidth]{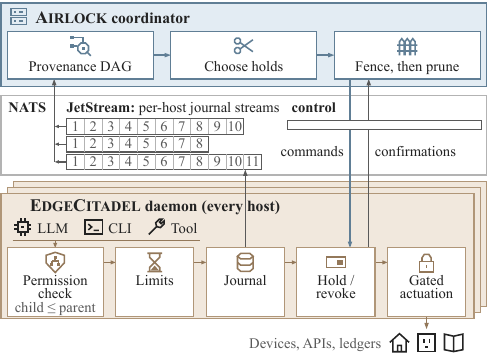}\par
\caption{Two layers. Every host runs the same \RT daemon; the \FB coordinator builds the provenance DAG from the journals and sends holds back (\cref{sec:arch}).}
\label{fig:arch}
\end{figure}

%% file: sections/s4-impl.tex
\section{Assured Containment: Minimum Cut and Fence-then-Prune}
\label{sec:design}

\input{figures/fig3-walk}

When tainted agents share execution paths with valuable ongoing sessions, or when network control delays approach effect deadlines, operators need a finer surgical boundary with a formal post-convergence guarantee.
Our assured tier, the optimized cut, projects live executions and per-object site switches into a control graph $\GC(\tau)$ to cast boundary selection as a deadline-aware minimum-cost $s$--$t$ cut (\cref{sec:design:choose}) and enforces it while the system keeps moving via the fence-then-prune protocol (\cref{sec:design:hold}; \cref{fig:walk}).

\subsection{Control Graph and Deadline-Aware Minimum Cut}
\label{sec:design:choose}

\paragraph{Projecting live executions and switches into $\GC(\tau)$.}
Because $\GP$ is an append-only historical log, containment never cuts $\GP$ directly; instead, at decision time $\tau$, the controller projects currently active executions (running invocations, dispatched-but-unstarted tasks, unread results, and pending effects) and site switches into the live control graph $\GC(\tau)$, plus one hop of potential next events permitted by each unit's capability grant and scope.
Historical completed edges of $\GP$ are omitted so $\GC(\tau)$ scales with active concurrency rather than run history.
Each live unit $u$ is expanded into a split node $u_{\mathrm{in}} \to u_{\mathrm{out}}$ (bridged by a task-hold edge), followed by an egress node $E_u$ (gated by an egress label) and per-credential hubs $H_\kappa \to b$ for each permitted target $b$ (gated by grant, admission, and credential revocations), mapping each switch $a \in \Act$ to an explicit removal set $R(a) \subseteq \GC(\tau)$.

\paragraph{Deadline-aware minimum-cost cut and collateral cost.}
Each edge $e \in \GC$ carries a deadline $d(e)$ (the earliest time it can actuate, derived from lower bounds on delivery/execution delays and scheduled actuation times), while each site $\sigma$ has a nominal control latency bound $L(\sigma)$ (\conf{400~ms}{D-87} by default, above the 99th-percentile site confirmation latency in pilot runs: \fact{F-T074}, \fact{F-T077}, \fact{F-T080}).
With planning allowance $\beta = $~\conf{50~ms}{config}, an action $a \in \Act$ with cost $c(a)$ at site $\sigma(a)$ is credited with its \emph{timely mask}:
\begin{equation}
\hat R(a) = \{\, e \in R(a) : t_{\mathrm{now}} + \beta + L(\sigma(a)) < d(e) \,\}.
\label{eq:mask}
\end{equation}
The controller selects a minimum-cost action set separating the live tainted frontier $s$ from all protected sinks $t$:
\begin{equation}
\min_{C \subseteq \Act} \sum_{a \in C} c(a) \quad\text{s.t.}\quad s \not\leadsto t \ \text{ in } \ \GC \setminus \textstyle\bigcup_{a \in C} \hat R(a).
\label{eq:cut}
\end{equation}
Here $\beta$ and $L(\sigma)$ are planning estimates used by~\eqref{eq:mask} to rank timely cuts; when bursty queueing inflates site confirmation latency (up to \fact{F-T163} p99 at the coordinator site) or graph projection plus solve time (\fact{F-T141}), Phase~1's frontier hold $\Bone$ and fenced $\tone$ confirmation (\cref{sec:design:hold}) prevent premature release and preserve post-$\tone$ safety (\cref{thm:safety}).
Crucially, \FB separates \emph{when to alert} (boundary and content detectors) from \emph{where to cut}: $c(a)$ is a heuristic proxy for collateral healthy work stopped by $a$, combining a tie-breaker $c_0 = $~\conf{0.01}{config}, the compute weight of live healthy units and waiting parents halted by $a$ (tainted work is weighted 0), \conf{0.5}{revoke\_weight} per live healthy invocation restricted by a revocation, and expected future request volume for long-lived services.
Safety never depends on $c(a)$.
In \HomeSvc at the late alert,~\eqref{eq:cut} revokes only the assistant's device-command grant (Step~\circled{4} in \cref{fig:walk}), keeping user query replies open while guaranteeing that no lamp command can fire.

\subsection{Tractability and the Laminar Max-Flow Fast Path}
\label{sec:design:fastpath}

\begin{proposition}[NP-hardness of multi-edge revocations]
\label{prop:hard}
Problem~\eqref{eq:cut} generalizes Minimum Label $s$--$t$ Cut; it is NP-hard and admits no polynomial $(1-\varepsilon)\ln n$-approximation unless $\mathrm{P}=\mathrm{NP}$, even on a depth-1 delegation tree with unit action costs~\citep{jha2002two,zhang2011label,fellows2010label,feige1998threshold,dinur2014parallel}.
\end{proposition}
Hardness arises when actions remove overlapping edge sets across distinct vertices.
In our gadget, however, single-vertex switches have nested removal sets (a task hold encloses an egress hold, which encloses a grant revocation):
\begin{theorem}[Laminar max-flow fast path]
\label{thm:fastpath}
If every finite-cost action is gated at an assigned vertex port and the gate removal sets at each port are laminar (nested or disjoint), the optimum of~\eqref{eq:cut} equals the minimum $s$--$t$ cut of a laminar gadget graph with at most $|V| + |\Act|$ vertices, solvable in one max-flow call.
\end{theorem}
\begin{proposition}[Exact fallback]
\label{prop:fallback}
If removing an exception set $Y \subseteq \Act$ of cross-vertex actions leaves a laminar instance, the exact optimum is found in $2^{|Y|}$ max-flow calls; for larger $|Y|$, a mixed-integer linear program over binary action variables and vertex potentials solves~\eqref{eq:cut} exactly.
\end{proposition}
Because revoking an agent-wide credential can remove edges across several live invocations, the controller checks vertex-locality and nestedness on every masked instance.
Timely masks~\eqref{eq:mask} preserve laminarity whenever site latency does not increase along nested actions (Proposition~4, \cref{app:theory}; verified by exhaustive enumeration across \fact{F-M005} gadget instances and \fact{F-M006} checks).
In our Python implementation (\sited: \conf{19k}{LoC, eval tag} LoC; coordinator/controller: \conf{6.7k}{LoC, eval tag} LoC), the solver runs a pure-Python Dinic max-flow on the laminar gadget, enumerates up to \conf{12}{config} exception actions, or invokes HiGHS~\citep{huangfu2018highs} as an exact MILP.

\subsection{The Fence-then-Prune Protocol and Safety}
\label{sec:design:hold}

\paragraph{Lock-fenced confirmations and fence-then-prune.}
When a \sited applies a versioned, HMAC-signed containment command under its site control lock, it replies with the current journal watermark read under that lock; the controller marks an action \emph{in force} once all target sites confirm, and \emph{fenced} once the coordinator has ingested each site's per-site journal up to that watermark (condition~\cond{F}, paired with the gateway's commit-time lock re-check).
To guarantee safety despite in-flight events and the absence of a global cross-site journal order, the optimized cut executes fence-then-prune (Steps~\circled{5}--\circled{7} in \cref{fig:walk}):
\begin{enumerate}[leftmargin=*,nosep]
\item \textbf{Fence (At-alert cut, frontier hold, \& source labeling):} At the alert, solve the initial cut $\Ba$ and push $\Ba$ with egress labels covering its source side $S(\Ba)$ (all units reachable from taint via edges not timely removed by $\Ba$; \cond{AI}). If $\Ba$ leaves any protected sink untimely, if the earliest tainted deadline lies within \conf{two}{phase1\_margin\_rounds} control rounds, or if a convergence check fails, also push a conservative frontier hold $\Bone$ on all cuttable edges leaving the taint's unremovable closure.
\item \textbf{Fenced convergence ($\tone$):} The fence is established at the earliest instant $\tone$ where four conditions hold simultaneously:
(K1) every action in force for the incident is confirmed and journal-fenced \cond{F};
(K2) egress labels are confirmed across the entire source side \cond{AI};
(K3) taint $X(\tau)$ is conservatively reconciled across all ingested per-site journals up to the fence watermarks regardless of inter-site arrival order \cond{J}; and
(K4) fenced actions $A(\tau)$ separate $X(\tau)$ from every pending protected effect, every protected class of a tainted shared-memory context, and every other separable protected sink.
\item \textbf{Prune (Refinement \& guarded release):} On the reconciled graph at $\tone$, refine the boundary to the minimum cut $\Btwo$ (pushing any new actions $\Btwo \setminus (\Ba \cup \Bone)$ with source-side labels and awaiting their fenced confirmations); keep cut actions and release other temporary holds only after $\tone$, while all actions in force are fenced and removing the hold preserves complete $s$--$t$ separation across all sinks in (K4) \cond{R}; revocations are permanent.
\end{enumerate}
A decision is marked contained only once $\tone$ holds; otherwise (or if an alert's source is not yet ingested) it remains indeterminate and releases no holds.

\setcounter{theorem}{4}
\begin{theorem}[Safety of fence-then-prune]
\label{thm:safety}
Under fenced confirmation \cond{F}, source-side labels and admission \cond{AI}, guarded release \cond{R}, journal reconciliation \cond{J}, complete mediation, and true initial sources $X_0$, let $\tone$ be the first instant at which \textup{(K1)}--\textup{(K4)} hold. Then for all $\tau \ge \tone$, $A(\tau)$ separates $X(\tau)$ from every sink checked in \textup{(K4)} in $\GC(\tau)$, so zero tainted effects commit at any such sink at or after $\tone$, including during boundary refinement and pruning.
\end{theorem}
The proof proceeds by induction over ingested journal events maintaining $t \notin S(\tau)$ and $\mathrm{labels}(\tau) \supseteq S(\tau)$ across arbitrary inter-site journal arrival orders (\cref{app:theory}).
Exhaustive model checking of a discrete protocol model confirms that omitting source-side labels, reconciliation (K3), or fenced release each allows post-$\tone$ leaks (e.g., omitting source-side labels leaked in \fact{F-M016} of \fact{F-M015} runs); effects committed between the alert and $\tone$ count as pre-containment.

%% file: figures/fig3-walk.tex
\begin{figure*}[t]
\centering
\includegraphics[width=\textwidth]{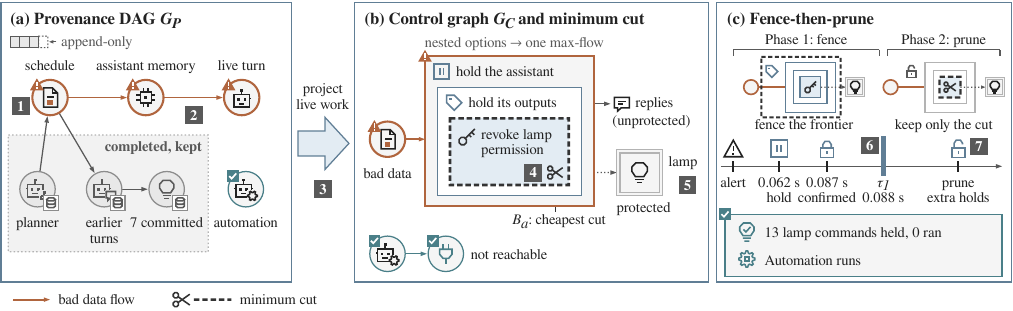}\par
\caption{One \HomeSvc run at the late alert. (a)~The trace follows the bad schedule \circled{1} to the live assistant \circled{2}; projection keeps only live work \circled{3}. (b)~The minimum cut \circled{4} revokes only the lamp permission and holds the lamp \circled{5}. (c)~Fence until $\tone$ \circled{6}, then prune \circled{7} (\cref{sec:design}).}
\label{fig:walk}
\end{figure*}

%% file: sections/s5-eval.tex
\section{Evaluation}
\label{sec:eval}

\input{figures/fig-llm-butterfly}
\input{figures/fig5-eval}

We evaluate \FB across four questions (\cref{fig:llm-butterfly,fig:eval-charts}):
(1)~In a physical smart-home application across six LLM deployments, does \FB prevent wrong actuations while preserving unaffected room and assistant operations (\cref{sec:eval:app})?
(2)~At the \emph{system layer}, how does \FB compare against native controls in Google A2A and LangGraph (\cref{sec:eval:systems})?
(3)~At the \emph{strategy layer} on \RT, how do \FB's two tiers compare against prior safeguards, and when does the optimized cut outperform full quarantine (\cref{sec:eval:strategy})?
(4)~How do boundary detectors, deadline-aware selection under delay, and mediation overhead behave (\cref{sec:eval:mech})?

\subsection{Experimental Setup and Metrics}
\label{sec:eval:setup}

\paragraph{Testbed, simulator, and workloads.}
Our distributed testbed spans three sites across two NTP-synchronized Linux hosts connected via NATS JetStream (site-to-site RTTs of \fact{F-T013} and \fact{F-T014}) and physical smart-home devices actuated through Home Assistant (median state confirmation \fact{F-T031}).
We evaluate three testbed workload families (\cref{app:workloads}):
(i)~the four-room physical smart-home LLM application (\cref{sec:eval:app});
(ii)~the long-running \HomeSvc smart-home service (\conf{three}{svc.json} agents, \fact{F-T360} tasks, early/late alerts at \conf{0~s}{D-79}/\conf{15~s}{svc.json}; \cref{sec:app}); and
(iii)~seven multi-agent workflow benchmarks of \conf{seven}{c2.json} agents across three sites (C2 cross-team fault, C3 session-wide stale state, N2 single reader, N3 root-plan fault, and three random DAGs with \conf{9--34}{rand-n7} tasks; early/late alerts at \conf{0.8~s}{x2}/\conf{10.8~s}{x2}), plus C2 under dynamic spread at rogue rates 0.3 and 1.0 (where tainted invocations emit undeclared in-scope effects and spawn child tasks).
In the controlled grids (\conf{10.5k}{LoC, eval tag} LoC harness), deterministic workers execute identical compiled plans and seeded faults against exact ground truth; for broad topology and spread sweeps, a discrete-event simulator executes the same controller code across \fact{F-S145} runs (\cref{app:sim}).

\paragraph{Evaluated controls and metrics.}
At the \emph{system layer}, thin adapters run the same workload plans over Google's A2A SDK~\citep{a2a2025} (\emph{cancel agent}, \emph{cancel all}, \emph{block input}) and LangGraph~\citep{langgraph2026} (\emph{cancel run}, \emph{block and resume}, \emph{time travel} checkpoint rollback).
At the \emph{strategy layer} on \RT, we compare \FB's two tiers and ablations (\cref{tab:strategy}) against topological and coarse safeguards~\citep{wang2025gsafeguard,xie2026spark,zhou2025guardian} and a sink-only \emph{\EGc} (\cref{app:baselines}).
Comparison grids trigger identical ground-truth alerts (\emph{early}, median \fact{F-T398} after faulty output production, or \emph{late}, after downstream readers start); dedicated grids evaluate \RT's built-in boundary detectors (\cref{sec:eval:mech}) and symptom-only backward root-cause resolution (\cref{app:baselines}).
Across \emph{all} controls, a run is \emph{clean after the alert} if zero tainted effects commit at any protected sink from the alert until the end of the run (an identical post-alert window) without timeout; for cut policies, we additionally audit \emph{post-$\tone$ leaks} (effects committed at or after $\tone$) to separate reaction-window leaks from post-convergence safety (\cref{thm:safety}).
\emph{Completed work} measures the compute-weighted fraction of healthy at-risk tasks that finish, while \emph{work kept without redo} (\emph{kept work}) excludes tasks re-executed after rollback, scored from independent destination ledgers (\cref{app:prereg}).

\subsection{Physical Smart-Home Application with LLMs}
\label{sec:eval:app}

\paragraph{Four-room smart-home application.}
We first evaluate \FB end-to-end on a physical smart-home application (\cref{fig:llm-butterfly}a) where a tree of LLM agents prepares three meeting rooms and a lobby from a shared calendar across three hosts: Host~1 runs the root planner, calendar service, and user assistant; Host~2 runs zone-a, room-r1, and room-r2; and Host~3 runs zone-b, room-r3, and a deterministic lobby agent.
Room agents configure physical lamps (via \RT's Home Assistant gateway, monitored by lab camera and state recorder) and post room announcements (protected effects), while lobby drives lobby lamps on a fixed schedule without reading the calendar and assistant answers six user questions about the day's schedule.
We evaluate \fact{F-T704} scored runs across six LLM deployments: local Qwen~3.8 27B (\fact{F-T563} runs across root and single-room calendar fault injections plus no-fault references), local Gemma~4 31B, three frontier cloud APIs (GPT~6.1~sol, Gemini~4, and Opus~5.5), and a heterogeneous tree (root Opus~5.5 planner with downstream Qwen~3.8 27B agents).
In each faulted run, a tool-boundary injection corrupts a calendar booking either at the \emph{root planner's first read} (affecting all meeting rooms) or at \emph{one room agent's local read} (room-r2).

\paragraph{Fault adoption and containment.}
Across all six model deployments, the LLM reading the corrupted calendar entry adopted the faulty booking in all faulted runs (e.g., \fact{F-T564} of 80 faulted Qwen runs, with at least \fact{F-T686} downstream agents computing on the corrupted plan under root injection).
Without containment, Qwen runs committed \fact{F-T565} (root) and \fact{F-T570} (local) wrong protected effects after the alert (\fact{F-T812}/\fact{F-T817} lamp commands, \fact{F-T813}/\fact{F-T818} announcements), while Gemma~4 and the cloud API models committed \fact{F-T707} wrong protected effects across their 20 uncontained runs (\fact{F-T839} lamp, \fact{F-T840} announcements).
Under both \FB tiers, \textbf{\fact{F-T567}} wrong protected effects committed after the alert across all Qwen runs at either injection point, \textbf{\fact{F-T588}} across Gemma~4 and all three frontier API models, and \textbf{\fact{F-T651}} in the mixed Opus--Qwen deployment, with \fact{F-T676} post-$\tone$ leaks and median $\tone$ of \fact{F-T568} (root) and \fact{F-T573} (local) on Qwen (\fact{F-T589}--\fact{F-T599} on other models).

\paragraph{Selective service retention and physical state audit.}
While both tiers achieved complete post-alert containment, they separated cleanly on service retention (\cref{fig:llm-butterfly}b).
Under root injection, full quarantine halted the assistant completely (all \fact{F-T669} of 6 user questions unanswered), whereas the optimized cut revoked only protected effect classes and kept the assistant answering \fact{F-T684} of 6 questions correctly per run across every model (only the single question about the corrupted room was affected).
Under local single-room injection, room-r2 committed only \fact{F-T693} lamp commands per run (vs.\ \fact{F-T694} without a fault, as its post-alert commands were held), while uninjected rooms completed all \fact{F-T690} scheduled lamp actuations and lobby preserved all of its automation commands.
A post-hoc audit against Home Assistant's physical state recorder across all \fact{F-T875} runs verified complete physical mediation: every physical lamp state transition (\fact{F-T876} changes) matched a journaled gateway commit (\fact{F-T877} unmatched; median lag \fact{F-T895}), and \textbf{\fact{F-T888}} of the \fact{F-T887} gateway-held lamp commands actuated a physical lamp.
On the controlled \HomeSvc replica (\cref{tab:svc}; \cref{sec:app:obs}), early alerts are bimodal depending on whether the schedule hold lands before the assistant's first read (\RejSrc blocked the read in \fact{F-T245} of 5 runs, \OC in \fact{F-T243}, \FQ in \fact{F-T244}, averaging \fact{F-T241} and \fact{F-T242} of \fact{F-T239} correct answers; pricing tainted replies at zero in simulation makes \OC block the read in all \fact{F-S111} paired runs), whereas across all \fact{F-T250} late-alert runs (5 per strategy), the schedule was already in shared memory (\fact{F-T249} correct answers), as repairing absorbed conversational state requires orthogonal rollback.

\subsection{System Layer: Agent Framework Controls}
\label{sec:eval:systems}

\textls[-10]{\Cref{tab:systems} and \cref{fig:eval-charts}a compare \RT and \FB against the native controls of Google A2A and LangGraph on the same workload plans and faults.}
Because A2A tracks tasks per agent without cross-agent provenance, cancelling the faulty agent is clean in only \fact{F-T312} of \fact{F-T310} content-fault runs (\fact{F-T557}) and blocking its output in \fact{F-T311} (\fact{F-T556}): every leaked effect is emitted by a downstream peer that consumed the tainted data before the alert.
Conversely, A2A's workflow-wide \emph{cancel all} stops all leaks but retains \fact{F-T314} of at-risk work.
Both \FB tiers are clean across all content and rogue runs (\fact{F-T315}/\fact{F-T310} and \fact{F-T172}/10; one-sided McNemar $p = $~\fact{F-T320} and \fact{F-T319} vs.\ A2A per-agent controls, Fisher $p = $~\fact{F-T336} under rogue spread) while retaining a median of \fact{F-T757} (full quarantine) and \fact{F-T316} (optimized cut) of at-risk work without redo (differing by \fact{F-T324} over \fact{F-T318} pairs, $p = $~\fact{F-T325}).

LangGraph exhibits the complementary trade-off of state-graph frameworks: \emph{block-and-resume} leaks in two-thirds of runs (clean in \fact{F-T479} of \fact{F-T474}, McNemar $p = $~\fact{F-T484}), while \emph{cancel-run} is clean (\fact{F-T475} of \fact{F-T474}) at the cost of discarding all at-risk work (median \fact{F-T476} kept).
LangGraph's \emph{time travel} rolls the graph back to a pre-fault checkpoint and re-executes: while re-execution eventually completes a median of \fact{F-T488} of healthy at-risk tasks, it retains only \fact{F-T487} of work \emph{without redo}, re-running \fact{F-T491} tasks, repeating \fact{F-T492} external actuations, and inflating late-alert makespan to \fact{F-T490} of its no-fault reference.
On paired runs, the optimized cut retains \fact{F-T485}, \fact{F-T477}, and \fact{F-T493} more work without redo than block-and-resume, cancel-run ($p = $~\fact{F-T478}), and time travel.

\subsection{Strategy Layer: Quarantine vs.\ Optimized Cut}
\label{sec:eval:strategy}

\paragraph{Why provenance reachability is necessary.}
Fixing \RT's enforcement gates across 11 strategies (\cref{tab:strategy}) isolates the impact of provenance tracing.
Source-only policies leak whenever taint has crossed to peers: halting the faulty agent's static team (\StopTeam~\citep{zhou2025guardian}) is clean in \fact{F-T130} of 54 runs, severing its future outgoing edges (\GSafe~\citep{wang2025gsafeguard}) in \fact{F-T131}, and blocking only its direct outputs (\RejSrc~\citep{xie2026spark}) in \fact{F-T132}.
Conversely, coarse global rules (blocking all protected effect classes in \SinkG or halting statically downstream tasks in \StopDown) achieve 54/54 cleanliness only by halting healthy work (\StopDown retains a median of \fact{F-T763} of at-risk work).
A sink-only \EGc (holding only tainted units' protected effects) ties the optimized cut when faults do not spread (\fact{F-S136} of \fact{F-S128} simulation runs), but trails by \fact{F-S134} points in kept work and \fact{F-S135} points in clean runs under active spread as unhalted units taint peers.

\paragraph{Two tiers and mechanism attribution.}
Both full quarantine and the optimized cut are clean after the alert in all \fact{F-T122} faulted batch runs (including all 12 rogue cells) and in all \fact{F-T172} runs at each of rogue rates 0.3 and 1.0 (as is the \TS ablation, which holds lineage-tainted artifacts and effects without pausing invocations).
Across all \fact{F-T800} main-grid runs that reached $\tone$, our fence audit confirms that all enforced actions were journal-fenced (\fact{F-T799} unfenced) and \textbf{\fact{F-T782} tainted effects committed after $\tone$} (median $\tone$ of \fact{F-T138}, 90th percentile \fact{F-T139}, solve time \fact{F-T141}).
On static batch workflows and moderate spread (rogue rate 0.3), both tiers (and \Greedy, which selects switches greedily by cost per covered edge) retain equivalent healthy work (pooled difference \fact{F-T152} over \fact{F-T151} matched runs, $p = $~\fact{F-T155}), making full quarantine an effective low-message default (\fact{F-T384} vs.\ \fact{F-T382} control messages/run).
Under \emph{saturated spread} (rogue rate 1.0, \cref{fig:eval-charts}b), however, full quarantine and \StopTeam both retain only \fact{F-T263} of at-risk work, whereas the optimized cut retains \textbf{\fact{F-T266}} in every run, a paired gain of \fact{F-T283} across all \fact{F-T285} pairs (sign-flip $p = $~\fact{F-T284}); in large-scale simulation (\fact{F-S122} paired runs under spread), it is also \fact{F-S112} percentage points cleaner.
This retention gain comes from the selective-control tier as a whole (source-side egress labels, finer switches, and guarded admission), while laminar max-flow solves vertex-local nested switches in one max-flow call (\cref{thm:fastpath}) and falls back to exact enumeration or MILP under cross-vertex coupling (Proposition~\ref{prop:fallback}).

\subsection{Detectors, Delay, Overhead, and Scale}
\label{sec:eval:mech}
\label{sec:eval:scale}

\paragraph{Boundary detectors and deadline-aware selection.}
Without external monitors, \RT's authority boundary detector fired spontaneously on out-of-scope dispatches in \fact{F-T179} of \fact{F-T180} saturated rogue runs at a median latency of \fact{F-T181} (\fact{F-T203} into the run, ahead of the \fact{F-T205} injected alert), and in a dedicated detector-triggered grid spanning authority-overreach and lease-exhaustion faults, built-in detectors caught \fact{F-T295} of \fact{F-T294} faulted runs (median latencies \fact{F-T297} and \fact{F-T296}; at most \fact{F-T298}), with both \FB tiers achieving clean containment and $\tone \le $~\fact{F-T304}.
Under injected control delays of \conf{1~s}{H14} and \conf{4~s}{H14} ($L$ of \conf{1.4}{H14} and \conf{4.4~s}{H14} on C2 and N2), all \fact{F-T404} cut runs still achieve fenced $\tone$ convergence with \fact{F-T406} post-$\tone$ leaks: while deadline-aware \OCc stays clean after the alert across all delays (\fact{F-T409} leaks; median $\tone$ scaling from \fact{F-T415} to \fact{F-T416} and \fact{F-T417}), \textbf{\CutND} (solving the minimum cut ignoring $L$ and $d(e)$) \textbf{suffers a pre-$\tone$ leak} at \conf{4~s}{H14} delay (clean after the alert in \fact{F-T412} of 12 runs, \cref{fig:eval-charts}c) by crediting slow actions that arrive after an effect commits at \fact{F-T413}.
Separately, while safety is independent of action costs $c(a)$ (\cref{thm:safety}), pricing revocations by an agent's expected work within its remaining resource budget raises Spearman correlation with lost compute from \fact{F-T206} (live-work heuristic) to \fact{F-T229} in an offline check (\cref{app:limitations}).\looseness=-1

\paragraph{Runtime overhead and ingestion scaling.}
While per-hop transactional journaling adds \fact{F-T438}--\fact{F-T446} on micro-handoffs (vs.\ \fact{F-T439}--\fact{F-T447} bare and \fact{F-T441}--\fact{F-T449} on A2A-SQLite), this cost is amortized whenever agents perform real computation: across all seven workflows (\cref{fig:eval-charts}d), \FB's end-to-end makespan is within \textbf{\fact{F-T468} to \fact{F-T469}} of A2A's (\fact{F-T466} to \fact{F-T467} above bare; laminar max-flow solve: \fact{F-L112}; coordinator peak memory \fact{F-T524}).
When stress-testing our single-process SQLite coordinator on a 63-agent workflow across 16, 32, and 63 logical sites on two hosts, centralized journal ingestion saturates, growing ingest lag to \fact{F-E165}--\fact{F-E166}~s at 63 sites (\cref{app:limitations}).
In that regime, \FB's fail-safe contract holds, marking un-ingested alerts as indeterminate, never claiming $\tone$ (\fact{F-E169} of \fact{F-E168}), and releasing \fact{F-E173} holds prematurely, though effects that commit before journals arrive cannot be intercepted (\fact{F-E171} of \fact{F-E168} runs clean after the alert), motivating sharded coordinator ingestion for larger clusters.

%% file: figures/fig-llm-butterfly.tex
\begin{figure}[!t]
\centering
\includegraphics[width=\columnwidth]{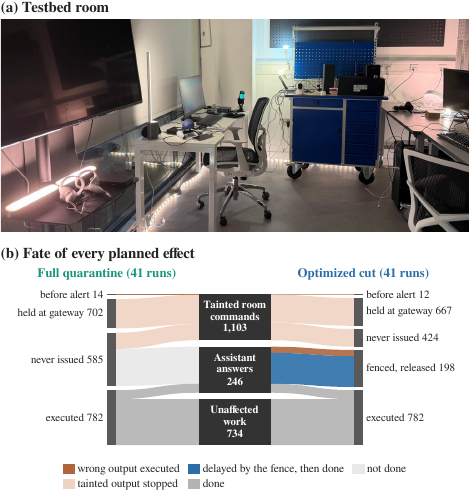}\par
\caption{Physical smart-home application across six LLM deployments (\cref{sec:eval:app}). (a)~Testbed room. (b)~Fate of planned effects: both tiers stop tainted commands; only the optimized cut keeps the assistant answering.}
\label{fig:llm-butterfly}
\end{figure}

%% file: figures/fig5-eval.tex
\begin{figure*}[!t]
\centering
\includegraphics[width=\textwidth]{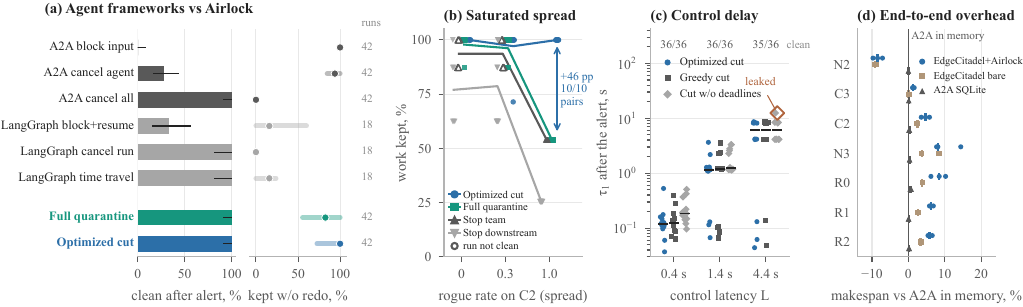}\par
\caption{Testbed workflows. (a)~Agent frameworks vs.\ \FB (\cref{sec:eval:systems}). (b)~Kept work as spread grows; hollow marks are runs that leaked (\cref{sec:eval:strategy}). (c)~$\tone$ under control delay (\cref{sec:eval:mech}). (d)~Makespan relative to A2A (\cref{sec:eval:scale}).}
\label{fig:eval-charts}

\vspace{-2pt}
{\makeatletter\def\@captype{table}\makeatother
\begin{minipage}[t]{0.485\textwidth}
\centering\footnotesize\setlength{\tabcolsep}{2.2pt}
\caption{System-layer comparison (\fact{F-T310} content, 10 rogue runs; $^\ast$LangGraph: \fact{F-T474} content, 6 rogue; $^\dagger$paired diff.\ in text). Best \textbf{bolded}, worst \underline{underlined}.}
\label{tab:systems}
\vspace{1pt}
\fitcol{%
\begin{tabular}{@{}ll|cc|c@{}}
\toprule
\multicolumn{2}{@{}l}{System / Control} & \multicolumn{2}{c}{Clean Runs $\uparrow$} & \multicolumn{1}{c@{}}{Retained $\uparrow$} \\
\cmidrule(lr){3-4}\cmidrule(l){5-5}
Framework & \multicolumn{1}{l}{Policy} & Content (\fact{F-T310}) & \multicolumn{1}{c}{Rogue (10)} & No redo (med.) \\
\midrule
\multicolumn{2}{@{}l|}{No control} & \fact{F-T751} & \fact{F-T752} & \fact{F-T753} \\
\midrule\midrule
\multirow{3}{*}{\textbf{A2A}}
 & - cancel agent & \fact{F-T312} & \underline{\fact{F-T332}} & \fact{F-T754} \\
 & - cancel all & \textbf{\fact{F-T313}} & \textbf{\fact{F-T333}} & \underline{\fact{F-T314}} \\
 & - block input & \underline{\fact{F-T311}} & \underline{\fact{F-T331}} & \textbf{\fact{F-T755}} \\
\cmidrule(r){1-2}
\multirow{3}{*}{\textbf{LangGraph$^\ast$}}
 & - cancel run & \textbf{\fact{F-T475} of \fact{F-T474}} & \textbf{\fact{F-T499} of 6} & \underline{\fact{F-T476}} \\
 & - block, resume & \underline{\fact{F-T479} of \fact{F-T474}} & \underline{\fact{F-T496} of 6} & --$^\dagger$ \\
 & - time travel & \textbf{\fact{F-T486} of \fact{F-T474}} & \textbf{\fact{F-T500} of 6} & \fact{F-T487} \\
\cmidrule(r){1-2}
\multirow{2}{*}{\textbf{\FB}}
 & - \FQ & \textbf{\fact{F-T756}} & \textbf{\fact{F-T258}} & \fact{F-T757} \\
 & - \OC & \textbf{\fact{F-T315}} & \textbf{\fact{F-T172}} & \textbf{\fact{F-T316}} \\
\bottomrule
\end{tabular}}
\end{minipage}\hfill
\begin{minipage}[t]{0.495\textwidth}
\centering\footnotesize\setlength{\tabcolsep}{1.8pt}
\caption{\textls[-20]{Strategy-layer comparison on \RT (\fact{F-T122} batch runs; saturated: C2 at rogue rate 1.0, 10 runs). Best (excl.\ \NoC) \textbf{bolded}, worst \underline{underlined}.}}
\label{tab:strategy}
\vspace{1pt}
\fitcol{%
\begin{tabular}{@{}ll|cc|c|c@{}}
\toprule
\multicolumn{2}{@{}l}{Strategy on \RT} & \multicolumn{2}{c}{Main Grid (54)} & \multicolumn{1}{c}{Cost} & \multicolumn{1}{c@{}}{Saturated} \\
\cmidrule(lr){3-4}\cmidrule(lr){5-5}\cmidrule(l){6-6}
Category & \multicolumn{1}{l}{Policy} & Clean $\uparrow$ & \multicolumn{1}{c}{Kept (med.) $\uparrow$} & \multicolumn{1}{c}{Msgs} & Clean, Kept $\uparrow$ \\
\midrule
\multicolumn{2}{@{}l|}{\NoC} & \fact{F-T133} & -- & 0 & 0, -- \\
\midrule\midrule
\multirow{3}{*}{\textbf{Source}}
 & - \GSafe & \fact{F-T131} & \textbf{\fact{F-T774}} & \fact{F-T764} & not run \\
 & - \RejSrc & \underline{\fact{F-T132}} & \textbf{\fact{F-T775}} & \textbf{\fact{F-T765}} & \underline{0}, -- \\
 & - \StopTeam & \fact{F-T130} & \fact{F-T758} & \fact{F-T766} & \textbf{10}, \fact{F-T190} \\
\cmidrule(r){1-2}
\multirow{2}{*}{\textbf{Coarse}}
 & - \StopDown & \textbf{\fact{F-T128}} & \underline{\fact{F-T759}} & \underline{\fact{F-T388}} & \textbf{10}, \underline{\fact{F-T191}} \\
 & - \SinkG & \textbf{\fact{F-T129}} & \fact{F-T760} & \fact{F-T767} & not run \\
\cmidrule(r){1-2}
\multirow{2}{*}{\textbf{\FB}}
 & - \FQ & \textbf{\fact{F-T126}} & \fact{F-T762} & \fact{F-T384} & \textbf{\fact{F-T259}}, \fact{F-T263} \\
 & - \OC & \textbf{\fact{F-T123}} & \textbf{\fact{F-T776}} & \fact{F-T382} & \textbf{\fact{F-T183}}, \textbf{\fact{F-T189}} \\
\cmidrule(r){1-2}
\multirow{3}{*}{\textbf{Ablation}}
 & - \TS & \textbf{\fact{F-T127}} & \fact{F-T761} & \fact{F-T768} & not run \\
 & - \Greedy & \textbf{\fact{F-T125}} & \textbf{\fact{F-T777}} & \fact{F-T386} & not run \\
 & - \CutND & \textbf{\fact{F-T124}} & \textbf{\fact{F-T778}} & \fact{F-T769} & not run \\
\bottomrule
\end{tabular}}
\end{minipage}}
\vspace{-8pt}
\end{figure*}

%% file: sections/s6-related.tex
\section{Related Work and Discussion}
\label{sec:related}

\paragraph{Agent guardrails and topological defenses.}
Single-agent guardrails and planner-level information-flow control restrict tool calls inside one agent~\citep{costa2025fides,debenedetti2025camel}, complementing \FB without white-box access.
Across multi-agent topologies, errors cascade rapidly~\citep{cemri2025mast,huang2025resilience,xie2026spark}, and prior defenses pair anomaly detection~\citep{forough2023anomaly,li2026adr} with source-local remediation: G-Safeguard severs a flagged agent's future outgoing edges~\citep{wang2025gsafeguard}, GUARDIAN isolates the flagged agent~\citep{zhou2025guardian}, and From Spark to Fire filters unverified claims at the source~\citep{xie2026spark}.
Severing future outgoing edges misses peers that already read the output (\cref{sec:eval:strategy}); \FB traces forward from such alerts on $\GP$ to fence consumers.

\paragraph{Framework controls, revocation, and recovery.}
Production frameworks provide per-task cancellation~\citep{a2a2025} or whole-graph interrupt and checkpoint rollback~\citep{langgraph2026}, and SOUNDGATE~\citep{stopmeansstop2026} fences in-flight side effects on cancellation; \RT adds daemon-mediated provenance for selective boundaries without workflow halts or rollback.
\textls[-12]{\RT's delegation tree builds on capability attenuation, actuation leases, and secret isolation~\citep{birgisson2014macaroons,zhan2026hearthnet,zhan2026edgecitadel,forough2026agents,prakash2026aip,muruaga2026bounded,syros2026saga,deochake2026hbhc}.}
Whereas VERA models authority loss under edge revocation~\citep{liu2026vera} and RSQ retires an authorization root via epoch barriers~\citep{zhu2026rsq}, \FB contains \emph{data taint} propagating within valid credentials under site deadlines.
Similarly, whereas intrusion recovery traces provenance backward for offline repair~\citep{king2003backtracking,goel2005taser,kim2010retro,yu2026rollback} and transactional runtimes undo single-agent actions~\citep{mohammadi2026atomix,patil2024goex}, \FB contains faults \emph{online} before commit.\looseness=-1

\paragraph{Limitations and future work.}
\cref{app:limitations} documents our prototype's scope: \FB contains external effects online rather than rolling back agent memory~\citep{yu2026rollback}; beyond roughly ten agents across three hosts, our single-process coordinator's journal ingestion saturates (\cref{sec:eval:scale}), motivating sharded ingestion, budget-aware pricing (\cref{sec:eval:mech}), and post-incident endorsement.

\section{Conclusion}
\label{sec:conclusion}

\textls[-6]{We presented \FB, a multi-agent runtime and guardrail co-designing \RT's vertical permission hierarchy and lock-fenced site switches with an online provenance DAG and fence-then-prune minimum-cut containment.
On our main testbed and smart-home grids, \FB blocks all post-alert faulty actuations with zero leaks after fenced convergence, retaining \fact{F-T485} to \fact{F-T477} more healthy work than A2A and LangGraph (\fact{F-T283} over full quarantine under dense spread).
These results demonstrate that uniting runtime authority boundaries with online provenance breaks the all-or-nothing trade-off between safety and availability, providing a practical foundation for deploying multi-agent systems over real-world services and physical devices.}\looseness=-1\par

%% file: sections/app-a-theory.tex
\newenvironment{stmt}[2]{\par\smallskip\noindent\textbf{#1~#2.}\ \itshape}{\par\smallskip}
\newenvironment{rmk}[2]{\par\smallskip\noindent\textit{#1~#2.}\ }{\par\smallskip}
\newcommand{\qedh}{\hfill$\square$}

\section{Containment as a Grouped $s$--$t$ Cut}
\label{app:theory}

This appendix states the problem the controller solves and proves the results the main text uses; statement numbers match the main text.
A.1 defines the live graph, the actions and the grouped cut.
A.2 shows that the problem is hard in general (Proposition~1).
A.3 identifies the case the runtime meets in practice, nested switches at each vertex, where one max-flow is exact (Theorem~2); Proposition~2.1 shows that the runtime's switches fall in it, and A.4 gives the exact fallback for the rest (Proposition~3).
A.5 adds deadlines (Proposition~4).
A.6 proves that fence-then-prune lets no tainted effect commit after $\tone$ (Theorem~5) and lists what the theorem does not cover (\cref{tab:scope}).
A.7 lists the machine checks behind every constructive statement.

\subsection{Problem statement}

\paragraph{Live graph.}
The controller works on a directed graph $G = (V, E)$ (parallel edges allowed).
Vertices are live invocations $u$ (each holding one credential $\kappa_u$ minted for it), targets $b$ (gateway-mediated endpoints; a \emph{sink} target has an uncuttable edge $b \to t$), a super-source $s$ joined by uncuttable edges to the tainted invocations, and the super-sink $t$.
Edges are the pending one-hop events: immediate effects $(u, b, \kappa_u)$, scheduled effects $(u, b, \kappa_u, d)$ with earliest commit time $d$, messages and spawns $(u, u')$, and results $(u, \mathrm{parent}(u))$.
Each edge has an individual cost $c(e) \in \mathbb{R}_{\ge 0} \cup \{\infty\}$ ($\infty$ = uncuttable).
Only edges on some $s \to t$ path matter.

\paragraph{Actions.}
A finite set $\Act$ of actions, each with a cost $c(a) \in \mathbb{R}_{\ge 0} \cup \{\infty\}$ and a removal set $R(a) \subseteq E$.
Selecting $C \subseteq \Act$ removes $\Rem(C) = \bigcup_{a \in C} R(a)$ (ANY semantics).
Individual edge cuts are the actions with $R = \{e\}$.
The runtime's vocabulary is listed with its removal sets in \cref{tab:vocab}.
No action revokes a delegation subtree; descendants are covered by their own actions and by taint-aware admission (\cref{app:safety}).

\paragraph{Delegation tree (special case).}
When actions are credential revocations with the semantics ``revoking $w$ revokes every descendant'', the removal sets $R(w) = \bigcup_{u \in \mathrm{sub}(w)} E(u)$, with $\mathrm{sub}(w)$ the subtree of $w$, form a laminar family; we call this the \emph{tree model}.
It is the action model with a particular $R$; nothing below depends on it except where stated.

\paragraph{Problem GC (grouped cut).}
Minimise $\sum_{a \in C} c(a)$ over $C \subseteq \Act$ such that $G \setminus \Rem(C)$ has no $s \to t$ path; $\mathrm{OPT} = \infty$ if no cut exists.
$B \subseteq \Act$ is a \emph{cut for $(G, X)$} if $G \setminus \Rem(B)$ has no path from $X \subseteq V$ to $t$.
Removing more edges never creates a path, so cuts are closed upward.
The objective is additive in actions; it is the proxy the solver optimises, and real work preserved is measured separately.

\paragraph{Scope.}
ANY semantics only.
The ALL variant, in which an edge survives while any of several authorities is intact, is not used by the runtime; it needs set-level enforcement invariants and special handling of unwitnessed edges, and is not treated.
GC is the weighted minimum label $s$--$t$ cut problem~\citep{jha2002two,zhang2011label} with one label per action and multi-labelled edges; the tree model, the single-label case and the action model are cost-preservingly inter-reducible with linear blow-up (private-chain and series-split constructions; brute-force checked).

\subsection{Hardness}

\begin{stmt}{Proposition}{1 (cited: Set Cover hardness)}
GC is NP-hard and admits no polynomial-time $(1-\varepsilon)\ln n$-approximation for any $\varepsilon > 0$ unless P = NP, even in the tree model with a depth-1 tree, unit action costs, every edge in exactly one removal set, no individually cuttable edge, and $G$ a parallel composition of $s \to t$ paths (series--parallel, planar, acyclic). It remains Set-Cover-hard when every removal set is a subset of the out-edges of a single vertex, provided the removal sets may cross.
\end{stmt}

\noindent\emph{Proof.}
(i) From Set Cover ($U = \{1..n\}$, sets $S_1, \dots, S_m$): one $s \to t$ path $P_i$ per element, with one uncuttable edge per set $S_j \ni i$, in $R(a_j)$, $c(a_j) = 1$.
$P_i$ is destroyed iff some selected $S_j$ contains $i$; $s, t$ are disconnected iff the selected sets cover $U$; costs are equal, so ratios are preserved.
(ii) Nodes $s, v, y_1..y_n, t$, uncuttable $s \to v$, $v \to y_i$, $y_i \to t$, $R(a_j) = \{v \to y_i : i \in S_j\}$: all of $\delta^+(v)$ must go, which is again Set Cover.\qedh{}
Both reductions were checked by exhaustive enumeration (\fact{F-M001} and \fact{F-M002} instances).
Set-Cover hardness: \citet{feige1998threshold}; \citet{dinur2014parallel} under P $\neq$ NP.
GC is also W[2]-hard in the number of selected actions~\citep{fellows2010label}, and APX-hard even when every removal set is the boundary of a disjoint path-shaped region of the graph (reduction from Vertex Cover; the regions then have several entry vertices).

\begin{rmk}{Remark}{}
In the instances of (i) every action is \emph{gated} in the sense of \cref{app:fastpath} (at $(s, \mathrm{out})$ by the first edges of its paths); their gate sets cross.
Gatedness alone therefore does not help (\fact{F-M014} such instances checked); the operative condition of Theorem~2 is laminarity per port.
Instance (ii) is the ``star property'' hardness of shared-risk group cuts~\citep{coudert2007srlg} and shows the condition cannot be dropped.
\end{rmk}

\subsection{The fast path: gated, per-port laminar instances}
\label{app:fastpath}

\paragraph{Definition (gate).}
A \emph{port} is $(v, \sigma)$, $\sigma \in \{\mathrm{out}, \mathrm{in}\}$, with star $\delta^\sigma(v)$.
Action $a$ is \emph{gated at $(v, \sigma)$ by $\Pi_a \subseteq \delta^\sigma(v)$} if every simple $s \to t$ path meets $R(a)$ iff it meets $\Pi_a$.
A \emph{vertex-local} action ($R(a) \subseteq \delta^\sigma(v)$) is gated by $\Pi_a = R(a)$.

\begin{stmt}{Lemma}{A.1 (substitution)}
$a$ is gated by $\Pi_a$ iff for every $F \subseteq E$, $G \setminus (F \cup R(a))$ is $s$--$t$ disconnected exactly when $G \setminus (F \cup \Pi_a)$ is.
\end{stmt}
\noindent\emph{Proof.}
($\Rightarrow$) a simple path survives $F \cup R(a)$ iff it avoids $F$ and $R(a)$ iff it avoids $F$ and $\Pi_a$.
($\Leftarrow$) if a simple path $P$ meets $R(a)$ but not $\Pi_a$, take $F = E \setminus P$.\qedh

\paragraph{Chain gadget $G^\gamma$.}
Let $\Act'$ be the finite-cost actions, each assigned to one port $\pi(a)$ at which it is gated by $\Pi_a$, such that at every port the family $\mathcal{L}_p = \{\Pi_a : \pi(a) = p\}$ is laminar (equal sets merged, keeping the minimum cost).
For each port $(v, \mathrm{out})$: one node $a_S$ per set $S \in \mathcal{L}_p$, and an edge $\mathrm{anchor}(S) \to a_S$ of capacity $c(S)$, where $\mathrm{anchor}(S)$ is the node of the smallest $S' \supsetneq S$ in $\mathcal{L}_p$, or $v$ if none (mirrored at in-ports: $a_S \to \mathrm{anchor}(S)$).
For each edge $e = (x, y)$, let $S_e$ be the smallest set of $\mathcal{L}_{(x,\mathrm{out})}$ containing $e$ and $T_e$ the smallest set of $\mathcal{L}_{(y,\mathrm{in})}$ containing $e$ (either may not exist); add \emph{one} edge $\mathrm{src}(e) \to \mathrm{dst}(e)$ of capacity $c(e)$, with $\mathrm{src}(e) = a_{S_e}$ or $x$, $\mathrm{dst}(e) = a_{T_e}$ or $y$.
$G^\gamma$ has $|V| + |\Act'|$ nodes and $|E| + |\Act'|$ edges.

\begin{stmt}{Theorem}{2 (fast path)}
If every finite-cost action is gated at its assigned port and the gate families are laminar at every port, then $\mathrm{OPT}$ equals the minimum $s$--$t$ cut of $G^\gamma$, and an optimal action set is read off the cut (an action is selected iff its chain edge is cut; an edge is cut iff its gadget edge is). One max-flow on $|V| + |\Act'|$ nodes.
\end{stmt}
\noindent\emph{Proof sketch.}
By Lemma~A.1 applied action by action (it holds for every $F$), $C$ is a cut iff $\bigcup_{a \in C} \Pi_a \cup C_E$ is an edge cut, so each action may be replaced by its gate set.
A simple $s \to t$ path $P$ of $G$ maps to the path $\gamma(P)$ of $G^\gamma$ that, at each edge $e = (x, y)$, descends the out-chain at $x$ to $a_{S_e}$ (if $S_e$ exists), crosses the gadget edge of $e$, and ascends the in-chain at $y$ from $a_{T_e}$ (if $T_e$ exists).
Conversely a gadget node at $(x, \mathrm{out})$ is entered only from $x$ or its parent and left only towards a child or along a gadget edge of an edge in its set, and symmetrically at in-ports, so every simple $s \to t$ path of $G^\gamma$ is $\gamma(P)$ for a unique $P$: no spurious connectivity.
Cutting the chain edge of $\Pi_a$ separates exactly the paths $\gamma(P)$ with $P \cap \Pi_a \neq \emptyset$ (by laminarity, the chain from $x$ to $a_{S_e}$ passes through every set at the port that contains $e$), and cutting the gadget edge of $e$ separates exactly the $\gamma(P)$ with $e \in P$.
Hence cuts of $G^\gamma$ of finite capacity and cuts of the substituted instance correspond with equal cost.\qedh{}
Brute-force checks: \fact{F-M003} gated instances, \fact{F-M005} laminar-forest instances, and \fact{F-M007} flat instances with families at both edge ends.

\begin{table}[t]
\centering\small
\caption{The runtime vocabulary and its removal sets.}
\label{tab:vocab}
\begin{tabular}{@{}p{2.3cm}p{1.35cm}p{2.3cm}l@{}}
\toprule
Action & Site & $R(a)$ & Port \\
\midrule
task hold on $u$ & host($u$) & all pending out-edges of $u$ & $(u,\mathrm{out})$ \\
egress hold on $u$ & gateway & immediate effect edges out of $u$ & $(u,\mathrm{out})$ \\
revoke $\kappa_u$ & verifier & effect edges out of $u$ under $\kappa_u$ & $(u,\mathrm{out})$ \\
filter $(\kappa_u, b)$ & ingress($b$) & effect and scheduled edges $(u, b)$ under $\kappa_u$ & $(b,\mathrm{in})$ \\
admission block at $b$ & ingress($b$) & all effect and scheduled edges into $b$ & $(b,\mathrm{in})$ \\
scheduled-effect, result or admission hold on $e$ & scheduler, host or admitting site & $\{e\}$ & either \\
\bottomrule
\end{tabular}
\end{table}

\begin{stmt}{Proposition}{2.1 (runtime mapping)}
Under single credentials (each credential held by one invocation, each effect edge under one credential) and the port assignment of \cref{tab:vocab}, every action is vertex-local and the families are laminar at every port: at $(u, \mathrm{out})$, $R(\text{task hold}) \supseteq R(\text{egress hold}) \supseteq R(\text{revoke } \kappa_u)$ and singletons; at $(b, \mathrm{in})$, $R(\text{block } b) \supseteq R(\text{filter}(\kappa, b))$ with filters for different credentials disjoint, and singletons. Hence Theorem~2 applies.
\end{stmt}
Configurations that violate the hypotheses, and are detected by the check: a credential shared by several invocations (the revocation is not vertex-local); an effect that requires several credentials (crossing revocation sets); deadline masks under non-monotone latencies (Proposition~4(c)); a two-port action (filter, singleton hold) placed at the wrong port.
Of \fact{F-M013} runtime-vocabulary instances, all \fact{F-M008} with monotone latencies and \fact{F-M010} of \fact{F-M011} with random latencies passed the check; the other \fact{F-M012} were correctly routed to the exact fallback.

\paragraph{The check.}
On every instance (each graph, source or deadline update) the solver computes the removal sets (masks), verifies vertex-locality syntactically, assigns each action to a port (actions with two candidate ports, parallel $u \to b$ edges only, are assigned by 2-SAT over crossing pairs), and verifies laminarity per port, at cost $O(|\Act'|^2 \max_a |R(a)|)$.
If the check passes, one max-flow; otherwise Proposition~3.
Exactness is claimed only for instances that pass.

\subsection{Fallback}

\begin{stmt}{Proposition}{3 (exact fallbacks)}
(a) Exception set. Let $Y \subseteq \Act'$ be a declared set such that $\Act' \setminus Y$ satisfies the hypotheses of Theorem~2 on $G$. Then $\mathrm{OPT}$ is computed by at most $2^{|Y|}$ max-flows: for each $C_Y \subseteq Y$, delete $\Rem(C_Y)$, solve the residual instance by Theorem~2, add $c(C_Y)$, and take the minimum.
(b) ILP. With binaries $y_a$ and potentials $\pi_v \in [0,1]$,
\begin{align*}
\min\ & \textstyle\sum_a c(a)\, y_a \quad \text{s.t.}\ \ \pi_s = 1,\ \pi_t = 0,\\
& \textstyle\pi_{\mathrm{tail}(e)} - \pi_{\mathrm{head}(e)} \le \sum_{a:\, e \in R(a)} y_a \quad (\forall e \in E),
\end{align*}
is exact, with $|E| + 2$ constraints and $\sum_e |\{a : e \in R(a)\}|$ nonzeros.
(c) Approximation. Giving each edge the cost of its cheapest covering action and taking an ordinary min cut is an $f$-approximation, $f = \max_a |R(a)|$.
\end{stmt}
\noindent\emph{Proof.}
(a) The branch $C_Y = C \cap Y$ of an optimal $C$ is enumerated; in it the best completion is the residual optimum.
Gates survive edge deletion (the gate condition is per simple path and the residual has fewer), and laminarity survives intersecting every set with a fixed edge set.
(b) For integral $y$ the potential system is feasible iff $\Rem(\{a : y_a = 1\})$ is a cut (set $\pi_v = 1$ iff $v$ is reachable from $s$ in the surviving graph); an uncuttable edge covered by no action yields $\pi_{\mathrm{tail}} \le \pi_{\mathrm{head}}$.
(c) The optimal grouped solution induces an edge cut of modified cost at most $f \cdot \mathrm{OPT}$.\qedh{}
The enumeration (a) was checked by brute force on \fact{F-M006} random instances with 1--3 exceptions.

\begin{rmk}{Remark}{(why $Y$ is an exception set and not ``the ungated actions'')}
Three parallel paths $s \to a \to t$, $s \to b \to t$, $s \to c \to t$ with actions $A = \{s{\to}a, s{\to}b\}$, $B = \{b{\to}t, s{\to}c\}$, $C = \{a{\to}t, c{\to}t\}$, unit costs, all edges uncuttable: every action is gated (at $(s, \mathrm{out})$ and at $(t, \mathrm{in})$), yet the gate sets pairwise cross and no port assignment is laminar (\fact{F-M025} laminar assignments exist); $\mathrm{OPT} = $~\fact{F-M026}.
Theorem~2 does not apply with $Y = \emptyset$; it does with $Y = \{A\}$.
The runtime declares $Y$ by vocabulary (the configurations listed under Proposition~2.1); it never searches for a minimum one, and the ILP is exact regardless.
\end{rmk}

\subsection{Deadlines}

\paragraph{Model.}
Decision time $t_{\mathrm{now}}$, planning allowance $\beta$, enactment latency upper bound $L(\sigma(a))$ of the site at which $a$ is enacted, and for each edge a conservative lower bound $d(e)$ on its earliest commit time ($\infty$ for an edge currently removed by a confirmed in-force action; exact under gateway not-before scheduling; $t_{\mathrm{now}}$ for an unheld immediate effect).
The \emph{timely mask} of $a$ is $\hat R(a) = \{ e \in R(a) : t_{\mathrm{now}} + \beta + L(\sigma(a)) < d(e) \}$, and an action in force takes its mask from confirmed state.
For a mask to certify timely enactment, each credited command must be issued within $\beta$ of $t_{\mathrm{now}}$, and $L$ then bounds its remaining enactment time, including queueing at the site; the runtime uses $\beta$ as a configured prediction, not a solver timeout, and neither measures nor enforces the first bound. Exceeding either bound invalidates the timing prediction, and convergence still requires the fenced-state check below.
There is no tree semantics for timely removal: in the tree model each witness has its own mask and no antichain convention.

\begin{stmt}{Proposition}{4 (deadlines)}
(a) $I_\beta = (G, \Act, c, \hat R)$ is an instance of the action model; $C$ is a feasible cut iff $\hat\Rem(C)$ is an edge cut; every hardness result holds for $I_\beta$; Theorem~2 applies to $I_\beta$ iff the masks pass the check, and Proposition~3 applies always.
(b) Ancestor redundancy fails. Vertex $v$ with out-edges $e_1, e_2, e_3$ to $t$ of individual cost 100; parent $p$, $c(p) = 1$, $R(p) = \{e_1, e_2, e_3\}$, $\hat R(p) = \{e_1, e_3\}$; child $u$, $c(u) = 10$, $\hat R(u) = R(u) = \{e_1, e_2\}$. $\mathrm{OPT}(I_\beta) = $~\fact{F-M022} by selecting both $p$ and $u$; the antichain convention yields \fact{F-M023}, and the untruncated tree semantics ($r_u \ge r_p$) yields \fact{F-M024} while falsely reporting $e_2$ removed.
(c) If $R(a) \subseteq R(b)$ and $L(\sigma(b)) \le L(\sigma(a))$ then $\hat R(a) \subseteq \hat R(b)$, and disjoint sets stay disjoint; hence laminarity at a port survives truncation when latency is non-increasing along inclusion (always when the nested actions share a site, as filters and blocks do).
(d) Vertex-locality survives truncation; a single-edge gate survives if the gate edge stays in the mask (sufficient, not necessary); a sealed region may leak.
(e) $\hat R_{\beta'}(a) \subseteq \hat R_\beta(a)$ for $\beta \le \beta'$, so a larger budget never lowers $\mathrm{OPT}$ and may make the instance infeasible.
\end{stmt}
\noindent\emph{Proof.}
(a), (c), (d) and (e) are immediate from the definition; (b) is exhaustive enumeration, also against the literal ILP of Proposition~3(b).\qedh

The masks are predictions; the fenced confirmations of \cref{app:safety} are ground truth.
An action whose mask is a proper subset of its removal set is paid in full and must be modelled per site.
During refinement, an edge already held by a confirmed in-force action has $d = \infty$, and every other edge keeps its literal deadline mask. The distinction matters for adaptive phase~1, whose source side may contain uncovered edges: a replacement action is credited only where it can take effect before the edge commits, and guarded release checks the resulting cut. Under the full-frontier freeze every pending source-side edge is held, so the literal masks are safe but conservative there, and truncation genuinely constrains only a solve that no in-force cut protects (a one-shot deadline-aware cut, or a partial phase~1).

\subsection{Safety of the two-phase protocol}
\label{app:safety}

\paragraph{Model.}
Every site keeps an ordered journal of commits, admissions and confirmations; the controller ingests journals, and ``time'' below is its reconciled state.
The live graph $G(\tau)$ holds the pending one-hop edges; a \emph{commit} of $e = (x, y)$ removes $e$ from $G$ and, if $x$ is tainted, taints $y$; a commit into $t$ with a tainted tail is a \emph{leak} ($t$ itself is never tainted).
$X(\tau)$ is the taint set, grown only by journaled commits from the declared source set $X(\tau_0)$.
$A(\tau)$ is the set of actions confirmed at \emph{every} site that enforces them (a partially released action is not in $A$).
The assumptions are:
\begin{itemize}
\item \cond{F} \emph{Fenced confirmation and commit-time re-check.} A site confirms action $a$ with a journal position $j$ such that every commit of an edge of $R(a)$ at that site precedes $j$, and no edge of $R(a)$ commits there after $j$ while $a$ is in force (the gateway re-validates holds and revocations at the commit point of every effect); the controller counts $a$ as confirmed at the site only after ingesting its journal through $j$. \emph{Mediation completeness:} every edge of $G$ commits through a site that enforces the actions covering it.
\item \cond{AI} \emph{Admission invariant.} The controller pushes a quarantine label set $Q(\tau)$ to the sites. A new edge whose tail is labelled is held at admission by an admission hold that is in force before the edge can commit (the site journals the admission and the hold as one event). The controller keeps $Q(\tau) \supseteq S(\tau) := \Reach_{G(\tau) \setminus \Rem(A(\tau))}(X(\tau))$, the source side of the cut in force, and confirms a label push before any release that can enlarge $S$.
\item \cond{R} \emph{Guarded release.} An action $a$ (including an admission hold) is released only if $A(\tau) \setminus \{a\}$ is a cut for $(G(\tau), X(\tau))$ on the reconciled state.
\item \cond{J} \emph{Consistent reconciliation.} Journals are per site; a commit's taint effect depends on its tail's taint at commit time, possibly decided by a commit on a slower site, and the fence positions of \cond{F} cover only commits of covered edges. A solve or release check therefore uses either a consistent snapshot (every labelled site's journal ingested through a common watermark, commits processed in timestamp order) or the conservative rule (J$'$): ``a known commit taints its head if its tail is on the controller's source side when the commit is reconciled''.
\end{itemize}

\paragraph{Messages.}
A message or spawn edge $(u, u')$ is admitted at the sender's site, where \cond{AI} applies to $u$, and commits at the receiver's admission point (the claim of the new task, or delivery to a live $u'$).
In between it is a pending edge of $G(\tau)$, also after $u$ finishes, so $G(\tau)$ contains every in-flight message (the case of \citet{zhu2026rsq}, §2.4); by mediation completeness an action covering it is enforced at the receiver.
Its content, and so its taint, is fixed at the send; taking the tail's taint at the commit, as the model does, over-approximates it.

\paragraph{Trust and faults.}
Sites, their journals, the gateway and the controller are trusted and non-Byzantine: they fail only by crashing or partition, and a journal records what its site did.
Faults are non-adversarial: an agent may act on corrupted data or malfunction, but does not target the enforcement mechanism; adversarial agents are out of scope.
This is why journals and confirmations are taken as reported, not authenticated (contrast the authenticated evidence of \citealt{zhu2026rsq}).

\paragraph{Effect linearization and end-to-end scoring.}
At the gateway, effect admission and action enforcement acquire the same per-site control lock, linearizing every effect commit against fence positions (\cond{F}).
While Theorem~5 bounds effects linearized from $\tau_1$ onward (classifying effects accepted before $\tau_1$ as pre-convergence, S3), our empirical evaluation applies a strictly stronger end-to-end criterion: outcomes are scored directly from external destination ledgers and physical device state (\cref{app:prereg}), counting a run as \emph{clean after the alert} only if zero tainted effects reach a protected sink at any time after the alert timestamp, including during the pre-$\tau_1$ convergence window.

\paragraph{Phase 1.}
Let $U$ be the set of vertices reachable from $X$ by edges that no action can remove (untruncated removal sets; with \cref{tab:vocab} typically $U = X$), and $B_1$ a set of actions whose removal sets cover $\delta^+(U)$ (task holds on every invocation in $U$, scheduled-effect holds on frontier scheduled effects, admission blocks at targets in $U$), applied at once with the labels $Q := U$.
If $t \notin U$ then, once $B_1$ is confirmed, every pending edge of $\delta^+(U)$ is held, so $B_1$ is a cut for $(G, U) \supseteq (G, X)$; if $t \in U$ no cut exists (every edge leaving $U$ is coverable, else its head would be in $U$).
Phase~1 \emph{converges} at the instant $\tau_1$ when the convergence check K1--K4 of Theorem~5 holds; otherwise $U$ and $B_1$ are recomputed.
Confirming all of $B_1$ everywhere is a sufficient, stricter form.
Deadline masks play no role in phase~1: every covering action is applied, and the journal decides which edges crossed.

\paragraph{Phase 2.}
Solve $(G(\tau_1), X(\tau_1))$ with the masks of the previous subsection (in-force actions with confirmed masks) by Theorem~2 or Proposition~3, obtaining $B_2$; apply $B_2 \setminus B_1$; wait for fenced confirmations; re-check that $A \setminus (B_1 \setminus B_2)$ is a cut on the reconciled state; release $B_1 \setminus B_2$ under \cond{R}.

\begin{stmt}{Theorem}{5 (two-phase safety, dynamic 1-hop runtime)}
Assume \cond{F}, \cond{AI}, \cond{R}, \cond{J}, mediation completeness, and that $X(\tau_0)$ contains the true sources. Let $\tau_1$ be the first instant at which the convergence check holds literally: (K1) every action the check relies on is confirmed with its fence position at every enforcing site, and every site's journal is ingested through those positions; (K2) labels are confirmed on the whole source side $S(A(\tau_1))$, including tainted contexts with no live invocation; (K3) $X(\tau_1)$ is reconciled from the journaled commits under \cond{J}; (K4) $A(\tau_1)$ is a cut for $(G(\tau_1), X(\tau_1))$ against every protected sink. (Under the full-frontier $B_1$, once all of it is confirmed, $S = U$, so K2 is the labels $Q := U$.) Then for every $\tau \ge \tau_1$, $A(\tau)$ is a cut for $(G(\tau), X(\tau))$. Consequently no tainted-tail effect commits at a protected sink after $\tau_1$: during phase~2, during the site-by-site release of $B_1 \setminus B_2$, and afterwards, whatever new events the agents produce.
\end{stmt}
\noindent\emph{Proof sketch.}
Invariant: $t \notin S(\tau)$ and $Q(\tau) \supseteq S(\tau)$; it holds at $\tau_1$.
Induction over the ingested events.
\emph{Commit of $e = (x, y)$:} by \cond{F} and mediation completeness $e \notin \Rem(A)$; if $x \in X \subseteq S$ then $y \in S$ already, so $S$ is unchanged and $y \neq t$; if $x \notin X$ nothing is tainted and $e$ leaves $G$.
\emph{Admission of $e = (x, y)$:} if $x \in Q$ the edge enters held \cond{AI}; if $x \notin Q \supseteq S$ it is unreachable from $X$; $S$ is unchanged.
\emph{Confirmation of an action or label push:} $A$ or $Q$ grows; $S$ shrinks or stays.
\emph{Guarded release of $a$:} by \cond{R} the new source side avoids $t$ and its labels were pushed first; while $a$ is released site by site, $A \supseteq A \setminus \{a\}$, so the invariant holds throughout.
\emph{Withdrawal of a label from $x$:} a label is withdrawn only for a vertex outside the source side after the release, which \cond{R} with taint-aware admission guarantees (the labels for that side are pushed and confirmed before the release); $S$ stays inside that side during the release ($A \supseteq A \setminus (B_1 \setminus B_2)$), so $x \notin S$ throughout and $Q \supseteq S$ is kept.
A profile that disables taint-aware admission while keeping the phase-1 freeze can withdraw labels on the post-release source side; it is outside the theorem's hypotheses.
\emph{Removal of executed edges or finished invocations:} $S$ shrinks.\qedh

\paragraph{Protocol remarks and scope.}
The convergence check evaluates K1--K4 on a candidate action set to three outcomes: \textsc{contained} when K1--K4 hold, \textsc{not\_contained} when $t \in U$ ($\mathrm{OPT} = \infty$) or a path from $X$ to a protected sink survives in $G \setminus \Rem(B)$, and \textsc{indeterminate} when any required fence position, label confirmation, or journal watermark is missing (e.g., at a partitioned or lagging site).
Only \textsc{contained} licenses a phase-2 release under \cond{R}; \textsc{not\_contained} retains all holds and adds actions, while \textsc{indeterminate} retains all holds and waits.
Under the consistent-snapshot rule of \cond{J}, the reconciled taint set $X_J(\mathcal{C})$ is the least fixpoint over timestamp-ordered commits through the watermark and is independent of inter-site ingestion order; under the conservative rule (J$'$), $X(\tau) \supseteq X_J(\mathcal{C})$ over-approximates taint so Theorem~5 holds under every ingestion order.
Once the phase-2 release of $B_1 \setminus B_2$ completes at $\tau_r$, only edges in $\Rem(B_2)$, irreversible actions from phase~1, or admission holds remain in force, and any admission hold whose tail lies outside $S(\tau)$ can be released immediately under \cond{R}.
Finally, if phase~1 issues only the at-alert deadline-aware cut $B_a$ (adding the frontier freeze only when $B_a$ is infeasible or within a slack margin), Theorem~5 still holds after $\tau_1$ provided K1--K4 are checked on the full source side $S(B_a)$ under \cond{J} and literal masks (verified across \fact{F-M015} discrete-model runs per variant: \fact{F-M017} post-$\tau_1$ leaks under the full conditions vs.\ \fact{F-M016} leaks when labelling $U$ only and \fact{F-M018} benign re-check refusals without literal masks).
\cref{tab:scope} summarizes the scope boundaries of Theorem~5 (S1--S10).

\begin{table}[t]
\centering\footnotesize
\caption{Scope boundaries of Theorem~5.}
\label{tab:scope}
\begin{tabular}{@{}l>{\raggedright\arraybackslash}p{0.86\columnwidth}@{}}
\toprule
S1 & \emph{Liveness.} A partitioned or lagging site can prevent $\tone$ from arriving; fail-closed site leases keep the site safe by holding new and pending effects upon expiry, at the cost of liveness. \\
S2 & \emph{Executed effects.} Committed external effects are not rolled back; Theorem~5 governs commits after $\tone$. \\
S3 & \emph{Pre-containment effects.} Effects that commit between the alert and $\tone$ (or are accepted under the gateway lock before $\tone$) are counted as pre-$\tone$ commits, separate from post-$\tone$ safety. \\
S4 & \emph{Declared sources.} Safety is relative to the declared initial source set $X_0$. \\
S5 & \emph{Cost optimality.} $B_2$ is cost-optimal for the $\tone$ snapshot and mask model; any valid cut is equally safe under Theorem~5. \\
S6 & \emph{Mediation.} External effects must pass through a site or gateway that enforces $\Act$. \\
S7 & \emph{Masks vs.\ fences.} Deadline masks $\hat R(a)$ are planning predictions; convergence and release rely on fenced journal positions. \\
S8 & \emph{Irreversible actions.} Irreversible phase-1 actions (e.g., credential revocation) cannot be lifted in phase~2 and are priced accordingly. \\
S9 & \emph{Healthy work.} Theorem~5 is a safety guarantee; healthy work retention is evaluated empirically (Section~5). \\
S10 & \emph{Persistent shared-memory services.} In task-scoped workflows, invocation labels $Q(\tau)$ satisfy K2 and \cond{AI} directly. For persistent multi-turn agents sharing state across external user requests (\HomeSvc), \FQ pauses the agent to block all subsequent turns and transitive dispatches, whereas \OC revokes the agent's protected sink classes before $\tone$ so that no protected actuation can commit while read-only user replies remain available (if a shared-memory agent can also dispatch to downstream agents that hold protected sink authority, K2 requires either pausing the shared context or propagating taint on its outbound dispatches). \\
\bottomrule
\end{tabular}
\end{table}

\subsection{Verification summary}
All constructive reductions and gadget claims were verified by exhaustive enumeration on random instances: Set-Cover and Vertex-Cover reductions (\fact{F-M001} and \fact{F-M002}), the chain gadget (\fact{F-M005} forest instances, \fact{F-M007} flat instances with two-sided edges, and the runtime-vocabulary instances in Section~A.3), exception-set enumeration (\fact{F-M006}), the deadline and crossing-gate counterexamples, and the two-phase protocol in a discrete runtime model (\fact{F-M015} runs per variant plus two deterministic scenarios; \fact{F-M020} checks pass, \fact{F-M021} fail).

%% file: sections/app-b-baselines.tex
\paragraph{Open-source runtime and controls.}
The open-source \RT distributed agent runtime~\citep{zhan2026edgecitadel} underlying \FB is publicly available at
\mbox{\scriptsize\url{https://github.com/EdgeCitadelTeam/EdgeCitadel}}.
\cref{tab:baselines} lists the evaluated controls. The paper names \FB's two settings by what they do; the registrations in \cref{app:prereg} use the names under which the experiments ran.

\begin{table}[ht]
\centering\footnotesize\setlength{\tabcolsep}{2.5pt}
\caption{Evaluated controls across the strategy layer (sharing \RT, alert timing, and enforcement gates) and system layer (comparing end-to-end agent frameworks).}
\label{tab:baselines}
\fitcol{%
\begin{tabular}{@{}>{\raggedright}p{0.31\columnwidth}>{\raggedright}p{0.19\columnwidth}>{\raggedright\arraybackslash}p{0.44\columnwidth}@{}}
\toprule
Control & Origin & Enforcement scope \\
\midrule
\multicolumn{3}{@{}l}{\emph{Strategy layer} (on \RT)} \\
\GSafe & topological \citep{wang2025gsafeguard} & flagged agent: egress revoked, results parked \\
\RejSrc & source filter \citep{xie2026spark} & faulty source's outputs blocked \\
\StopTeam & team halt \citep{zhou2025guardian} & faulty agent's entire team stopped \\
\StopDown & graph halt & all tasks downstream of source stopped \\
\SinkG & class gate & all protected effect classes blocked globally \\
\EGc & sink-only rule & tainted units' protected effects held \\
\TS & \FB ablation & lineage-touched artifacts/effects held only \\
\FQc & \FB (default) & \TS{} + pause and egress-revoke tainted units \\
\OCc & \FB (assured) & deadline-aware min-cost cut, two-phase protocol \\
\Greedy; \CutND & \FB ablations & greedy cut; exact cut ignoring deadlines \\
\midrule
\multicolumn{3}{@{}l}{\emph{System layer (framework native controls)}} \\
A2A \citep{a2a2025} & native SDK & cancel agent, cancel all, block input \\
LangGraph \citep{langgraph2026} & native SDK & cancel run, block and resume, time travel \\
\bottomrule
\end{tabular}}
\end{table}
\paragraph{Per-cell results.}
With \fact{F-T192} repetitions per cell, we summarize the main grid's cells by one rule instead of tabulating them: a setting wins a cell if its mean kept work is at least 5 points higher, and the win is clear if the 95\% bootstrap intervals do not overlap.
Of the \fact{F-T145} cells with at-risk healthy work, the optimized cut won \fact{F-T146} against full quarantine (\fact{F-T147} clear), tied \fact{F-T148} and lost \fact{F-T149} (\fact{F-T150} clear); pooled over \fact{F-T151} matched runs the difference is \fact{F-T152} (95\% interval \fact{F-T153} to \fact{F-T154} points, $p = $~\fact{F-T155}).
The clear win is rand-n7-s1 at the late alert (\fact{F-T197} points); the two losses are C3 at the early alert (\fact{F-T198}) and rand-n7-s0 at the late alert (\fact{F-T199}).
By the same rule the optimized cut won \fact{F-T156} cells against Cut without deadlines (\fact{F-T157} clear), tied \fact{F-T158} and lost \fact{F-T159}, and against Greedy cut won \fact{F-T160}, tied \fact{F-T161} and lost \fact{F-T162}.
On these seven-agent DAGs, individual tainted branches typically expose only one or two candidate cut points, so greedy selection often coincides with the exact minimum cut at native latency.
However, because the laminar max-flow fast path (Theorem~2) solves the exact cut in 0.022--0.031~ms (\S5.5)---orders of magnitude below network and journaling latency---\FB achieves exact polynomial-time optimality at negligible cost, avoiding both the $\Theta(\ln n)$ worst-case approximation gap of greedy selection on shared removal sets (Proposition~1) and the pre-$\tone$ safety leak of Cut without deadlines under 4~s control delay (Proposition~4, \S5.5), while retaining \fact{F-T283} more healthy work than \FQ under dense rogue spread (\S5.4) and \fact{F-T197} points more on \texttt{rand-n7-s1}.

\paragraph{Symptom-only root-cause narrowing.}
When an alert reports only a downstream symptom rather than the originating invocation, the controller traverses $\GP$ backward from the symptom to rank candidate upstream root causes before tracing forward.
In simulation (under stress latency with a late alert on the first contaminated downstream artifact), backward ranking placed the true faulty invocation first in \fact{F-S053} of instances and within the top three in \fact{F-S054}.
In a 36-run testbed grid (\fact{F-T709} runs across C2, N2, N3, and one random workflow $\times$ \OC, \Greedy, and \TS), each control received a single symptom-only alert and initiated containment from the single top-ranked candidate.
With one symptom alert and no healthy contrast outputs, all upstream ancestors on the symptom path share the same evidence score and are ordered by graph distance: the symptom invocation itself ranks first (distance~0) and the true upstream origin ranks second in all \fact{F-T713} of \fact{F-T709} runs (tied on score with earlier ancestors).
Starting containment from the distance-0 symptom alone suffices when all downstream effects lie within the symptom's closure (\fact{F-T715} of 12 runs clean for \OC, \fact{F-T716} for \Greedy, \fact{F-T717} for \TS), but misses sibling branches fed by the upstream origin in N3 (\fact{F-T718} of 9 clean).
Thus, a single symptom alert narrows candidate origins to the upstream ancestor chain on $\GP$, whereas isolating a single upstream ancestor without ties requires either healthy contrast outputs or seeding $X_0$ with the candidate ancestor set.

%% file: sections/app-c-summary.tex
\begin{table*}[!t]
\centering\footnotesize\renewcommand{\arraystretch}{0.93}
\caption{Registered hypotheses and their outcomes, using the paper's terminology (the system's internal name in registration logs denotes \OC; ``Trace \& quarantine'' denotes \FQ). Dates are in 2026; each registration preceded its grid's first run. ``Sim.'' marks outcomes evaluated in simulation.}
\label{tab:hyp}
\begin{tabular}{@{}l>{\raggedright}p{0.27\textwidth}l>{\raggedright}p{0.385\textwidth}>{\raggedright\arraybackslash}p{0.10\textwidth}@{}}
\toprule
ID & Prediction & Registered & Outcome & Main text \\
\midrule
H1 & With rogue spread, the optimized cut contains more runs than Trace \& stop & by Oct~22 & Not supported: both clean in every rogue run of the main grid (\fact{F-T143} and \fact{F-T144} of 12) & Table~3 \\
H2 & Without rogue spread, Trace \& stop keeps at least as much work & by Oct~22 & Not supported: Trace \& stop keeps less & Table~3 \\
H3 & As control latency grows, Cut without deadlines leaks more than the optimized cut & by Oct~22 & Holds weakly: one run leaked at 4~s (\fact{F-T412} of 12 clean), none of the optimized cut's & \S5.5 \\
H4 & The optimized cut issues fewer control operations than Greedy cut & by Oct~22 & Not supported: \fact{F-T381} vs.\ \fact{F-T385} logical actions, \fact{F-T382} vs.\ \fact{F-T386} messages (medians) & Table~3 \\
H5 & $\GC$ stays flat with live workload while $\GP$ grows & by Oct~22 & Holds (sim.): over 40 workflow instances $\GP$ grows from \fact{F-S082} to \fact{F-S083} nodes, while $\GC$ has \fact{F-S085} and then \fact{F-S084} & --- \\
H6 & Sink guard contains about as much but keeps less work & Oct~21 & Holds (sim.) & --- \\
H7 & Full quarantine contains less at higher latency & Oct~21 & Refuted on the random sweep; holds only on C2 (sim.) & --- \\
H8 & HomeSvc: no tainted lamp command and $\ge$90\% of requests answered; full quarantine and Stop team answer none & Oct~21 & Fails on one clause: full quarantine answered all \fact{F-T251} requests in the one early-alert run that blocked the read; the other clauses hold & \S2.2, \S5.2 \\
H9 & HomeSvc: report the optimized cut's tainted replies and added latency & Oct~21 & Reported: \fact{F-T246} and \fact{F-T247} tainted replies per run (full quarantine \fact{F-T248}); added latency \fact{F-T349} and \fact{F-T350} & Table~1 \\
H10 & Containment against A2A's native controls & Oct~21 & Holds in all four parts & \S5.3 \\
H10-LG & Containment against LangGraph's native controls & Oct~21 & (a) and (b) hold; (c) in part: time travel's late-alert makespan is \fact{F-T490} its reference, below the registered 1.2$\times$ & \S5.3 \\
H11 & Overhead without faults within fixed bounds & Oct~21 & Fails as registered: per-hop hand-offs exceed the bounds (e.g., \fact{F-T450} A2A-SQLite's, bound 3$\times$); makespan stays within its $\pm$10\% bound & \S5.5 \\
H12 & Scaled testbed: no tainted commit after $\tone$ at any site count & Oct~24 & Fails: no run reached $\tone$ (\fact{F-E169} of \fact{F-E168}); \fact{F-E171} of \fact{F-E168} clean after the alert & \S5.5 \\
H12-sim & The simulator agrees with the scaled testbed at 32 sites & Oct~24 & Not borne out (\cref{app:sim}) & --- \\
H13 & Built-in detectors trigger containment & Oct~23 & H13a fails as written: \fact{F-T295} of \fact{F-T294} runs detected, the miss a hand-off failure; the other clauses hold & \S5.5 \\
H14 & Containment under added control delay & Oct~24 & H14a holds; H14b holds except the predicted rise in freezes; H14c null & \S5.5 \\
H15 & Full quarantine against the optimized cut under rogue spread & Oct~24 & H15a: a tie, both clean in every run; H15b: the cut keeps more work at rogue rate 1.0 & \S5.4 \\
H16 & Meeting-room application & Oct~24 & H16a and H16b hold; H16g not decided as registered, checked post hoc against Home Assistant's history & \S5.2 \\
\bottomrule
\end{tabular}
\end{table*}

%% file: sections/app-c-prereg.tex
\paragraph{Scoring and statistical protocol.}
All safety and work-retention metrics are scored from independent destination ledgers at each sink rather than from runtime journals.
A run is scored as \emph{clean after the alert} iff zero tainted protected effects reach a sink after the alert timestamp, and as achieving \emph{fenced $\tone$} iff every action in force for the incident across all controller decisions is confirmed with its site journal fence position (verified across all \fact{F-T509} converged runs in the main, \HomeSvc, system-comparison, and control-delay grids, with \fact{F-T799} unfenced runs outside the saturated 63-agent ingest stress test).
\emph{Completed work} measures the compute-weighted fraction of healthy at-risk tasks that finish; \emph{work kept without redo} (\emph{kept work}) excludes tasks re-executed after checkpoint rollback (so LangGraph time-travel re-executions count as redone), and refused hand-offs into the declared tainted closure do not penalize the healthy upstream sender.
In the system comparisons (H10, H10-LG, H15), paired clean-rate comparisons use one-sided exact McNemar tests (content faults) and one-sided Fisher exact tests (rogue faults) at $\alpha = 0.05$, while paired kept-work comparisons (H10, H10-LG, H14, H15) use one-sided sign-flip permutation tests at $\alpha = 0.05$.

\paragraph{Protocol amendments.}
Five pre-scoring protocol refinements were recorded during testbed bring-up:
(i)~after initial pilot tail latencies exceeded the $150$~ms planning estimate, the site control-latency bound was fixed to $L = 400$~ms across all native testbed grids (and $L = \text{added delay} + 400$~ms in the H14 delay sweep);
(ii)~the LangGraph comparison (H10-LG) was executed on a 129-run representative subset (C2, C3, and rand-n7-s1, three repetitions per cell) with unchanged statistical thresholds;
(iii)~in saturated rogue cells ($\rho = 1.0$), where \RT's built-in authority boundary detector fires before the harness's scheduled alert, runs are scored from the earliest alert on the fault lineage and reported separately;
(iv)~in \HomeSvc (H8), service availability is scored on \emph{correct} (uncontaminated) replies rather than raw reply count; and
(v)~the 63-agent scaling sweep (H12) was evaluated at 16, 32, and 63 logical sites after default daemon listen-backlog saturation at 3 and 8 sites, and re-verified on the released controller build (H12-A2): the registered replay check flags \fact{F-E175} post-run releases (\fact{F-E177} during the run), while a post-hoc check confirms that all \fact{F-E185} of \fact{F-E184} actions in those releases were covered by the converged record's fence positions before release (\fact{F-E186} uncovered).

%% file: sections/app-d-workloads.tex
\input{figures/fig1-c2}

\paragraph{Workloads.}
\begin{table}[ht]
\centering\footnotesize\setlength{\tabcolsep}{2.5pt}
\caption{Evaluated workloads. Counts are workload design constants; alerts fire at fixed wall-clock delays after the faulty output is produced (\emph{early}) or after its last consumer starts (\emph{late}).}
\label{tab:workloads}
\fitcol{%
\begin{tabular}{@{}lccp{2.2cm}l@{}}
\toprule
Workload & Agents & Tasks & Fault & Alert (early / late) \\
\midrule
Meeting rooms & \multicolumn{4}{p{0.74\columnwidth}}{LLM delegation tree; one corrupted booking injected at calendar read (root or local room)} \\
\HomeSvc (\texttt{svc}) & \conf{3}{svc.json} & \fact{F-T360} & wrong device schedule; shared memory & \conf{0~s}{D-79} / \conf{15~s}{svc.json} \\
\midrule
C2 & \conf{7}{c2.json} & \conf{17}{c2.json} & wrong value crosses teams & \conf{0.8~s}{x2} / \conf{10.8~s}{x2} \\
C3 & \conf{7}{c3.json} & \conf{16}{c3.json} & stale version, session-wide & \conf{0.8~s}{x2} / \conf{0.8~s}{x2} \\
N2 & \conf{7}{n2.json} & \conf{9}{n2.json} & single downstream reader & \conf{0.8~s}{x2} / \conf{0.8~s}{x2} \\
N3 & \conf{7}{n3.json} & \conf{14}{n3.json} & corrupted root plan & \conf{0.8~s}{x2} / \conf{0.8~s}{x2} \\
rand (3) & \conf{7}{rand-n7} & \conf{27--34}{rand-n7} & 3 random DAG workflows & \conf{0.8~s}{x2} / \conf{0.8~s}{x2} \\
\bottomrule
\end{tabular}}
\end{table}
A workflow benchmark is a plan of seven agents on three sites that a deterministic worker executes step by step through the runtime; faults and rogue draws are seeded by task identifiers, so the same plan produces the same messages and effects under every control (\cref{tab:workloads}).
\emph{C2} (\cref{fig:c2}): a leaf in one team produces a wrong artifact that its teammate passes to both leaves of the other team, where one commands a device; the root's final report is also protected, and the other team has healthy device work. It tests cross-team spread into a physical action. At rogue rates 0.3 and 1.0 tainted invocations also take undeclared actions and spawn tasks (at most \conf{200}{spawn cap}); at rate 1.0 the faulty invocation also dispatches outside its authority scope.
\emph{C3}: a stale configuration version is read inside a long conversation, and later turns still carry it after the source is fixed: session-wide taint, the hard case for precise blocking.
\emph{N2}: a single bad output read by one downstream agent that commands a device.
\emph{N3}: a corrupted root plan that reaches every downstream agent.
\emph{rand-n7-s0}--\emph{s2}: three random seven-agent DAG workflows from the simulator's generator; \emph{rand-n63-s0} scales the same generator to 63 agents across 16, 32, and 63 logical site daemons on two hosts (\cref{tab:scale}).
\HomeSvc and the meeting-room application are detailed in Sections~2 and~5.2.

\paragraph{Alert timing.}
Controlled comparison grids trigger identical ground-truth alerts naming the faulty origin at fixed wall-clock delays (\cref{tab:workloads}) after faulty output production (\emph{early}, before downstream readers consume it) or after the last reader starts (\emph{late}; \conf{20~s}{H16} after the injected calendar read in the meeting-room application).
From this origin alert, \FB traces $\GP$ forward online to determine reachable taint, select containment switches, and preserve healthy branches.
Dedicated grids evaluate \RT's built-in authority and resource boundary detectors without injected alerts (Section~5.5) and symptom-only backward root-cause narrowing (\cref{app:baselines}).
Configured delays scale with each grid's execution time scale (time scale~2 for the workflow, system-comparison, and detector grids; time scale~1 for \HomeSvc).

%% file: figures/fig1-c2.tex
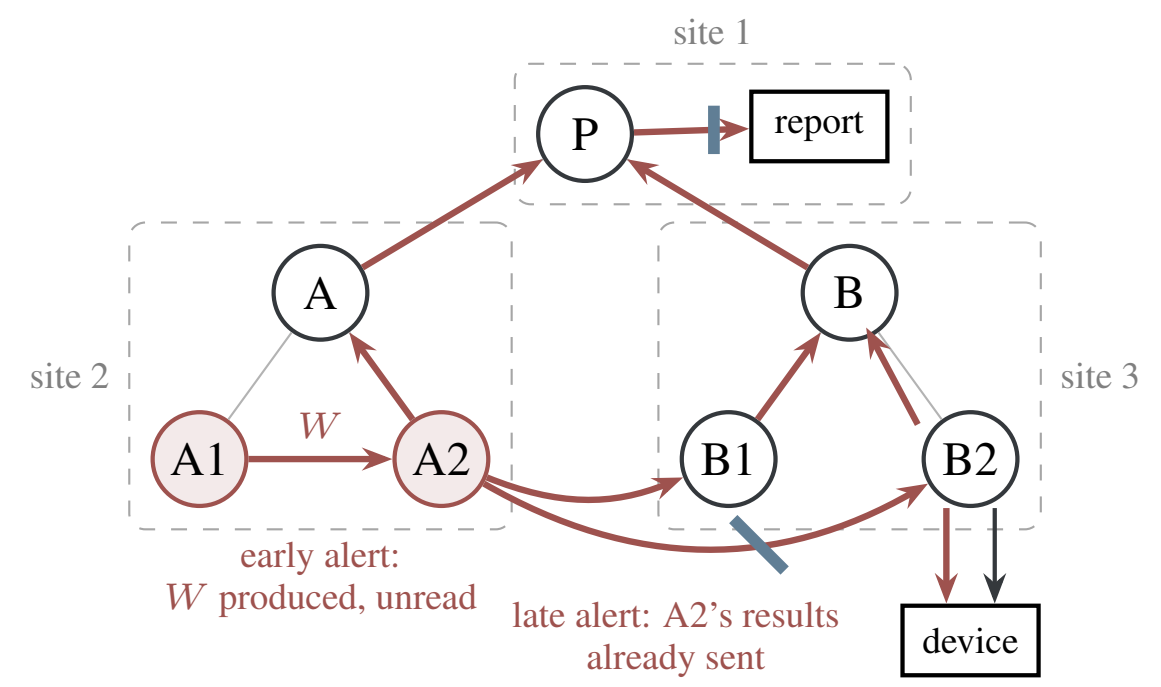
\begin{figure}[t]
\centering
\begin{tikzpicture}[
  ag/.style={circle,draw=clean,thick,minimum size=6.2mm,inner sep=0pt,font=\footnotesize},
  bad/.style={ag,draw=taint,fill=taint!12},
  sink/.style={rectangle,draw=black,thick,minimum width=9mm,minimum height=4.6mm,font=\scriptsize,align=center,inner sep=1pt},
  site/.style={draw=gray!70,dashed,rounded corners=3pt,inner sep=4pt},
  tree/.style={draw=gray!60,thin},
  flow/.style={-{Stealth[length=2mm]},draw=taint,very thick},
  ok/.style={-{Stealth[length=2mm]},draw=clean,thick},
  cutbar/.style={draw=ctrl,line width=2.2pt},
  lbl/.style={font=\scriptsize,align=center}
]
\node[ag]  (P)  at (0,0)        {P};
\node[ag]  (A)  at (-1.75,-1.05) {A};
\node[ag]  (B)  at (1.75,-1.05)  {B};
\node[bad] (A1) at (-2.55,-2.15) {A1};
\node[bad] (A2) at (-0.95,-2.15) {A2};
\node[ag]  (B1) at (0.95,-2.15)  {B1};
\node[ag]  (B2) at (2.55,-2.15)  {B2};
\node[sink] (rep) at (1.55,0.05) {report};
\node[sink] (dev) at (2.55,-3.35) {device};
\draw[tree] (P)--(A) (P)--(B) (A)--(A1) (A)--(A2) (B)--(B1) (B)--(B2);
\draw[flow] (A1) -- node[above,lbl,text=taint]{$W$} (A2);
\draw[flow] (A2) to[bend right=22] (B1);
\draw[flow] (A2) to[bend right=30] coordinate[pos=0.62] (inflight) (B2);
\draw[flow] (A2) -- (A);
\draw[flow] (A) -- (P);
\draw[flow] (B1) -- (B);
\draw[flow] ([xshift=-1.2mm]B2.north west) -- ([xshift=-1.2mm]B.south east);
\draw[flow] (B) -- (P);
\draw[flow] (P) -- (rep);
\draw[flow] ([xshift=-1.6mm]B2.south) -- ([xshift=-1.6mm]dev.north);
\draw[ok]   ([xshift=1.6mm]B2.south) -- ([xshift=1.6mm]dev.north);
\draw[cutbar] ($(inflight)+(-0.17,0.17)$) -- ($(inflight)+(0.17,-0.17)$);
\draw[cutbar] ($(P)!0.55!(rep)+(0,-1.6mm)$) -- ($(P)!0.55!(rep)+(0,1.6mm)$);
\begin{scope}[on background layer]
\node[site,fit=(P)(rep),label={[lbl,text=gray]above:site 1}] {};
\node[site,fit=(A)(A1)(A2),label={[lbl,text=gray]left:site 2}] {};
\node[site,fit=(B)(B1)(B2),label={[lbl,text=gray]right:site 3}] {};
\end{scope}
\node[lbl,text=taint,anchor=north] at ($(A1)!0.5!(A2)+(0,-0.42)$) {early alert:\\ $W$ produced, unread};
\node[lbl,text=taint,anchor=north] at ($(inflight)+(-0.55,-0.25)$) {late alert: A2's results\\ already sent};
\end{tikzpicture}
\caption{Authority and influence in the C2 workload: gray tree edges show delegation, red arrows data influence across branches.
Leaf A1 produces a wrong artifact $W$, which reaches the final report and, through B2's tainted task (red), the team-B device that B2 also commands from healthy tasks (gray).
Blue bars: one valid cut at the late alert.}
\label{fig:c2}
\end{figure}

%% file: sections/app-e-simulation.tex
The discrete-event simulator executes the same containment controller code over generated multi-agent workflows.
It models scheduled effect deadlines, additive action costs, and seeded rogue propagation over a single time-ordered event log (so inter-site journal ingestion lag does not arise; unless marked realistic, control latencies use stress settings).

\paragraph{Broad topology and spread sweeps.}
Across \fact{F-S145} runs over the random-workflow sweep and C2, zero tainted protected effects committed after confirmed containment (\fact{F-S030} in the random sweep, \fact{F-S031} on C2).
In a paired \fact{F-S127}-run batch comparing \OC and \FQ across spread regimes, under rogue spread (\fact{F-S122} runs) \OC changes clean rate by \fact{F-S112} percentage points (95\% CI \fact{F-S113} to \fact{F-S114}) and kept work by \fact{F-S117} points (95\% CI \fact{F-S118} to \fact{F-S119}); without rogue spread (\fact{F-S128} runs), both settings achieve identical clean rates in all \fact{F-S128} runs.
Comparing \EGc against \OC over the same batch, \EGc matches \OC's clean outcome in \fact{F-S136} of \fact{F-S128} non-rogue runs, while under rogue spread \EGc trails \OC by \fact{F-S135} points in clean rate and \fact{F-S134} points in kept work (narrowing to \fact{F-S138} clean and \fact{F-S137} kept work when paired with egress revocation).
In the \HomeSvc scenario under realistic latency at the early alert, pricing tainted replies at zero causes \OC to block the tainted schedule read (\fact{F-S108} tainted reads, \fact{F-S107} wrong replies, \fact{F-S109} of requests answered across all \fact{F-S111} paired runs, matching \RejSrc), whereas crediting tainted replies as kept work yields \fact{F-S110} wrong replies.

\paragraph{Comparison with the 63-agent scaled testbed (H12-sim).}
Because the simulator's single event log assumes instantaneous journal visibility, H12-sim predicted that all paired runs at 32 sites would be clean after the alert; on the two-host testbed, centralized SQLite journal ingestion lag caused \fact{F-E125} of \fact{F-E120} runs to remain clean (\cref{tab:scale}, \cref{app:limitations}).

%% file: sections/app-f-limitations.tex
\begin{table*}[!t]
\renewcommand{\thetable}{G.1}\renewcommand{\theHtable}{G.1}
\centering\footnotesize\setlength{\tabcolsep}{4pt}\renewcommand{\arraystretch}{0.93}
\caption{Native containment controls of agent frameworks and protocols, from official documentation and released SDKs. None tracks or acts on downstream consumers of a faulty output.}
\label{tab:framework-controls}\addtocounter{table}{-1}
\begin{tabular}{@{}>{\raggedright}p{0.135\textwidth}>{\raggedright}p{0.20\textwidth}>{\raggedright}p{0.21\textwidth}>{\raggedright}p{0.23\textwidth}>{\raggedright\arraybackslash}p{0.14\textwidth}@{}}
\toprule
Framework & Stop or cancel & Fence before an effect & Effects already under way & Rollback or replay \\
\midrule
A2A 1.0.x, a2a-sdk 1.2.3~\citep{a2a2025} & Cancel one task; no cancel-all (the operator lists and cancels tasks per server) & None for the operator; an agent may reject a task itself & Cancellation is attempted, ``but success is not guaranteed'' & None \\
LangGraph 1.2.16~\citep{langgraph2026} & Cancel a run (interrupt), also in bulk & \texttt{interrupt()} or interrupt-before/after per node; edit the state, then resume & Stops the run's worker; a tool call may have started & Rollback of graph state; time travel re-runs every later node \\
AutoGen 0.7.5~\citep{autogen2026} & Cancellation token per run; termination conditions after the current turn; stop the runtime & Approval of code execution; an intervention handler can drop a message before delivery & Async tools are cancelled and code executors killed; sync tools and the current message finish & Save and load of state \\
CrewAI 1.15.28~\citep{crewai2026} & No stop or cancel; time and iteration limits & Before-tool-call hook (other hook errors fail open) & Guardrails and human review run after execution & Checkpoints; replay re-runs every later task \\
OpenAI Agents SDK 0.24.0~\citep{openaiagents2026} & Cancel a streamed run, now or after the turn; turn limit & Tool approval; tool input guardrail (function tools only); blocking input guardrail & Async tools are cancelled at their next await; sync tools are not stopped & Restore of run state or session history \\
MCP 2026-07-28, SDK 2.3.0~\citep{mcp2026} & Cancel a request or a task & The host confirms tool calls; consent is advisory & The receiver may ignore a cancel and ``is not obligated to actually stop the work'' & None (stateless) \\
Semantic Kernel 1.45.0~\citep{semantickernel2026} & Cancel an orchestration (experimental); stop the runtime & Function invocation filter; an intervention handler can drop a message & Agents that already received messages continue; the current message finishes & Snapshot and restore of process state \\
\bottomrule
\end{tabular}
\end{table*}
\paragraph{Cost model and budget-aware pricing.}
The cut objective $c(a)$ is an additive proxy for collateral disruption; safety is independent of $c(a)$ (Theorem~5), whereas work retention among valid cuts depends on how well $c(a)$ reflects downstream compute.
Across the main grid, \OC achieves 100\% median kept work (zero lost healthy tasks in the vast majority of runs) because whenever fine-grained artifact or effect holds (priced at the base floor $c_0$) suffice to sever tainted paths, the solver selects them over broad agent pauses or revocations.
Because most runs cluster at $(c_0, 0\text{ lost compute})$, variance compression under the default live-work heuristic yields a modest overall Spearman rank correlation of \fact{F-T206} (95\% bootstrap CI \fact{F-T207} to \fact{F-T208} across \fact{F-T209} cut runs; \fact{F-T214} for action count).
In the few runs where healthy tasks are lost (\fact{F-T216} of \OC's \fact{F-T215} lost tasks), the loss arises from three structural factors in the live-work heuristic:
(i)~it prices a credential revocation only by currently running healthy invocations of that agent, so revoking a temporarily idle agent that has future planned tasks or pending hand-offs is priced at $c_0$;
(ii)~holding a scheduled effect whose issuer has already finished stops no live invocation and is priced at $c_0$; and
(iii)~conservatively chaining an agent's invocations within a request marks downstream healthy invocations as tainted and prices pausing them at $c_0$.
Pricing revocations instead by an agent's expected work within its remaining resource budget (bounded by planned tasks) raises the per-run Spearman correlation with lost compute in an offline check to \fact{F-T229} (\fact{F-T230} on \OC), cleanly separating zero-loss $c_0$ actions (\fact{F-T232} lost tasks across \fact{F-T231} revocations) from higher-disruption revocations (\fact{F-T234} of \fact{F-T235} lost tasks across \fact{F-T233} revocations priced $> 1$).

\paragraph{Coordinator ingestion scaling.}
\begin{table}[t]
\centering\footnotesize\setlength{\tabcolsep}{3pt}
\caption{63-agent scaling stress test across 16, 32, and 63 logical sites on two hosts (\OC, H12). Decision: alert to first decision, median (max). Confirmation: per-run median, median over runs (max 95th percentile). Ingest lag: per-run median journal lag from site to coordinator, range (max). $^\ast$Includes one deferred alert (\fact{F-E178}~s without it).}
\label{tab:scale}
\fitcol{%
\begin{tabular}{@{}lccc@{}}
\toprule
& 16 sites & 32 sites & 63 sites \\
\midrule
Runs & \fact{F-E096} & \fact{F-E120} & \fact{F-E144} \\
Runs reaching $\tone$ & \fact{F-E097} & \fact{F-E121} & \fact{F-E145} \\
Clean after the alert & \fact{F-E101} & \fact{F-E125} & \fact{F-E149} \\
Tainted effects after the alert & \fact{F-E102} & \fact{F-E126} & \fact{F-E150} \\
Decision latency (s) & \fact{F-E112} (\fact{F-E113}) & \fact{F-E136} (\fact{F-E137}) & \fact{F-E160} (\fact{F-E161}$^\ast$) \\
Confirmation latency (ms) & \fact{F-E114} (\fact{F-E115}) & \fact{F-E138} (\fact{F-E139}) & \fact{F-E162} (\fact{F-E163}) \\
Control messages & \fact{F-E116} & \fact{F-E140} & \fact{F-E164} \\
Ingest lag (s) & \fact{F-E117}--\fact{F-E118} (\fact{F-E119}) & \fact{F-E141}--\fact{F-E142} (\fact{F-E143}) & \fact{F-E165}--\fact{F-E166} (\fact{F-E167}) \\
Kept work (mean) & \fact{F-E109} & \fact{F-E133} & \fact{F-E157} \\
\bottomrule
\end{tabular}}
\end{table}
Across our three-host workflow and smart-home grids (\S5.2--\S5.4), \FB converges with sub-second $\tone$ at native latency (\fact{F-T387} 90th percentile on the main grid) and zero post-$\tone$ safety leaks.
To isolate the scaling bottleneck of a single-process SQLite coordinator, we stress-tested a 63-agent workflow across 16, 32, and 63 logical site daemons on two hosts (\cref{tab:scale}).
Even at 63 sites, laminar max-flow solving (0.022--0.031~ms) and per-site action confirmation (\fact{F-E114}--\fact{F-E162}~ms median) remain fast, pinpointing centralized single-process SQLite journal ingestion as the sole bottleneck (median ingest lag \fact{F-E020} to \fact{F-E021} at 16 sites, \fact{F-E026} to \fact{F-E027} at 32 sites, and \fact{F-E049} to \fact{F-E050} at 63 sites).
Under this overload, \FB's fail-closed contract holds---the controller marks un-ingested alerts as \textsc{indeterminate}, never falsely claims $\tone$ (\fact{F-E169} of \fact{F-E168} runs), and releases \fact{F-E173} holds prematurely---while effects that commit before their upstream journals arrive cannot be intercepted (\fact{F-E171} of \fact{F-E168} runs clean after the alert), motivating sharded journal ingestion for larger clusters.

\paragraph{Multi-agent coding collaboration case study.}
We additionally evaluated \FB in observe mode on a live multi-agent document-revision workflow with three third-party CLI coding agents: OpenAI Codex and Google Antigravity CLI as reviewers and Claude Code as author, coordinating through an append-only board and a versioned document store with gateway-mediated publication as the protected sink.
When a lagging store replica served a stale document version to a reviewer three times, the controller opened \fact{F-L109} incidents and in all \fact{F-L110} cases selected a single hold on the pending board publication (\fact{F-L111} decision latency, \fact{F-L112} laminar max-flow solve, \fact{F-L113} ahead of the publication deadline), with runtime mediation overhead of \fact{F-L118} of reviewer CLI time and \fact{F-L119} of author CLI time (median hop \fact{F-L120}).
The case study also highlights two operational boundaries that motivate future work: because provenance tracks data exposure rather than semantic influence (\fact{F-L116} of \fact{F-L115} board entries marked tainted from \fact{F-L117} stale-text writes) and incidents do not automatically clear when an agent reads a tainted artifact to repair it (\fact{F-L104} re-solves during repair), practical human-in-the-loop workflows benefit from explicit post-incident endorsement to declassify repaired artifacts.

\paragraph{Prototype scope.}
Consistent with Theorem~5 (\cref{tab:scope}), \FB contains external side effects online rather than rolling back internal agent memory: in \HomeSvc, \OC blocks all tainted lamp actuations and answers $\ge 96\%$ of user requests (with \fact{F-S109} correct replies and \fact{F-S107} wrong replies in simulation when tainted replies are priced at zero so the schedule read is blocked, \cref{app:sim}), whereas once the assistant's shared memory has ingested the corrupted schedule (\fact{F-T249} correct replies in late-alert runs, Table~1), restoring uncontaminated text replies requires pairing \FB with orthogonal memory or checkpoint rollback. Enforced containment covers channels mediated through \RT's site daemons and gateway.

%% file: sections/app-e-frameworks.tex
\cref{tab:framework-controls} summarizes what each framework offers an operator.
Beyond stopping everything, controls act on one call, message, task, run or handoff edge, and none finds the consumers of a bad output: CrewAI's task context declares inputs, and traces and checkpoint histories exist, but no control uses them for that.
Fences before an effect exist only in process, as per-call hooks, approvals and per-node interrupts; none sits at the actuator.
Stopping work already under way is mostly cooperative; only killing a process (an AutoGen executor, an MCP server) is forced.
Recovery relies on checkpoint rollback, replay, or state restore, which re-executes subsequent nodes rather than containing in-flight downstream side effects online.

%% file: main.bbl
\begin{thebibliography}{45}
\providecommand{\natexlab}[1]{#1}
\providecommand{\url}[1]{\texttt{#1}}
\expandafter\ifx\csname urlstyle\endcsname\relax
  \providecommand{\doi}[1]{doi: #1}\else
  \providecommand{\doi}{doi: \begingroup \urlstyle{rm}\Url}\fi

\bibitem[{A2A Project}(2026)]{a2a2025}
{A2A Project}.
\newblock {Agent2Agent (A2A)} protocol specification and {Python} {SDK}
  (a2a-sdk 1.2.3).
\newblock \url{https://a2a-protocol.org}, 2026.

\bibitem[Birgisson et~al.(2014)Birgisson, Politz, Erlingsson, Taly, Vrable, and
  Lentczner]{birgisson2014macaroons}
Birgisson, A., Politz, J.~G., Erlingsson, {\'U}., Taly, A., Vrable, M., and
  Lentczner, M.
\newblock Macaroons: Cookies with contextual caveats for decentralized
  authorization in the cloud.
\newblock In \emph{Network and Distributed System Security Symposium (NDSS)},
  2014.

\bibitem[Cemri et~al.(2025)Cemri, Pan, Yang, Agrawal, Chopra, Tiwari, Keutzer,
  Parameswaran, Klein, Ramchandran, Zaharia, Gonzalez, and
  Stoica]{cemri2025mast}
Cemri, M., Pan, M.~Z., Yang, S., Agrawal, L.~A., Chopra, B., Tiwari, R.,
  Keutzer, K., Parameswaran, A., Klein, D., Ramchandran, K., Zaharia, M.,
  Gonzalez, J.~E., and Stoica, I.
\newblock Why do multi-agent {LLM} systems fail?
\newblock In \emph{Advances in Neural Information Processing Systems
  (NeurIPS)}, 2025.

\bibitem[Costa et~al.(2025)Costa, K{\"o}pf, Kolluri, Paverd, Russinovich,
  Salem, Tople, Wutschitz, and Zanella-B{\'e}guelin]{costa2025fides}
Costa, M., K{\"o}pf, B., Kolluri, A., Paverd, A., Russinovich, M., Salem, A.,
  Tople, S., Wutschitz, L., and Zanella-B{\'e}guelin, S.
\newblock Securing {AI} agents with information-flow control, 2025.
\newblock URL \url{https://arxiv.org/abs/2505.23643}.
\newblock arXiv:2505.23643.

\bibitem[Coudert et~al.(2007)Coudert, Datta, P{\'e}rennes, Rivano, and
  Voge]{coudert2007srlg}
Coudert, D., Datta, P., P{\'e}rennes, S., Rivano, H., and Voge, M.-E.
\newblock Shared risk resource group: Complexity and approximability issues.
\newblock \emph{Parallel Processing Letters}, 17\penalty0 (2):\penalty0
  169--184, 2007.

\bibitem[{CrewAI}(2026)]{crewai2026}
{CrewAI}.
\newblock {CrewAI} 1.15.28.
\newblock \url{https://docs.crewai.com}, 2026.

\bibitem[Debenedetti et~al.(2025)Debenedetti, Shumailov, Fan, Hayes, Carlini,
  Fabian, Kern, Shi, Terzis, and Tram{\`e}r]{debenedetti2025camel}
Debenedetti, E., Shumailov, I., Fan, T., Hayes, J., Carlini, N., Fabian, D.,
  Kern, C., Shi, C., Terzis, A., and Tram{\`e}r, F.
\newblock Defeating prompt injections by design, 2025.
\newblock URL \url{https://arxiv.org/abs/2503.18813}.
\newblock arXiv:2503.18813.

\bibitem[Deochake(2026)]{deochake2026hbhc}
Deochake, S.
\newblock Heartbeat-bound hierarchical credentials: Cryptographic revocation
  for {AI} agent swarms, 2026.
\newblock URL \url{https://arxiv.org/abs/2605.20704}.
\newblock arXiv:2605.20704.

\bibitem[Dinur \& Steurer(2014)Dinur and Steurer]{dinur2014parallel}
Dinur, I. and Steurer, D.
\newblock Analytical approach to parallel repetition.
\newblock In \emph{Proceedings of the 46th ACM Symposium on Theory of Computing
  (STOC)}, 2014.

\bibitem[Feige(1998)]{feige1998threshold}
Feige, U.
\newblock A threshold of $\ln n$ for approximating set cover.
\newblock \emph{Journal of the ACM}, 45\penalty0 (4):\penalty0 634--652, 1998.

\bibitem[Fellows et~al.(2010)Fellows, Guo, and Kanj]{fellows2010label}
Fellows, M.~R., Guo, J., and Kanj, I.
\newblock The parameterized complexity of some minimum label problems.
\newblock \emph{Journal of Computer and System Sciences}, 2010.
\newblock \doi{10.1016/j.jcss.2010.02.012}.

\bibitem[Forough et~al.(2023)Forough, Bhuyan, and Elmroth]{forough2023anomaly}
Forough, J., Bhuyan, M., and Elmroth, E.
\newblock Anomaly detection and resolution on the edge: Solutions and future
  directions.
\newblock In \emph{2023 IEEE International Conference on Service-Oriented
  System Engineering (SOSE)}, pp.\  227--238. IEEE, 2023.
\newblock \doi{10.1109/SOSE58276.2023.00034}.

\bibitem[Forough et~al.(2026)Forough, Kogias, and Haddadi]{forough2026agents}
Forough, J., Kogias, M., and Haddadi, H.
\newblock When agents handle secrets: A survey of confidential computing for
  agentic {AI}, 2026.
\newblock URL \url{https://arxiv.org/abs/2605.03213}.
\newblock arXiv:2605.03213.

\bibitem[Goel et~al.(2005)Goel, Po, Farhadi, Li, and de~Lara]{goel2005taser}
Goel, A., Po, K., Farhadi, K., Li, Z., and de~Lara, E.
\newblock The {Taser} intrusion recovery system.
\newblock In \emph{Proceedings of the 20th ACM Symposium on Operating Systems
  Principles (SOSP)}, 2005.

\bibitem[Greenblatt et~al.(2026)Greenblatt, Cotra, and
  Wijk]{metr2026hfinvestigation}
Greenblatt, R., Cotra, A., and Wijk, H.
\newblock Brief independent investigation of agents' behavior, reasoning and
  collaboration in the {OpenAI} / {Hugging Face} hacking incident, 2026.
\newblock URL
  \url{https://metr.org/blog/2026-08-26-openai-hugging-face-incident-investigation/}.
\newblock METR blog post, 26 August 2026.

\bibitem[{Home Assistant}(2026)]{homeassistant2026}
{Home Assistant}.
\newblock {Home Assistant}: Open source home automation.
\newblock \url{https://www.home-assistant.io/}, 2026.
\newblock Accessed: 2026-10-30.

\bibitem[Huang et~al.(2025)Huang, Zhou, Jin, Zhou, Chen, Wang, Yuan, Lyu, and
  Sap]{huang2025resilience}
Huang, J.-T., Zhou, J., Jin, T., Zhou, X., Chen, Z., Wang, W., Yuan, Y., Lyu,
  M.~R., and Sap, M.
\newblock On the resilience of {LLM}-based multi-agent collaboration with
  faulty agents.
\newblock In \emph{Proceedings of the 42nd International Conference on Machine
  Learning (ICML)}, pp.\  26202--26226, 2025.

\bibitem[Huangfu \& Hall(2018)Huangfu and Hall]{huangfu2018highs}
Huangfu, Q. and Hall, J. A.~J.
\newblock Parallelizing the dual revised simplex method.
\newblock \emph{Mathematical Programming Computation}, 10\penalty0
  (1):\penalty0 119--142, 2018.

\bibitem[Jha et~al.(2002)Jha, Sheyner, and Wing]{jha2002two}
Jha, S., Sheyner, O., and Wing, J.
\newblock Two formal analyses of attack graphs.
\newblock In \emph{Proceedings of the 15th IEEE Computer Security Foundations
  Workshop (CSFW)}, 2002.

\bibitem[Khan(2026)]{stopmeansstop2026}
Khan, S.
\newblock Stop means stop: Measuring and repairing the enforcement gap in
  agent-framework control primitives, 2026.
\newblock URL \url{https://arxiv.org/abs/2607.14166}.
\newblock arXiv:2607.14166.

\bibitem[Kim et~al.(2010)Kim, Wang, Zeldovich, and Kaashoek]{kim2010retro}
Kim, T., Wang, X., Zeldovich, N., and Kaashoek, M.~F.
\newblock Intrusion recovery using selective re-execution.
\newblock In \emph{9th USENIX Symposium on Operating Systems Design and
  Implementation (OSDI)}, pp.\  89--104, 2010.

\bibitem[King \& Chen(2003)King and Chen]{king2003backtracking}
King, S.~T. and Chen, P.~M.
\newblock Backtracking intrusions.
\newblock In \emph{Proceedings of the 19th ACM Symposium on Operating Systems
  Principles (SOSP)}, 2003.

\bibitem[{LangChain}(2026)]{langgraph2026}
{LangChain}.
\newblock {LangGraph} 1.2.16.
\newblock \url{https://github.com/langchain-ai/langgraph}, 2026.

\bibitem[Li et~al.(2026)Li, Hu, Xu, Ozbas, Liu, Van, Li, Zhou, Alizadeh, Zhang,
  Sriramadhesikan, and Zhang]{li2026adr}
Li, C., Hu, P., Xu, J., Ozbas, B., Liu, O., Van, C., Li, M., Zhou, W.,
  Alizadeh, M., Zhang, P., Sriramadhesikan, K., and Zhang, M.
\newblock {ADR}: An agentic detection system for enterprise agentic {AI}
  security.
\newblock In \emph{Proceedings of Machine Learning and Systems}, volume~8, pp.\
   792--806, 2026.
\newblock URL
  \url{https://proceedings.mlsys.org/paper_files/paper/2026/hash/f03cb785864596fa5901f1359d23fd81-Abstract-Conference.html}.

\bibitem[Liu et~al.(2026)Liu, Yu, and Jiang]{liu2026vera}
Liu, L., Yu, H., and Jiang, X.
\newblock {VERA}: Authority-preserving edge revocation for federated {AI}-agent
  workflows, 2026.
\newblock URL \url{https://arxiv.org/abs/2608.30091}.
\newblock arXiv:2608.30091.

\bibitem[{Microsoft}(2026{\natexlab{a}})]{autogen2026}
{Microsoft}.
\newblock {AutoGen} 0.7.5 (agentchat, core, ext).
\newblock \url{https://github.com/microsoft/autogen}, 2026{\natexlab{a}}.

\bibitem[{Microsoft}(2026{\natexlab{b}})]{semantickernel2026}
{Microsoft}.
\newblock {Semantic Kernel} 1.45.0.
\newblock \url{https://learn.microsoft.com/en-us/semantic-kernel/},
  2026{\natexlab{b}}.

\bibitem[{Model Context Protocol}(2026)]{mcp2026}
{Model Context Protocol}.
\newblock {Model Context Protocol} specification, revision 2026-07-28, and
  {Python SDK} 2.3.0.
\newblock \url{https://modelcontextprotocol.io/specification/2026-07-28}, 2026.

\bibitem[Mohammadi et~al.(2026)Mohammadi, Potamitis, Klein, Arora, and
  Bindschaedler]{mohammadi2026atomix}
Mohammadi, B., Potamitis, N., Klein, L., Arora, A., and Bindschaedler, L.
\newblock Atomix: Timely, transactional tool use for reliable agentic
  workflows, 2026.
\newblock URL \url{https://arxiv.org/abs/2602.14849}.
\newblock arXiv:2602.14849.

\bibitem[Muruaga(2026)]{muruaga2026bounded}
Muruaga, X.
\newblock Bounded agents: Delegation security for multi-agent {AI} systems,
  2026.
\newblock URL \url{https://arxiv.org/abs/2608.15888}.
\newblock arXiv:2608.15888.

\bibitem[{NATS Authors}(2026)]{nats2026}
{NATS Authors}.
\newblock {NATS} and {JetStream}.
\newblock \url{https://nats.io}, 2026.

\bibitem[{OpenAI}(2026{\natexlab{a}})]{openai2026hfreport}
{OpenAI}.
\newblock {OpenAI--Hugging Face} incident technical report.
\newblock Technical report, OpenAI, 2026{\natexlab{a}}.
\newblock URL
  \url{https://cdn.openai.com/pdf/67869394-cb91-4c12-888c-5cbd85c7814c/OpenAI-Hugging-Face%20Incident-Technical-Report.pdf}.

\bibitem[{OpenAI}(2026{\natexlab{b}})]{openaiagents2026}
{OpenAI}.
\newblock {OpenAI Agents SDK} 0.24.0.
\newblock \url{https://openai.github.io/openai-agents-python/},
  2026{\natexlab{b}}.

\bibitem[Patil et~al.(2024)Patil, Zhang, Fang, {Noppapon C.}, Huang, Hao,
  Casado, Gonzalez, Popa, and Stoica]{patil2024goex}
Patil, S.~G., Zhang, T., Fang, V., {Noppapon C.}, Huang, R., Hao, A., Casado,
  M., Gonzalez, J.~E., Popa, R.~A., and Stoica, I.
\newblock {GoEX}: Perspectives and designs towards a runtime for autonomous
  {LLM} applications, 2024.
\newblock URL \url{https://arxiv.org/abs/2404.06921}.
\newblock arXiv:2404.06921.

\bibitem[Prakash(2026)]{prakash2026aip}
Prakash, S.
\newblock {AIP}: Agent identity protocol for verifiable delegation across {MCP}
  and {A2A}, 2026.
\newblock URL \url{https://arxiv.org/abs/2603.24775}.
\newblock arXiv:2603.24775.

\bibitem[Saltzer \& Schroeder(1975)Saltzer and
  Schroeder]{saltzer1975protection}
Saltzer, J.~H. and Schroeder, M.~D.
\newblock The protection of information in computer systems.
\newblock \emph{Proceedings of the IEEE}, 63\penalty0 (9):\penalty0 1278--1308,
  1975.

\bibitem[Syros et~al.(2026)Syros, Suri, Ginesin, Nita-Rotaru, and
  Oprea]{syros2026saga}
Syros, G., Suri, A., Ginesin, J., Nita-Rotaru, C., and Oprea, A.
\newblock {SAGA}: A security architecture for governing {AI} agentic systems.
\newblock In \emph{Network and Distributed System Security Symposium (NDSS)},
  2026.

\bibitem[Wang et~al.(2025)Wang, Zhang, Yu, Wan, Meng, Guo, Wang, and
  Wang]{wang2025gsafeguard}
Wang, S., Zhang, G., Yu, M., Wan, G., Meng, F., Guo, C., Wang, K., and Wang, Y.
\newblock {G-Safeguard}: A topology-guided security lens and treatment on
  {LLM}-based multi-agent systems.
\newblock In \emph{Proceedings of the 63rd Annual Meeting of the Association
  for Computational Linguistics (ACL)}, pp.\  7261--7276, 2025.
\newblock \doi{10.18653/v1/2025.acl-long.359}.

\bibitem[Wang et~al.(2026)Wang, Yu, Zhang, Duan, Zheng, Wu, Shi, and
  Cai]{yu2026rollback}
Wang, Y., Yu, C., Zhang, J., Duan, Y., Zheng, M., Wu, Z., Shi, K., and Cai, T.
\newblock From faulty memories to corrected actions: Dependency-guided rollback
  repair for memory-augmented agents, 2026.
\newblock URL \url{https://arxiv.org/abs/2608.10502}.
\newblock arXiv:2608.10502.

\bibitem[Xie et~al.(2026)Xie, Zhu, Zhang, Zhu, Ye, Qi, Chen, and
  Zhou]{xie2026spark}
Xie, Y., Zhu, C., Zhang, X., Zhu, T., Ye, D., Qi, M., Chen, H., and Zhou, W.
\newblock From spark to fire: Modeling and mitigating error cascades in
  {LLM}-based multi-agent collaboration, 2026.
\newblock URL \url{https://arxiv.org/abs/2603.04474}.
\newblock arXiv:2603.04474.

\bibitem[Zhan et~al.(2026{\natexlab{a}})Zhan, Li, Zhang, and
  Haddadi]{zhan2026hearthnet}
Zhan, Z., Li, K., Zhang, Y., and Haddadi, H.
\newblock {HearthNet}: Edge multi-agent orchestration for smart homes,
  2026{\natexlab{a}}.
\newblock URL \url{https://arxiv.org/abs/2604.09618}.
\newblock arXiv:2604.09618.

\bibitem[Zhan et~al.(2026{\natexlab{b}})Zhan, Zhang, and
  Haddadi]{zhan2026edgecitadel}
Zhan, Z., Zhang, Y., and Haddadi, H.
\newblock Poster: {EdgeCitadel} - {Hybrid} {NATS-MQTT} orchestration for edge
  multi-agent systems.
\newblock In \emph{Companion of the 2026 ACM International Conference on Mobile
  Systems, Applications, and Services}, MobiSys Companion '26, pp.\  43--44,
  New York, NY, USA, 2026{\natexlab{b}}. Association for Computing Machinery.
\newblock ISBN 9798400727115.
\newblock \doi{10.1145/3812835.3814871}.
\newblock URL \url{https://doi.org/10.1145/3812835.3814871}.

\bibitem[Zhang et~al.(2011)Zhang, Cai, Tang, and Zhao]{zhang2011label}
Zhang, P., Cai, J.-Y., Tang, L.-Q., and Zhao, W.-B.
\newblock Approximation and hardness results for label cut and related
  problems.
\newblock \emph{Journal of Combinatorial Optimization}, 21\penalty0
  (2):\penalty0 192--208, 2011.

\bibitem[Zhou et~al.(2025)Zhou, Wang, and Yang]{zhou2025guardian}
Zhou, J., Wang, L., and Yang, X.
\newblock {GUARDIAN}: Safeguarding {LLM} multi-agent collaborations with
  temporal graph modeling.
\newblock In \emph{Advances in Neural Information Processing Systems
  (NeurIPS)}, 2025.

\bibitem[Zhu \& Wang(2026)Zhu and Wang]{zhu2026rsq}
Zhu, G. and Wang, C.
\newblock Authorization revocation for long-running {AI} agents: Root-scoped
  quiescence under delegation and asynchronous execution, 2026.
\newblock URL \url{https://arxiv.org/abs/2609.21284}.
\newblock arXiv:2609.21284.

\end{thebibliography}
